\documentclass{article}
\usepackage{amsmath}
\usepackage{bm}
\usepackage{stmaryrd}
\usepackage{pifont}
\usepackage{tablefootnote}
\usepackage{bbm}

\PassOptionsToPackage{numbers}{natbib}
\usepackage[preprint]{neurips_2026}

\usepackage[utf8]{inputenc} 
\usepackage[T1]{fontenc}    
\usepackage{hyperref}       
\usepackage{url}            
\usepackage{booktabs}       
\usepackage{amsfonts}       
\usepackage{nicefrac}       
\usepackage{microtype}      
\usepackage{xcolor}         
\usepackage{graphicx}
\usepackage{algorithm}
\usepackage{algpseudocode}
\usepackage{subcaption}

\title{From Isolated feature to Orbits:\\Discovering Music Concepts via Multi-SAE Alignment}

\author{%
  Liwei~Lin \\
  Computer Science Department, NYU Shanghai\\
  Machine Learning Department, MBZUAI \\
  \texttt{liwei.lin@nyu.edu} \\
  \And
  Gus Xia \\
  Machine Learning Department, MBZUAI \\
  \texttt{Gus.Xia@mbzuai.ac.ae} \\
}

\begin{document}

\maketitle

\begin{abstract}

How can we understand what a music foundation model has learned \textit{internally}? Most interpretability approaches, such as probing and Sparse Autoencoders (SAEs), focus on identifying individual features with minimal structural assumptions. We argue that many concepts are better understood as \textit{structured relations} rather than isolated features. This is especially prominent in music, where tonal structures are organized in the space of pitch and time. For example, concepts such as chords or keys are naturally expressed as structured sets (e.g., the 12 transpositions of a chord or the diatonic system within a key), rather than isolated features. In this study, \textbf{we shift from feature identification to structure-based analysis}, asking whether the learned inner representations of music foundation model emerge as organized structures over features. To this end, we introduce a framework that uses pitch transposition as an inductive bias to induce ordered orbits via multi-view SAE alignment. Concretely, we generate pitch-shifted input pairs and align their SAE representations to discover structured groups of pitch-related features. Experimental results show that this approach recovers orbit structures corresponding to chords, keys, and melodic patterns across two state-of-the-art music foundation models, while requiring only minimal grounding (e.g., a few anchor examples) to interpret entire concept families. 
\end{abstract}

\section{Introduction}

Pretrained music audio foundation models~\cite{copet2023simple, yizhi2023mert, ghosh2025audio, wu2025clamp, won2024foundation, zhu2025muq} learn representations that support a wide range of downstream tasks~\cite{castellon2021codified}, yet what musical information these representations encode and how such information is organized in the architecture or activations remain difficult to interpret. 

Existing work has approached the interpretability of music audio representations from several directions. For self-supervised music audio models, their representation quality can be evaluated by downstream benchmarks across music understanding tasks~\cite{yuan2023marble}, providing indirect evidence of what musical information their representations may capture. Probing and intervention studies~\cite{castellon2021codified, koo2024understanding, wei2024music, papaioannou2025universal} have tested whether music audio generative models encode and use symbolic musical concepts. Recently, sparse autoencoders (SAEs)~\cite{makhzani2013k} have emerged as a promising tool for decomposing neural activations into sparse, potentially interpretable features~\cite{gao2024scaling}. In music, recent work has applied SAEs to generative models~\cite{singh2025discovering, aparin2026audiosae}, identifying features that can be linked to musical concepts. 

However, these approaches mostly stop at individual concept-feature correspondences. Probing detects predefined concepts using labeled examples, while SAE-based analyses typically assign meanings post hoc by matching individual features to external labels. This is limiting in the music domain, where concepts are often organized in relational structures. Chords, for example, form a cyclic ordered set under transposition along the pitch-class cycle\footnote{For readers unfamiliar with foundational music theory concepts (e.g., pitch classes, chords, pitch transposition, and the Circle of Fifths), we provide a brief primer in Appendix~\ref{sec:music_theory_primer}.}. If we can identify groups of related compound features with their relation structure such as 
\[
\textit{C major chord}\rightarrow \textit{C\(\sharp\) major chord}\rightarrow \textit{D major chord}\rightarrow \dots,
\]
then interpreting a small number of features in the group can help infer the semantics of the remaining features through that structure. 
Moreover, the invariant shared by all members of the group constitutes an abstract concept, in this case, \textit{major chord}. 
Our goal is therefore to recover not only learned features, but also the relational structure among them. 

\begin{figure}[t]
  \centering
\includegraphics[width=0.26\textwidth,
  trim=4.4cm 19cm 10cm 2.5cm,
  clip
  ]{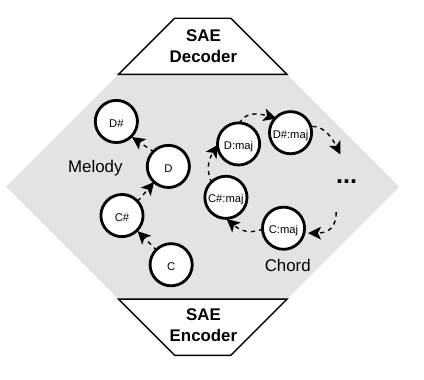}
  \hfill
  \includegraphics[
  width=.73\textwidth,
  trim=2.2cm 20.2cm 2.8cm 2.6cm,
  clip
  ]{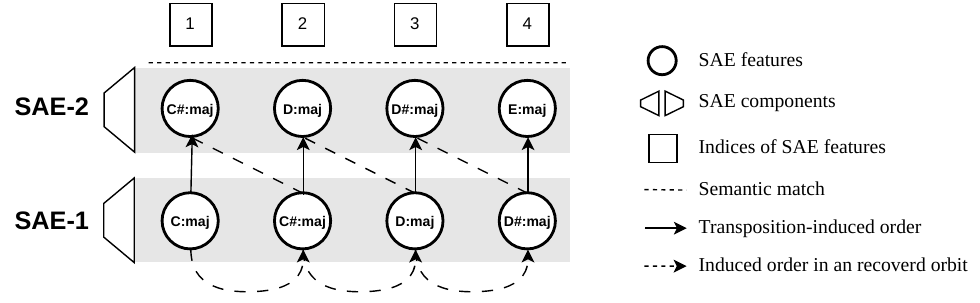}
  
\caption{
Recovered orbits over SAE features (left) and orbit recovery via two aligned SAEs (right). 
The \(i\)-th feature in SAE-2 is the one-semitone-up version of the \(i\)-th feature in SAE-1. 
We pair each feature $i$ in SAE-2 with its semantically closest counterpart in SAE-1, measured by decoder vector similarity.
This semantically matched feature in SAE-1 is defined to be the \textit{successor} of the original \(i\)-th feature. Iteratively applying this successor relation recovers ordered pitch orbits.
}

  \label{fig:intro}
\end{figure}

Towards this goal, we focus on pitch transposition as a fundamental transformation in music: many pitch-specific music concepts transform equivariantly under transposition. For example, transposing a \textit{C major chord} upward by one semitone yields a \textit{C\(\sharp\) major chord}, where the pitch concept component shifts but the chord quality concept component does not. We utilize this property of pitch transposition~\cite{omt2_transposition} to induce ordered \textbf{orbits}~\cite{chen2020group} over SAE features, thereby recovering pitch-related concepts and their relational structure, as illustrated in Figure~\ref{fig:intro} (left).

Our method constructs multiple pitch-shifted views of each sample, trains a dedicated SAE for each shift, and encourages views of the same sample to activate the same feature indices across SAEs via an alignment training strategy. Consequently, a feature in one SAE is expected to represent the one-semitone transposition of the same-index feature in the next SAE, e.g., from \textit{C major chord} to \textit{C\(\sharp\) major chord}. As illustrated in Figure~\ref{fig:intro} (right), after training, we recover orbits by turning this cross-SAE transposition relation into a successor relation of feature indices within a single SAE. Once an orbit and its order are recovered, a few labeled anchors can propagate semantics to the remaining features, while the invariant shared across the orbit reveals the underlying abstract concept, e.g., \textit{major chord}. In this way, interpretation shifts from isolated features to structured musical concepts.

Our contributions are as follows:\footnote{Source code is available at \url{https://github.com/Kikyo-16/sae-alignment}.}
\begin{itemize}
    \item Conceptually, we extend SAE interpretability from the identification of individual features to structure-based analysis. This shift enables the interpretation of higher-level concepts, even when such concepts do not emerge as single SAE features.
    
    \item Methodologically, we introduce a music-informed self-supervised framework for recovering pitch-related feature orbits through multi-SAE alignment. The method is highly label-efficient: instead of requiring extensive labeled data and manual annotation of individual features during post-hoc analysis, it can use one or a few anchor examples to interpret an entire orbit of related features, and in some cases supports one-shot transfer to downstream understanding tasks.
    
    \item Empirically, we demonstrate the framework on two state-of-the-art music foundation models and recover musically meaningful orbits associated with chords, keys, and melodic structure. In particular, our experimental results provide new evidence on how these models encode key-related concepts. 
\end{itemize}

\section{Related Work}

\subsection{Interpretability of Music Foundation Models}

The rapid deployment of music foundation models has motivated a shift toward understanding their internal representational dynamics. Early efforts focused on linear probing, where supervised classifiers are trained on hidden activations to detect musical attributes such as genre, emotion, or pitch \cite{castellon2021codified, koo2024understanding, wei2024music, papaioannou2025universal}. While probing confirms that these models encode a wealth of musical information, it is fundamentally limited by polysemanticity, where individual neurons represent multiple unrelated concepts \cite{cunningham2023sparse}. Mechanistic interpretability aims to disentangle these representations using Sparse Autoencoders (SAEs) \cite{makhzani2013k}. Recent work \cite{singh2025discovering} has applied SAEs to generative music models like MusicGen, discovering interpretable features such as timbre, texture, and stylistic markers. 
However, these studies focus on individual features, overlooking the structured relations present in polysemantic music concepts.

\subsection{Evolution of SAE Architectures and Sparsity}

The quality of discovered features is highly sensitive to the sparsity mechanism of the SAE. Traditional \(L_1\)-regularized autoencoders often suffer from the ``shrinkage problem'', where feature magnitudes are suppressed during reconstruction \cite{gao2024scaling}. To address this, Top-K SAEs \cite{makhzani2013k} have been introduced, which use an activation function that preserves exactly $K$ latents, thereby improving the reconstruction-sparsity frontier. Our work employs the same Top-K mechanism.

\subsection{SAE Alignment and Relational Interpretability}

Our work is situated at the intersection of mechanistic interpretability and representational alignment. Recent frameworks like Universal Sparse Autoencoders (USAE) \cite{thasarathan2025universal} jointly train dictionaries across multiple models to discover universal concept spaces. In a complementary direction, representational similarity metrics have been used to match features between models post-hoc \cite{cunningham2023sparse}. We extend this by recovering entire orbits of pitch-related features through multi-view SAE alignment. This shifts the focus from identifying what a model has learned to recovering the relational structures that define its understanding of music.


\section{Methodology}

We induce ordered orbits over SAE features via pitch transposition. Specifically, we train multiple SAEs per music foundation model. Each SAE is trained on an adjacent pitch-shifted view of the dataset. We ensure that feature indices remain compatible across SAEs by synchronizing the top-K selection during SAE training. 
This treatment implies a zig-zag path through features across two SAEs, which is further compressed into a successor map in one SAE. Using this successor map as a generator then identifies orbits or partial orbits consisting of SAE features.


\subsection{Transposition-Induced Orbits of Music Concepts}
\label{sec:concept_orbits}

Let $\mathcal{C}$ denote the space of musical concepts, and let
$\mathsf{T}_{\mathcal{C}} : \mathcal{C} \rightarrow \mathcal{C}$ denote
transposition upward by one semitone.
Repeatedly applying $\mathsf{T}_{\mathcal{C}}$ to a concept
$c \in \mathcal{C}$ induces its orbit:
\begin{equation}
\label{eq:concept_orbit}
\operatorname{Orb}(c)
=
\left\{
\mathsf{T}_{\mathcal{C}}^{i}(c)
\mid
i \in \mathbb{Z}
\right\}
\subseteq \mathcal{C}.
\end{equation}

The behavior of a concept under pitch transposition gives rise to two cases.
A \emph{pitch-invariant concept}, such as instrumentation or a rhythmic
pattern, remains unchanged under transposition:
\begin{equation}
\label{eq:pitch_invariant_concept}
\mathsf{T}_{\mathcal{C}}(c)=c,
\qquad
\operatorname{Orb}(c)=\{c\}.
\end{equation}
Such a concept is a fixed point of pitch transposition, and its orbit is a
singleton.

In contrast, a \emph{pitch-specific concept}, such as a melody, chord, or key,
transforms equivariantly under pitch transposition.
Repeated application of the transposition operator produces a non-trivial
ordered orbit:
\begin{equation}
\label{eq:pitch_specific_orbit}
\mathsf{T}_{\mathcal{C}}(c)\neq c,
\qquad
c
\rightarrow
\mathsf{T}_{\mathcal{C}}(c)
\rightarrow
\mathsf{T}_{\mathcal{C}}^{2}(c)
\rightarrow
\cdots.
\end{equation}
For pitch-class concepts in twelve-tone equal temperament, this orbit is
typically cyclic with period twelve. For example,
\begin{equation}\small
\label{eq:major_chord_orbit}
\text{C major chord}
\rightarrow
\text{C$\sharp$ major chord}
\rightarrow
\text{D major chord}
\rightarrow
\cdots
\rightarrow
\text{B major chord}
\rightarrow
\text{C major chord}.
\end{equation}
Along this orbit, the pitch component shifts while the chord-quality
component remains invariant.
The invariant shared across the orbit constitutes the underlying abstract
concept, in this case, \emph{major chord}.

\subsection{Multi-SAE Alignment with Shared Top-$K$ Support}
\label{sec:multi_sae_alignment}

Before recovering transposition-induced orbits within a single SAE, we first
establish an index-wise correspondence across multiple SAEs trained on
pitch-shifted views of the same audio.

Let $\mathcal{A}$ denote the audio space, and let
$\mathsf{T}_{\mathcal{A}}$ denote upward transposition by one semitone.
Given an audio segment $a \in \mathcal{A}$, we construct $S$ adjacent pitch-shifted views:
\begin{equation}
\label{eq:pitch_shifted_audio_views}
a^{(s)}
=
\mathsf{T}_{\mathcal{A}}^{s}(a),
\qquad
s=0,\ldots,S-1.
\end{equation}
A frozen music foundation model $H$ extracts a sequence of frame-level
representations from each view, yielding a matrix
\begin{equation}
\label{eq:pitch_shifted_representations}
X^{(s)}
=
\begin{bmatrix}
x^{(s)}_1, x^{(s)}_2, \dots, x^{(s)}_L
\end{bmatrix}^\top
=
H\!\left(a^{(s)}\right)
\in
\mathbb{R}^{L\times d},
\end{equation}
where $L$ is the number of frames and $d$ is the representation
dimensionality, and $x^{(s)}_t \in \mathbb{R}^d$ denotes the representation at frame $t$.
Because pitch transposition preserves temporal alignment, representations at
the same frame position across the $S$ views form a group of corresponding
transformed examples.

For each pitch-shifted view $s$, we instantiate a Top-$K$ SAE whose latent
feature set is
\begin{equation}
\label{eq:sae_feature_set}
\mathcal{F}^{(s)}
=
\left\{
f_1^{(s)},\ldots,f_M^{(s)}
\right\},
\end{equation}
where all SAEs have the same latent dimensionality $M$, while their encoder and
decoder weights are learned independently.
They additionally share a learnable pre-encoder bias
$b\in\mathbb{R}^{d}$, which provides a common center for all pitch-shifted
representation spaces. For clarity, we omit the frame index $t$ below.
Given a representation $x^{(s)}\in\mathbb{R}^{d}$, the $s$-th SAE produces
the non-negative pre-sparsification activation
\begin{equation}
\label{eq:pre_sparse_activation}
u^{(s)}
=
\operatorname{ReLU}
\left(
W_{\mathrm{enc}}^{(s)}
\left(x^{(s)}-b\right)
\right)
\in
\mathbb{R}_{\geq 0}^{M}.
\end{equation}
If Top-$K$ selection were performed independently for each view, the $S$
SAEs could assign arbitrary latent indices to corresponding concepts.
We instead perform support selection jointly across all transformed views.

Because activation scales may differ across SAEs, we first normalize each
activation vector by its root-mean-square magnitude:
\begin{equation}
\label{eq:activation_rms}
\alpha^{(s)}
=
\sqrt{
\frac{1}{M}
\sum_{i=1}^{M}
\left(u_i^{(s)}\right)^2
},
\qquad
\bar{u}^{(s)}
=
\frac{u^{(s)}}{\max\!\left( \epsilon, \alpha^{(s)} \right)}.
\end{equation}
Here, $\epsilon>0$ is a small constant used for numerical stability; the same
$\epsilon$ is used for decoder normalization later.

We then compute an index-wise shared score by taking the maximum normalized
activation across the view dimension:
\begin{equation}
\label{eq:shared_activation_score}
u_i^{\mathrm{shared}}
=
\max_{s\in\{0,\ldots,S-1\}}
\bar{u}_i^{(s)},
\qquad
i=1,\ldots,M.
\end{equation}
The $K$ largest shared scores define a common binary support mask:
\begin{equation}
\label{eq:shared_topk_mask}
m
=
\operatorname{TopKMask}
\left(
u^{\mathrm{shared}},K
\right),
\qquad
m\in\{0,1\}^{M},
\qquad
\sum_{i=1}^{M}m_i=K.
\end{equation}
The same mask is applied to the original, unnormalized activations of every
SAE:
\begin{equation}
\label{eq:shared_sparse_activation}
z^{(s)}
=
m\odot u^{(s)},
\qquad
s=0,\ldots,S-1.
\end{equation}
The normalization in Equation~\ref{eq:activation_rms} therefore affects only
which indices enter the shared support; the view-specific activation
magnitudes used for reconstruction are preserved.

Each SAE reconstructs the representation associated with its own pitch view:
\begin{equation}
\label{eq:sae_reconstruction}
\widehat{x}^{(s)}
=
W_{\mathrm{dec}}^{(s)}z^{(s)}+b.
\end{equation}
The shared pre-encoder bias is subtracted before encoding and restored after decoding,
so all SAEs model deviations from the same learned center.
The $S$ SAEs are jointly optimized using the mean reconstruction loss
\begin{equation}
\label{eq:multi_sae_reconstruction_loss}
\mathcal{L}_{\mathrm{rec}}
=
\frac{1}{S}
\sum_{s=0}^{S-1}
\mathbb{E}_{a,t}
\left[
\left\|
\widehat{x}_{t}^{(s)}
-
x_{t}^{(s)}
\right\|_2^2
\right].
\end{equation}

After each parameter update, every decoder column is normalized to unit norm.
Writing
\begin{equation}
W_{\mathrm{dec}}^{(s)}
=
\left[
w_1^{(s)},\ldots,w_M^{(s)}
\right],
\end{equation}
we apply
\begin{equation}
\label{eq:decoder_column_normalization}
w_i^{(s)}
\leftarrow
\frac{w_i^{(s)}}
{\max\!\left(
\epsilon, 
\left\|w_i^{(s)}\right\|_2
\right)}.
\end{equation}
This prevents arbitrary scaling of decoder vectors and enables their
directions to be compared after training.

\paragraph{How shared Top-$K$ support encourages alignment.}

At initialization, the shared mask imposes only index-level co-selection:
all $S$ SAEs reconstruct each group of pitch-shifted representations using
the same latent indices.
During training, whenever index $i$ is selected, its encoder and decoder
parameters across the SAEs are jointly updated on temporally aligned examples
related by pitch transposition.
This couples the otherwise independent permutations of the SAE dictionaries
and encourages same-index features $f_i^{(s)}$ and $f_i^{(s+1)}$ to encode
adjacent transposed realizations of a shared concept family. However, this alignment remains an inductive bias rather than a guarantee,
since reconstruction ambiguity, polysemanticity, and imperfect pitch
equivariance may still produce mismatches.
We therefore use a simple algorithm to recover the learned structures after training.

\subsection{Recovering Rings and Sequences}
\label{sec:recovering_orbits}

After training, we recover transposition-induced structures directly from the
learned decoder dictionaries, without using input examples or labels. The
first step is to construct a successor relation within a reference SAE, such
that the successor of a feature represents its one-semitone-up counterpart.

We first consider two adjacent models, \textbf{SAE-0} and \textbf{SAE-1}, and take \textbf{SAE-0} as the reference space.
Shared-support alignment encourages the same feature index across the two SAEs
to encode adjacent transpositions. Starting from $f^{(0)}_i$, we therefore
first move to the same-index feature $f^{(1)}_i$, which approximates its
one-semitone-up realization, and then match it back to a feature in SAE $0$:
\begin{equation}
\label{eq:two_sae_successor_path}
f^{(0)}_i
\;\xrightarrow{\mathrm{alignment}}\;
f^{(1)}_i
\;\xrightarrow{\mathrm{semantic{\ }match}}\;
f^{(0)}_j.
\end{equation}
The semantic match is computed via two constraints. Using the unit-normalized decoder vectors, we define the confidence of the
candidate transition $f^{(0)}_i\rightarrow f^{(0)}_j$ as
\begin{equation}
\label{eq:two_sae_successor_score}
v^{(0)}_\uparrow(i,j)
=
\cos
\left(
w^{(1)}_i,
w^{(0)}_j
\right).
\end{equation}
The tentative successor of $f^{(0)}_i$ is the feature with the highest
confidence:
\begin{equation}
\label{eq:two_sae_successor}
j^\star(i)
=
\arg\max_j v^{(0)}_\uparrow(i,j).
\end{equation}
The transition is accepted only if
$v^{(0)}_\uparrow(i,j^\star(i))\geq\tau$; otherwise, the successor of
$f^{(0)}_i$ is left undefined.

When at least three SAEs are available, we take an interior SAE $s$,
where $0<s<S-1$, as the reference space. A candidate transition
$f^{(s)}_i\rightarrow f^{(s)}_j$ can then be evaluated from both neighboring
pitch shifts. We define
\begin{align}
\label{eq:bidirectional_scores}
v^{(s)}_\uparrow(i,j)
&=
\cos
\left(
w^{(s+1)}_i,
w^{(s)}_j
\right),
\\
v^{(s)}_\downarrow(i,j)
&=
\cos
\left(
w^{(s)}_i,
w^{(s-1)}_j
\right).
\end{align}
Here, $v^{(s)}_\uparrow(i,j)$ measures whether the upward-transposed source
matches the candidate destination, while $v^{(s)}_\downarrow(i,j)$ measures
whether the downward-transposed destination matches the source.

We select the destination using the mean of the two directional similarities:
\begin{equation}
\label{eq:multi_sae_successor}
j^\star(i)
=
\arg\max_j
\frac{
v^{(s)}_\uparrow(i,j)+v^{(s)}_\downarrow(i,j)
}{2}.
\end{equation}
For the selected destination, its confidence is given by the stronger of the
two directional matches:
\begin{equation}
\label{eq:multi_sae_successor_confidence}
v^{(s)}(i)
=
\max
\left\{
v^{(s)}_\uparrow\!\left(i,j^\star(i)\right),
v^{(s)}_\downarrow\!\left(i,j^\star(i)\right)
\right\}.
\end{equation}
Accept the transition if $v^{(s)}(i)\geq\tau$; otherwise, reject it. In our experiments, we use $S=3$ and take the middle
SAE, $s=1$, as the reference space, which is the minimum setup for double-direction evaluation.

The accepted transitions initially define a partial successor relation.
When multiple transitions point to the same feature,
we retain only the transition with the highest confidence and reject the
others. The remaining transitions define the feature-space transposition map
\begin{equation}
\label{eq:feature_transposition_map}
\mathsf{T}_{\mathcal{F}}
\left(
f^{(s)}_i
\right)
=
f^{(s)}_{j^\star(i)}.
\end{equation}
Repeated application of $\mathsf{T}_{\mathcal{F}}$ produces an ordered
trajectory
\begin{equation}
\label{eq:recovered_feature_trajectory}
f,\;
\mathsf{T}_{\mathcal{F}}(f),\;
\mathsf{T}_{\mathcal{F}}^{2}(f),\;
\ldots.
\end{equation}
If the trajectory returns to its starting feature after more than one step, it
forms a non-trivial ring. If a feature maps directly to itself, it forms a
fixed point corresponding to a candidate pitch-invariant concept. If the
trajectory terminates because its next successor is undefined, it forms an
open sequence. Once one or a
few features in a recovered structure are grounded using interpretable
examples, the successor relation iteratively interprets the remaining features.

\section{Experiments}
In this section, we describe the experimental setup and present the concepts discovered by our method in the foundation-model representations.

\subsection{Datasets and Configuration}
We use 326 and 164 songs from the training split of Slakh2100~\cite{manilow2019cutting} as our training and validation sets, respectively. Slakh2100 is a synthetic multi-instrument dataset synthesized using high-quality sound fonts. See Appendix~\ref{app:datasets} for usage details. We extract frame-level audio representations from two pretrained music foundation models: MuQ~\cite{zhu2025muq} and MusicFM~\cite{won2024foundation}. Both models comprise 12 layers with a hidden dimensionality of \(d=1024\). Their frame rate is 25 fps. During training, we use 32 30-second representations per batch to train our SAEs, with a learning rate of 1e-3.

\begin{figure}[t]
  \centering

  \begin{subfigure}[t]{0.32\textwidth}
    \centering
\includegraphics[width=\linewidth]{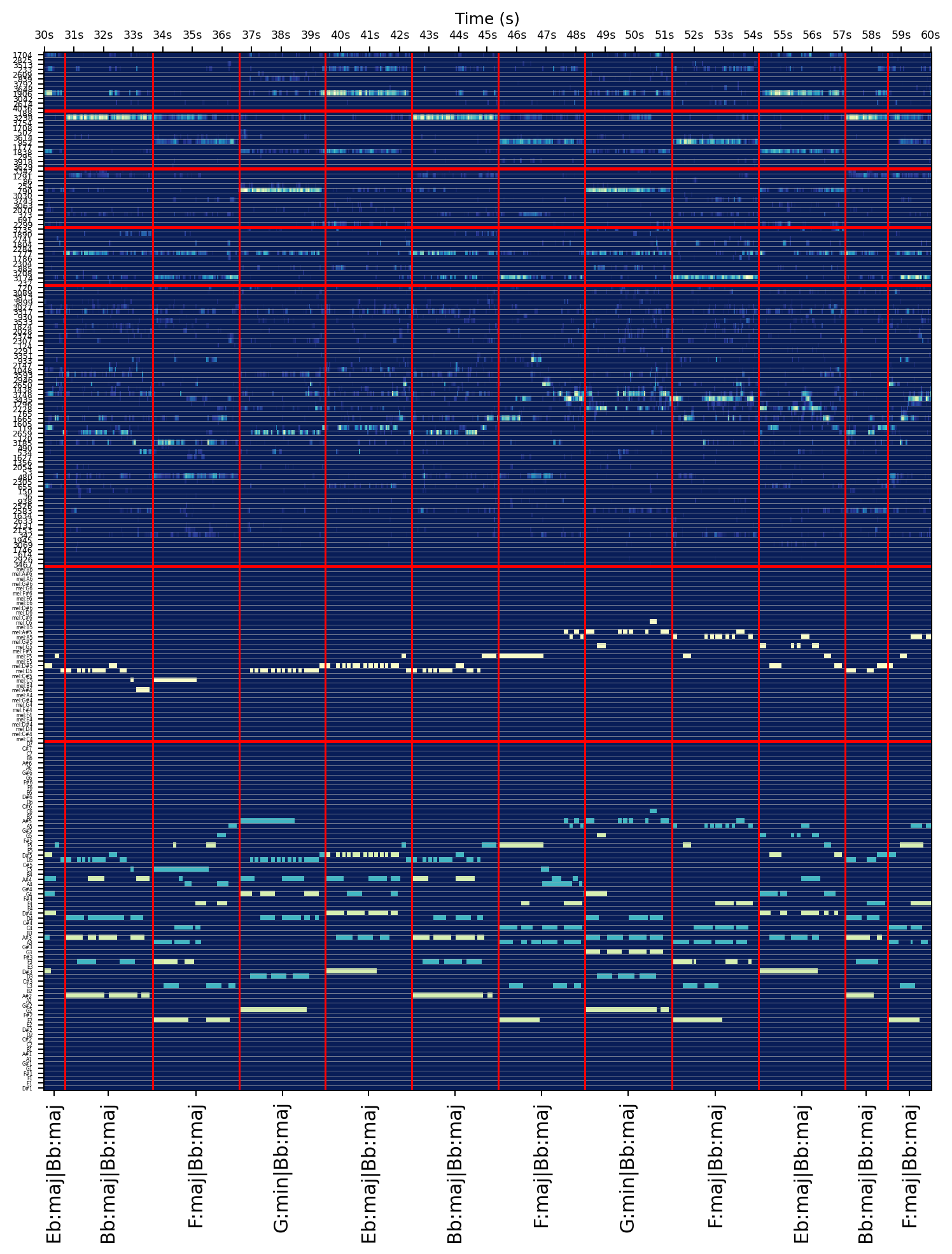}
    \caption{B$\flat$ major.}
    \label{fig:activation-muq-32}
  \end{subfigure}
  \hfill
  \begin{subfigure}[t]{0.32\textwidth}
    \centering
\includegraphics[width=\linewidth]{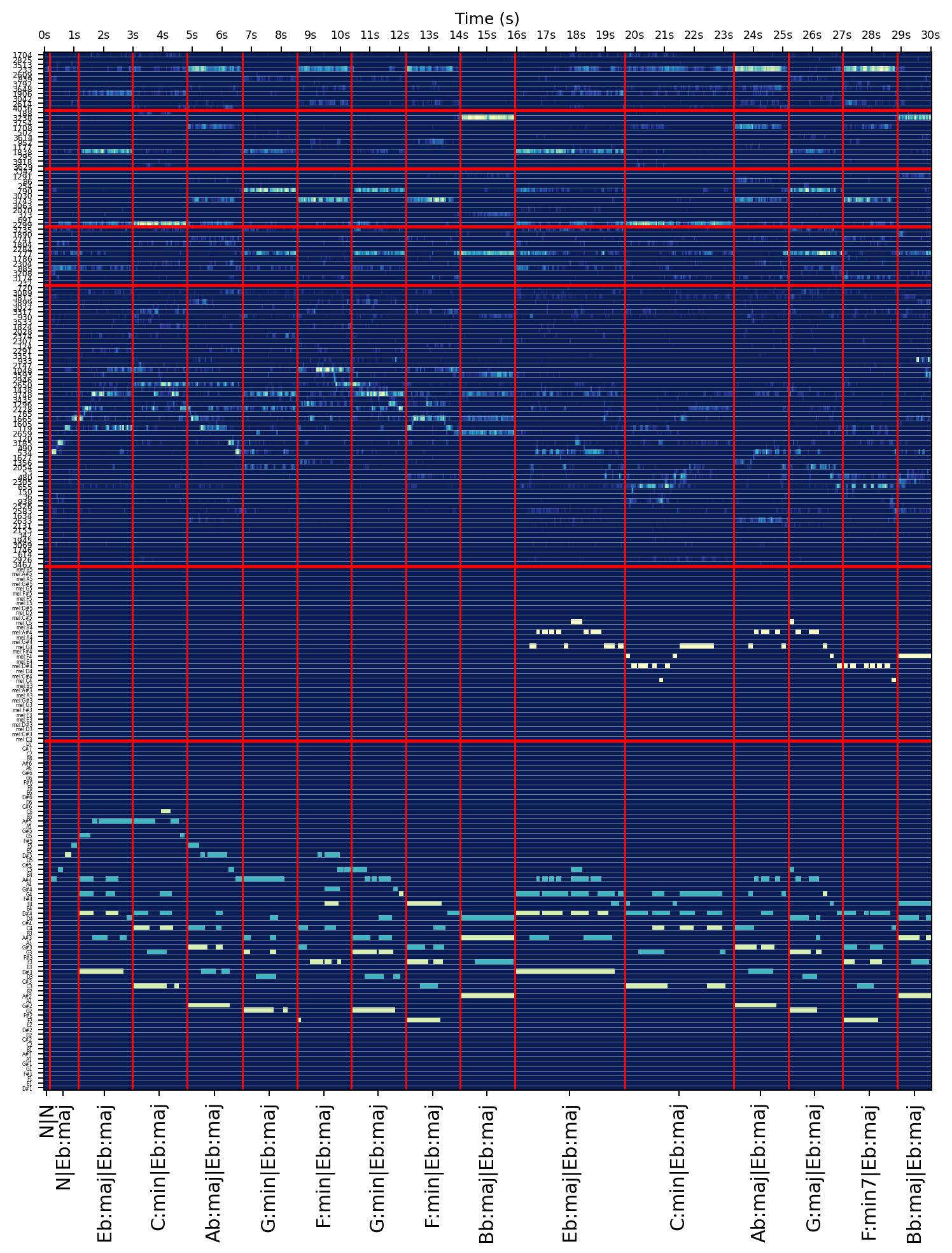}
    \caption{E$\flat$ major.}
    \label{fig:activation-musicfn-32}
  \end{subfigure}
    \hfill
 \begin{subfigure}[t]{0.32\textwidth}
    \centering
\includegraphics[width=\linewidth]{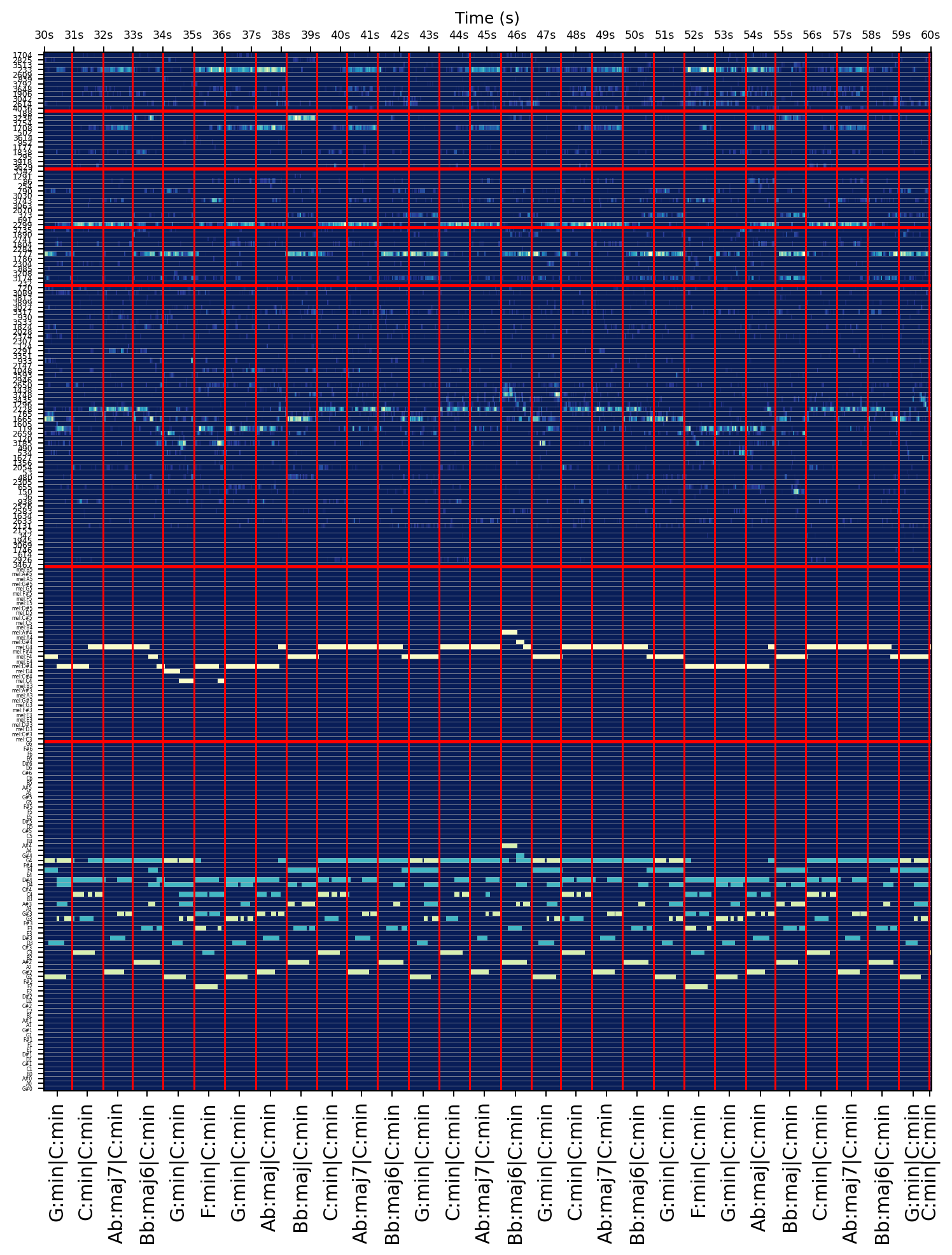}
    \caption{C minor.}
    \label{fig:activation-64}
  \end{subfigure}
  \caption{Activations of MuQ SAE features extracted from different-key songs, organized according to recovered orbits. The horizontal axis is song time with annotated chord roots. The vertical axis is feature indices. We append the the piano-roll ground truth as the bottom row. Vertical grid lines separate chord segments, and horizontal grid lines separate different orbits. Four 12-sized orbits are shown to correspond with the chord progression and one long chain mimics the melody.}
  \label{fig:activation}
  \vspace{-5pt}
\end{figure}




\subsection{Recovered Orbits}
We recover several pitch-invariant orbits and fixed points in the MuQ and MusicFM representations\footnote{Demo page with listening samples is available at \url{https://Kikyo-16.github.io/sae-alignment-demo-page}.}. We found that both models imply some similar 12-size rings and chains, such as the major and minor chord ring, and the pianoroll-like chain, as shown in Figure~\ref{fig:activation}.  We select these three realistic multi-instrument vocal recordings from POP909 dataset. More activation examples including failure expamles are included in Appendix~\ref{app:orbit_activation} and~\ref{app:genre_robustness}.

We uncover a previously unreported representation pattern: a subdominant-centered orbit 
encoding relative key structure.  
More specifically, we identify a 12-feature ring that appears to encode scale-degree information relative to the key. For example, as shown in Figure~\ref{fig:activation-muq-32}, the first row corresponds to this ring. In a piece in \textit{B$\flat$ major}, frames with chord root \textit{E$\flat$} (scale degree \(\hat{4}\) in \textit{B$\flat$ major}) activate one feature in the ring, while other frames do not. If we label this feature as \textit{E$\flat$}, then the remaining features are labeled following the cyclic order of the ring. We tested this ring on additional samples in different keys and observed similar behavior in the major keys. In minor keys, the ring tends to follow the relative major key rather than the tonic minor key. The occurrence of this relative-major subdominant provides new evidence for how a foundation model encodes the concept of key in local frames. 

Motivated by this discovery, we conduct additional experiments in Section~\ref{sec:downstream-task-evaluation} to assess the quality of rings discovered from different foundation models. These include the relative-major subdominant ring, the major chord ring, and the minor chord ring.



\section{Downstream Task Evaluation}\label{sec:downstream-task-evaluation}
We identified several prominent orbits in the SAE feature 
space that correspond to musically meaningful concepts, including major chords, 
minor chords, and musical keys, by inspecting the activations of selected anchor 
samples. In this section, we leverage these discovered orbits for two downstream 
music understanding tasks, with the goal of evaluating the quality of the discovered 
features and comparing the interpretability of the two foundation models.

\subsection{Baselines}
For key detection, we train a supervised single-layer linear probe on the same training dataset as the baseline. Motivated by the 
recovered relative-major subdominant orbit, the probe has 12 outputs and is trained 
to detect frames whose chord root matches the relative-major subdominant of the 
current key. For example, in an \textit{B$\flat$ major} sample, this root is 
\textit{E$\flat$}; frames with chords rooted in \textit{E$\flat$} receive a one-hot target at the 
\textit{E$\flat$} entry, while all others receive the zero vector. The probe is trained 
with binary cross-entropy loss. For chord recognition, we additionally train a supervised single-layer linear probe as a baseline. The probe predicts 25 classes, corresponding to 12 major chords, 12 minor chords, and a no-chord class. We further compare with two task-specific models, S-Key~\cite{kong2025s} for key detection and LVCR~\cite{jiang2019large} for 
chord recognition. 
For S-Key and LVCR, we directly use their open-source checkpoints, which are trained 
on external datasets. See Appendix~\ref{app:baselines} for baseline details.

\subsection{Metrics}
We report segment- and song-level key accuracy, excluding relative major–minor errors, for evaluating the subdominant rings. For the other key detection evaluations, we report standard key accuracy and relaxed MIREX accuracy. For chord recognition, we report root weighted recall over 13 classes and root-and-quality (MajMin) weighted recall over 25 classes. See Appendex~\ref{app:evaluation_metrics} for metric details.

\subsection{Ablations}
To evaluate whether our algorithm preserves the semantic quality of general SAE features, we train a standard SAE as an ablation baseline. Since its features do not have an explicit orbit-like 
structure, we use the weight vectors of the frozen linear probes to retrieve 
corresponding SAE encoder weight vectors by cosine similarity. These matched features are then 
used for key detection and chord recognition. In the result table, the standard SAE and our proposed orbit-based SAE are denoted as \textit{s-SAE} and \textit{o-SAE}, respectively. Additional ablation results on hidden states from different layers, varying sparsity $K$, different $\tau$, and different random seeds are provided in Appendix~\ref{app:ablations} and \ref{app:robustness}.


\begin{table}\small
\caption{Weighted Chord Recall (WCR) across Different Models.}\label{table:chord_rec}
\renewcommand{\arraystretch}{0.9} 
  \setlength{\aboverulesep}{1.5pt}   
  \setlength{\belowrulesep}{1.5pt}    
  \centering
   \begin{tabular}{llcccccc}
    \toprule
    &&
    \multicolumn{3}{c}{Root WCR $\uparrow$}   &
    \multicolumn{3}{c}{MajMin WCR $\uparrow$}    
    \\
    \cmidrule(r){3-5}
    \cmidrule(r){6-8}
    Models &
    Representations & POP909&RWC& Slakh2100& POP909 & RWC& Slakh2100 \\
    \midrule
    s-SAE   &  MusicFM (4) &71.08&67.56&70.32&65.93&64.68&68.39  \\
     (Est. Bd) & MuQ  (2)&77.42&78.08&78.10&73.11&76.92&76.53 \\
    \midrule
      
      o-SAE   &  MusicFM (4) & 74.01&69.65 &75.35&70.09&67.68&73.99 \\
     (Est. Bd)& MuQ (2) &80.03&79.79&78.45&73.68&77.40&75.61 \\
     \midrule
     s-SAE   &  MusicFM (4) &80.31&78.43&79.85&75.14&75.82&79.67\\
     (Oracle Bd)& MuQ (2) &83.84&84.91&84.08&79.45&83.97&82.47 \\
    \midrule
      o-SAE   & MusicFM (4) &80.39&77.49&81.11&76.52&75.63&80.05 \\
     (Oracle Bd)& MuQ (2) &\textbf{85.29}&86.21&84.67&78.96&83.76&82.06 \\
     \midrule
    Probe & MusicFM (4)&83.60&82.80&85.49&80.13&80.19&85.66\tablefootnote{Because root-with-quality evaluation excludes unsupported labels (e.g. suspended chords), it assesses fewer frames. As a result, its score can be lower than the root-only score.
} \\
    (Oracle Bd)& MuQ (2) &84.17&86.01&\textbf{86.05}&\textbf{81.49}&84.66&\textbf{86.42}\\
     \midrule
     LVCR &-&84.25&\textbf{87.96}&85.65&81.37&\textbf{87.47}&83.81 \\
    \bottomrule
  \end{tabular}
 
\end{table}

\begin{table}\small
  \caption{Key Detection Accuracy  (ignoring relative major/minor difference).}
  \label{table:subdom}
  \renewcommand{\arraystretch}{0.9} 
  \setlength{\aboverulesep}{1.5pt}   
  \setlength{\belowrulesep}{1.5pt}    
  \centering
  \begin{tabular}{llcccccc}
    \toprule
    &&
    \multicolumn{3}{c}{Song-level Accuracy $\uparrow$}   &
    \multicolumn{3}{c}{Segment-level Accuracy $\uparrow$}    
    \\
    \cmidrule(r){3-5}
    \cmidrule(r){6-8}
    Models &
    Representations & FMAKv2& Gtzan& GS & FMAKv2& Gtzan &GS\\

    \midrule
     
      Probe   &  MusicFM (3) & 50.92&48.31&40.89&44.92&48.31&36.39\\
      & MuQ (5)  &54.66&48.80&47.68&48.33&48.80&40.33 \\
    \midrule
      
      s-SAE   &  MusicFM (3) &35.09&30.60&25.99&32.07&30.60&25.50\\
     & MuQ (5)   &56.61&48.80&47.19&50.00&48.80&41.20 \\
     \midrule
    o-SAE  
     & MuQ (5) &\textbf{56.92}&\textbf{51.57}&\textbf{50.50}&\textbf{50.32}&\textbf{51.57}&\textbf{44.56}  \\
      \bottomrule
       \end{tabular}
       \vspace{-2pt}
\end{table}

\subsection{Results}
We discovered a subdominant ring in the MuQ representations, but not in the MusicFM representations, whereas major- and minor-chord rings emerged in both models. Across all evaluation settings, the orbits discovered in the MuQ representations were consistently of higher quality than those discovered in the MusicFM representations. All SAE results reported in the tables use a sparsity level of $K=48$.

\begin{figure}[t]
  \centering
  \begin{subfigure}[t]{0.49\textwidth}
    \centering
    
\includegraphics[width=\linewidth]{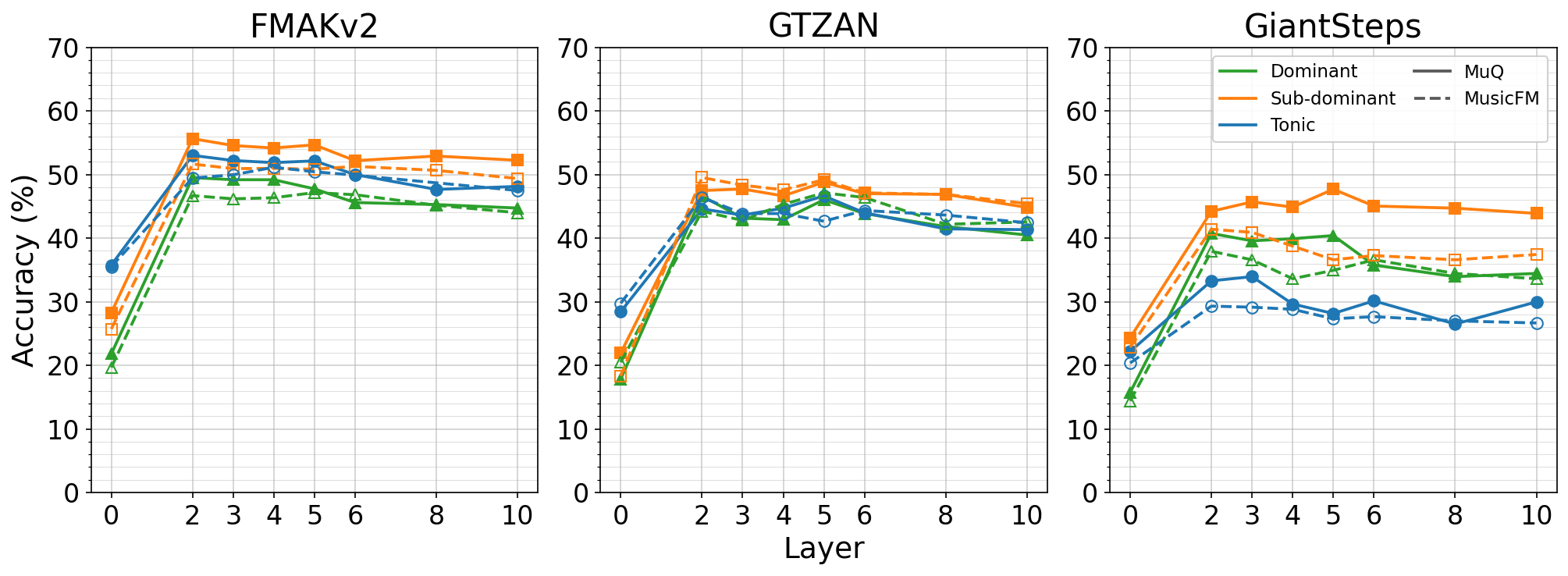}
    \caption{Probe}\label{fig:tonic_subdom_dom_a}

  \end{subfigure}
  \hfill
  \begin{subfigure}[t]{0.49\textwidth}
    \centering
    \includegraphics[width=\linewidth]{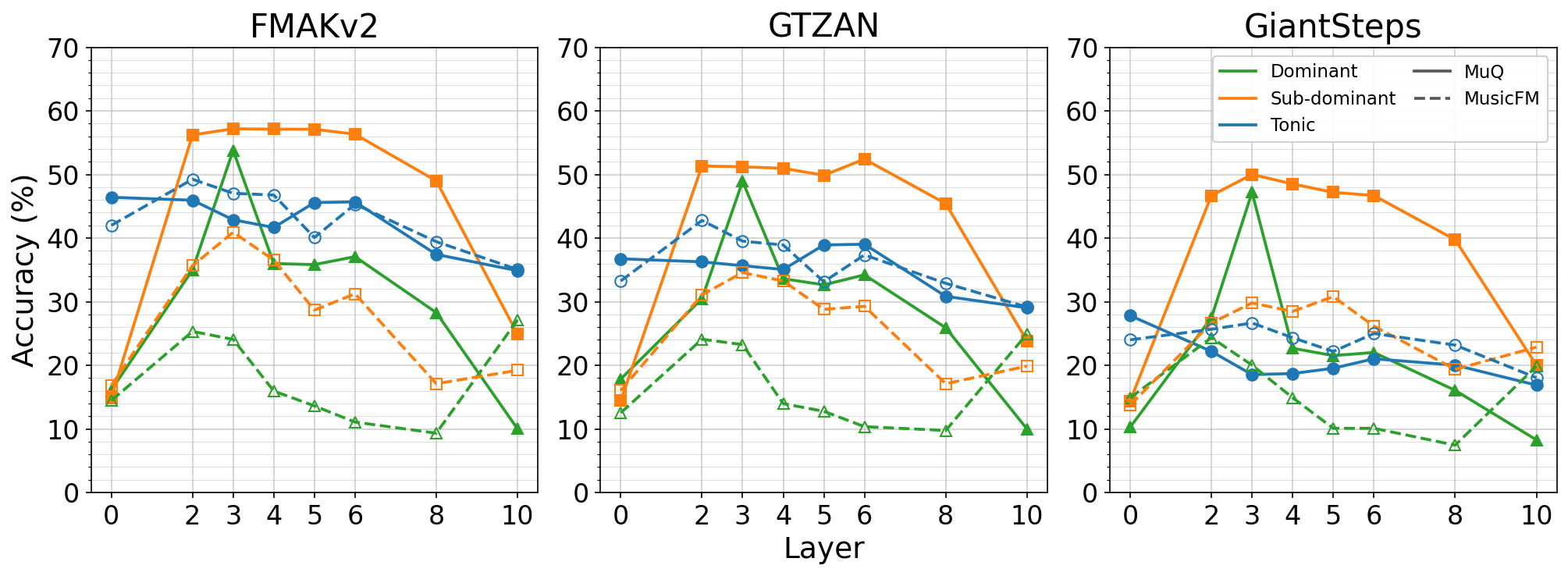}
    \caption{s-SAE}\label{fig:tonic_subdom_dom_b}

  \end{subfigure}
  \caption{Key detection accuracy comparing probing with standard SAE. Given labels, probing yields generally high accuracy regardless of base model and concept nature. In contrast, SAE uncovers high-contrast but consistent results that are specific to music concept.}
   \label{fig:tonic_subdom_dom}
    \vspace{-2pt}
\end{figure}

\begin{table}[t]\small
  \caption{Key Detection Results Derived from Chord-based Features.}\label{table:chord-key}
\renewcommand{\arraystretch}{0.9} 
  \setlength{\aboverulesep}{1.5pt}   
  \setlength{\belowrulesep}{1.5pt}    
  \centering
  \begin{tabular}{llcccccc}
    \toprule
    &&
    \multicolumn{3}{c}{MIREX Accuracy $\uparrow$}   &
    \multicolumn{3}{c}{Accuracy $\uparrow$}    
    \\
    \cmidrule(r){3-5}
    \cmidrule(r){6-8}
    Methods &
    Representations & FMAKv2& Gtzan& GS & FMAKv2& Gtzan &GS \\
    \midrule
    Probe&MusicFM (2) &62.75&60.29&65.98&53.26&49.76&56.95\\
     &MuQ (2) &66.13&63.22&69.70&56.90&52.89&61.92\\
     \midrule
     s-SAE& 
    MusicFM (2)&68.89&63.76&63.96&60.42&55.66&54.47 \\
    &MuQ (2)&68.87&66.63&69.34&61.08&58.92&62.42
    \\
    \midrule
    o-SAE&MusicFM (2)&69.89&64.40&61.76&61.76&56.51&52.32\\
     &MuQ (2)&70.17&67.09&71.14&64.07&58.67&62.35\\
     \midrule 
     S-KEY &-&\textbf{73.24}&\textbf{74.22}&\textbf{72.09}&\textbf{64.92}&\textbf{65.54}&\textbf{64.40}\\

    \bottomrule
  \end{tabular}
\end{table}

\subsubsection{Chord Rings}
Table~\ref{table:chord_rec} reports SAE-based results 
using chord boundaries estimated from SAE features (\textit{Est. Bd}) and reference chord boundaries (\textit{Oracle Bd}). The number in parentheses after each representation name indicates the layer index. Details of the chord-prediction post-processing procedure and additional layer-wise results are provided in Appendices~\ref{app:downstream_inference} and \ref{app:ablations}, respectively.

The discovered major- and minor-chord rings support effective chord recognition, achieving competitive performance against fully supervised methods despite requiring minimal label supervision. Moreover, rings derived from MuQ representations consistently outperform those derived from MusicFM representations.

\subsubsection{Subdominant Ring}
Key-detection performance in Table~\ref{table:subdom} demonstrates that our SAE discovers subdominant rings that capture musically meaningful key information. Across all evaluations, rings derived from MuQ representations consistently outperform those from MusicFM.

Intuitively, one would expect the algorithm to recover tonic or dominant rings alongside the subdominant, where in fact it does not. To verify whether tonic and dominant signals are present in the underlying representations, we conduct linear probing experiments. Figure~\ref{fig:tonic_subdom_dom_a} confirms the presence of both signals, albeit weaker than the subdominant signal. 

Are they dropped during sparse feature selection? We identify s-SAE encoder weight vectors closest to the linear probe weight vectors. Figure~\ref {fig:tonic_subdom_dom_b} shows that the subdominant features selected this way maintain strong performance, whereas the tonic and dominant features remain absent. This discrepancy indicates that while standard linear probes detect weak tonic and dominant signals, the SAE does not evaluate them as active or salient features.

\subsubsection{SAE Evaluates Features Saliency Beyond Linear Recoverability}\label{sec:tonic_dom_subdom}

The divergence between probing and SAE-based extraction suggests that tonic and dominant information may not exist as explicit, readily accessible concepts within the foundation model, but are instead resynthesized by the supervised linear probe.

To shed more light on this hypothesis, we examine whether key information can be inferred from readily represented chord progressions. We develop a music-informed rule-based algorithm (detailed in Appendix~\ref{app:chord_key_inference}) to infer major and minor keys by aggregating frame-level chord activations (from SAEs or probes) and weighting the corresponding tonic, subdominant, and dominant functions.

The chord-infer-key evaluation (Table~\ref{table:chord-key}) shows that keys can be effectively reconstructed through simple manual linear combination of chord activations. These results show that key information can be dispersed in other concepts. It implies that linear probes can predict tonic signals even when the underlying representation lacks a distinct, explicit tonic concept. In contrast, the unsupervised orbit-based SAE discovers only the salient structure: the subdominant ring. 

To generalize, linear recoverability does not guarantee that the corresponding information is organized as an explicit concept in the representation, while SAEs capture concepts that are salient and sparse. This motivates orbit-based analysis, which examines not only recoverability but also the structural organization of musical concepts.

\subsubsection{Global Recoverability vs. Local Structure}
Previous probing studies~\cite{castellon2021codified, koo2024understanding, wei2024music, papaioannou2025universal} often aggregate frame-level representations into a global representation using max or average pooling and then train a supervised probe to predict musical key. While this evaluates whether key information is recoverable, it does not reveal how key-related concepts are organized and encoded over time. In contrast, our method identifies an explicit subdominant concept, a local representation that implies the key. Notably, even linear probing shows that the subdominant is the most salient local concept for key perception (Figure~\ref{fig:tonic_subdom_dom_a}), an enlightening result that previous global-key studies missed.

\section{Limitations}
The orbits in this study are characterized by pitch transposition. Pitch-invariant concepts, such as rhythm, cannot be organized without additional inductive biases. 
In parallel, our method focuses on the frame level, and it remains unexplored how foundation models organize and aggregate frame-level concepts over time.
Additionally, although 12-sized rings emerge from unsupervised SAE alignment, it is potentially related to the pitch transposition step size being $1/12$ of an octave. That, combined with the distribution of data we use, yields mostly Western music concepts. How other transformations may interpret other forms of music is yet to be studied. 

\section{Conclusion and Future Work}

In conclusion, we presented a\textbf{ structure-aware interpretability} framework that leverages pitch transposition as an inductive bias to recover orbit structures over SAE features via multi-view alignment. These orbits reveal relational organization in learned representations, enabling the interpretation of higher-level musical concepts such as chords, keys, and melodic patterns with minimal supervision. Notably, we uncover a subdominant-centered orbit that encodes relative key structure—a previously unreported representation pattern that resembles the ``circle of fifth'' commonly used in tonal theory.

More broadly, we believe this structure-centric perspective points to several promising directions. First, the proposed framework is not limited to music: incorporating known transformations as inductive biases may enable structure-based interpretability in other domains, and can naturally complement existing feature-level methods. Second, while our approach does not eliminate polysemanticity at the level of individual features, it offers robustness to such ambiguity by shifting the unit of analysis to structured groups, where relational consistency reduces feature-level noise. Finally, the presence of coherent orbit structures suggests a potential criterion for evaluating learned representations: models that exhibit consistent transformation-induced structure may reflect a deeper level of conceptual organization.

We hope this work motivates a broader shift toward structure-aware interpretability, where understanding emerges not only from individual features, but from the relational systems they collectively form.

\section*{Acknowledgments}
We would like to thank \textbf{Daniel Chin} for his helpful research discussions and assistance with writing, and \textbf{Junyan Jang} for providing valuable music theory support. We also thank Mohamed bin Zayed University of Artificial Intelligence (MBZUAI), with special thanks to \textbf{Prof.~Qirong Ho} and his team for providing computational resources.

\bibliography{refs}
\newpage
\appendix

\section{Primer on Music Theory Fundamentals}
\label{sec:music_theory_primer}

To make this work self-contained for readers without a formal background in music theory, this section provides a brief primer on the core musical concepts referenced throughout the paper, including pitch classes, pitch transposition, chords, harmonic functions, relative keys, and the Circle of Fifths.

\subsection{Pitch, Pitch Classes, and Equal Temperament}
In Western tonal music, sound frequencies are discretized into musical pitches. Under the standard twelve-tone equal temperament (12-TET) system, an octave is divided into $12$ logarithmically equal intervals called semitones. Pitches that differ by one or more full octaves share the same chroma or harmonic quality. Abstracting away octave height yields the concept of a \textbf{pitch class}. The set of pitch classes is naturally represented modulo 12:
\begin{equation}
    \mathcal{P} = \{0, 1, 2, \dots, 11\} \cong \mathbb{Z}_{12},
\end{equation}
corresponding to the pitch names $\{\text{C}, \text{C}\sharp/\text{D}\flat, \text{D}, \dots, \text{B}\}$.

\subsection{Chords and Pitch Transposition}
A \textbf{chord} is a combination of two or more distinct pitch classes played simultaneously or sequentially. For instance, a \textit{C major triad} consists of the pitch classes $\{\text{C}, \text{E}, \text{G}\}$. 
\textbf{Pitch transposition} is an operator $T_k: \mathcal{P} \to \mathcal{P}$ that shifts every pitch class upward by $k$ semitones:
\begin{equation}
    T_k(p) = (p + k) \pmod{12}.
\end{equation}
When applied to a chord, pitch transposition shifts its pitch constituents while preserving its internal interval structure (i.e., chord quality such as major or minor). Repeatedly applying a 1-semitone transposition induces a cyclic ordered set (an orbit) of size 12 over pitch-class space.

\subsection{Harmonic Functions and the Circle of Fifths}
\label{subsec:circle_of_fifths}
In tonal harmony, pitch classes and chords revolve around functional roles relative to a central pitch called the \textbf{Tonic} ($\text{I}$, scale degree 1), which acts as the tonal "home" or resolution point. Two fundamental harmonic functions anchored to the tonic are:
\begin{itemize}
    \item \textbf{Dominant} ($\text{V}$, scale degree 5): Located $+7$ semitones (a perfect fifth) above the tonic (e.g., $\textit{G}$ in $\textit{C major}$). It creates strong harmonic tension that pulls directly toward the tonic.
    \item \textbf{Subdominant} ($\text{IV}$, scale degree 4): Located $+5$ semitones above (or $-7$ semitones below, a perfect fourth) the tonic (e.g., $\textit{F}$ in $\textit{C major}$). It provides a fundamental harmonic framework and expands the tonal center.
\end{itemize}

The \textbf{Circle of Fifths}\footnote{Image source: \url{https://en.wikipedia.org/wiki/Circle_of_fifths}.} as shown in Figure~\ref{fig:circle_of_fifths} organizes all 12 chromatic pitch classes geometrically according to the interval of a perfect fifth ($7$ semitones). Moving clockwise around the circle corresponds to stepping up by $+7$ semitones (toward the Dominant), whereas moving counter-clockwise corresponds to stepping down by $-7$ semitones / up by $+5$ semitones (toward the Subdominant). As illustrated in Figure~\ref{fig:circle_of_fifths}, the Tonic, Subdominant, and Dominant occupy three adjacent positions on the circle, reflecting maximum harmonic overlap and structural proximity.

\subsection{Relative Keys and Key Signatures}
\begin{figure}[t]
    \centering
    \includegraphics[trim=1cm 5cm 12cm 1.5cm, clip, width=0.6\linewidth]{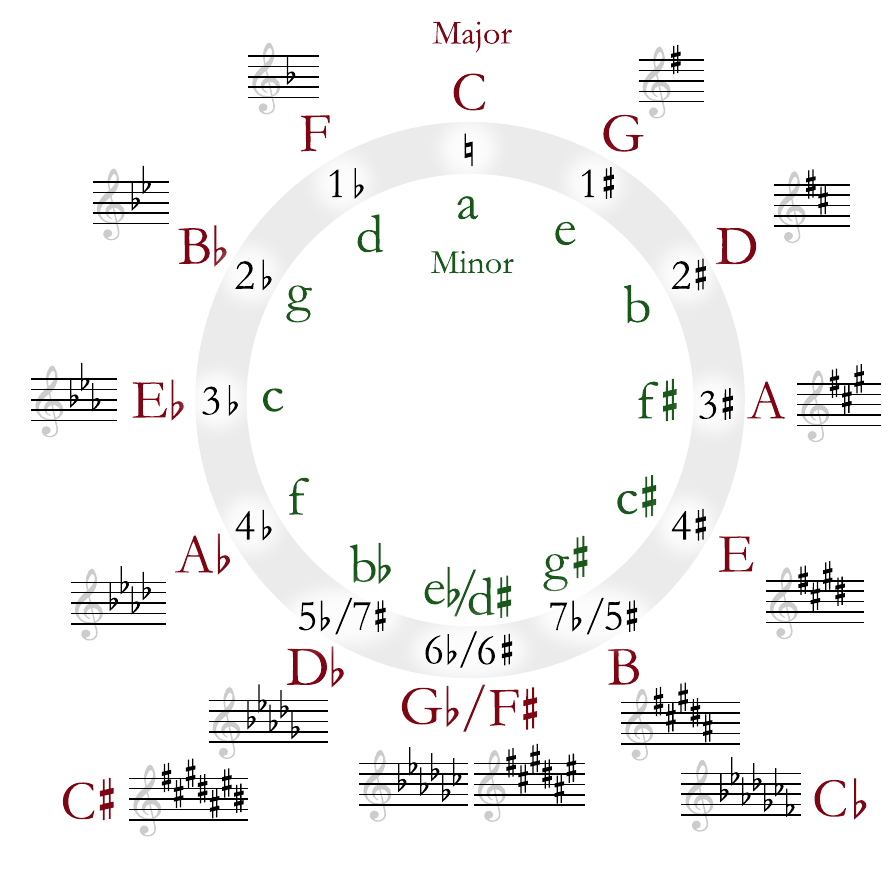}
    \caption{\textbf{The Circle of Fifths.} A geometric representation of the relationships among the 12 chromatic pitch classes, harmonic functions, and relative key signatures arranged by intervals of perfect fifths. Outer circle labels represent major keys; inner circle labels represent relative minor keys.}
    \label{fig:circle_of_fifths}
\end{figure}

A \textbf{key signature} designates the set of sharps or flats shared by a diatonic scale. Every major key shares its exact key signature and constituent pitch class set with a corresponding \textbf{relative minor key} (located $3$ semitones / a minor third below the major tonic, e.g., $\textit{C major}$ and $\textit{A minor}$). On the Circle of Fifths (Figure~\ref{fig:circle_of_fifths}), relative major and minor keys are geometrically paired together in the outer and inner rings, respectively.

\section{Datasets}
\label{app:datasets}

The SAEs and supervised probes are trained exclusively on Slakh2100 Dataset, while
all reported downstream results are obtained on realistic-recording datasets or on the
official Slakh2100 test split. No evaluation track is used for training or
model selection. Table~\ref{tab:dataset_overview} summarizes the datasets used
in our experiments.

\subsection{Training and Validation Data}
\label{app:training_data}

\paragraph{Slakh2100.}

We construct the training and validation sets from Slakh2100
\cite{manilow2019cutting}, a multitrack dataset synthesized from aligned MIDI
performances using high-quality sample-based soundfonts.
A notable difference from real-world recordings is the absence of vocals.
Because vocal signals contain strong phonetic and speech-related structure,
they may occupy substantial representational capacity while being unrelated
to the musical concepts studied here.
Slakh2100 therefore provides a controlled setting for analyzing
transposition-induced harmonic and melodic structure.

Nevertheless, synthesized audio does not capture the full acoustic variability
of real recordings, including expressive articulation, room acoustics,
microphone characteristics, mixing, and mastering.
Its MIDI arrangements, however, originate from realistic musical scores and
performances, preserving much of their symbolic and structural content.
The remaining domain gap is thus primarily acoustic, and we evaluate transfer
to real recordings in our downstream experiments.

\paragraph{Key-balanced split construction.}

We select subsets of the Slakh2100 training and validation splits to obtain
approximately balanced coverage of the 24 major and minor keys.
Tracks are grouped by their pseudo key labels, and up to 16 tracks are sampled
from each class.
For classes with fewer than 16 tracks, all available tracks are retained
without duplication.
Using a fixed random seed of 42 yields 326 training tracks and 164 validation
tracks.

Pseudo frame-level chord and key annotations are derived from the aligned MIDI
files using the MIDI-AutoLabel-Dataset toolkit.
They are used for split balancing, supervised probe training, and validation
metrics, but not as supervision for SAE training.

\paragraph{Segmentation and pitch shifting.}

Each track is divided into 30-second segments with a
10-second hop; silent segments are removed.
We generate five views at $\{-2,-1,0,+1,+2\}$ semitones using the
\texttt{torchaudio} phase vocoder with a Hann window,
$n_{\mathrm{fft}}=1024$, and hop length 256.
For a shift of $s$ semitones, the phase-vocoder rate is $2^{-s/12}$, after
which the waveform is linearly resampled to its original length to preserve
temporal alignment.
When aligning $S=3$ SAEs, three consecutive views are randomly sampled for each training
example. Validation tracks are divided into non-overlapping 30-second
segments.

\begin{table}[t]
\centering
\caption{
Overview of the datasets used in our experiments.
The reported numbers correspond to tracks retained after label-validity filtering.
}
\label{tab:dataset_overview}
\small
\begin{tabular}{lllr}
\toprule
Dataset & Usage & Audio domain & Tracks \\
\midrule
Slakh2100 & Training & Synthesized, instrumental & 326 \\
Slakh2100 & Validation & Synthesized, instrumental & 164 \\
\midrule
FMAKv2 & Key detection & Real recordings & 5,434 \\
GTZAN-Key & Key detection & Real recordings & 830 \\
GiantSteps & Key detection & Real recordings & 604 \\
\midrule
POP909 & Chord recognition & Real recordings & 810 \\
RWC-POP-100 & Chord recognition & Real recordings & 100 \\
Slakh2100 test & Chord recognition & Synthesized, instrumental & 151 \\
\bottomrule
\end{tabular}
\end{table}

\subsection{Evaluation Data}
\label{app:evaluation_data}

We evaluate both in-domain performance and transfer from the synthesized,
vocal-free training data to real recordings.
Chord recognition is evaluated on Chinese, Japanese, and western popular music, while
key detection is evaluated across genre-diverse and electronic music
collections.

\subsubsection{Chord-Recognition Datasets}
\label{app:chord_datasets}

\paragraph{POP909.}
POP909 \cite{pop909-ismir2020} contains Chinese popular songs with MIDI
arrangements aligned to the original audio.
We use tracks 100--909, yielding 810 songs, and evaluate the original vocal
mixes using the provided time-aligned chord annotations.

\paragraph{RWC-POP-100.}
We use all 100 chord-annotated songs from RWC-POP-100
\cite{Gotoetal2026}.
The professionally produced Japanese popular-music recordings contain vocals,
expressive performances, and production effects absent from Slakh2100.
Together, POP909 and RWC-POP-100 test whether the discovered chord concepts
transfer from synthesized instrumental audio to real recordings.

\paragraph{Slakh2100 test split.}
We use all 151 songs in the official Slakh2100 test split.
These songs are disjoint from the training and validation data and provide an
in-domain evaluation under the same synthesized, instrumental conditions.

\subsubsection{Key-Detection Datasets}
\label{app:key_datasets}

We evaluate on three commonly used key-detection datasets.
Table~\ref{tab:key_dataset_distribution} reports their major--minor
distributions, which provide important context for interpreting the results,
particularly for the strongly minor-dominated GiantSteps dataset.

\begin{table}[t]
\centering
\caption{Major--minor distribution of the key-detection datasets.}
\label{tab:key_dataset_distribution}
\small
\begin{tabular}{lrrr}
\toprule
Dataset & Major & Minor & Total \\
\midrule
FMAKv2     & 2,777 & 2,657 & 5,434 \\
GTZAN-Key  &   357 &   473 &   830 \\
GiantSteps &    93 &   511 &   604 \\
\bottomrule
\end{tabular}
\end{table}

\paragraph{FMAKv2.}
FMAKv2 contains genre-diverse real recordings with song-level major and minor
key labels.
We retain 5,434 tracks with matched labels and successfully processed audio,
including 2,777 major- and 2,657 minor-key tracks.

\paragraph{GTZAN-Key.}
GTZAN-Key extends the GTZAN genre collection
\cite{tzanetakis2002musical} with global key annotations.
Following the nine-genre evaluation protocol of S-KEY
\cite{kong2025s}, we exclude the classical subset and evaluate only tracks
with a unique global key, enabling direct comparison with this baseline.
After audio preprocessing, 830 tracks remain across the nine genres listed in
Table~\ref{tab:gtzan_genre_distribution}.

\begin{table}[t]
\centering
\caption{Genre distribution of GTZAN tracks.}
\label{tab:gtzan_genre_distribution}
\small
\begin{tabular}{lr@{\hspace{1.5em}}lr}
\toprule
Genre & Tracks & Genre & Tracks \\
\midrule
Blues   & 98 & Jazz   & 79 \\
Country & 96 & Metal  & 93 \\
Disco   & 97 & Pop    & 94 \\
Hip-hop & 79 & Reggae & 97 \\
Rock    & 97 &        &    \\
\midrule
Total   & 830 &        &    \\
\bottomrule
\end{tabular}
\end{table}

\paragraph{GiantSteps.}
GiantSteps \cite{knees2015two} contains 604 electronic dance music excerpts
with global key labels.
We divide each excerpt into non-overlapping 30-second segments for evaluation.
Its production-heavy timbres and strongly imbalanced mode distribution provide
a distinct test of acoustic and genre transfer.

\subsection{Evaluation Preprocessing and Exclusion Criteria}
\label{app:dataset_filtering}

For evaluation, audio is divided into non-overlapping 30-second segments, and
annotations are aligned to the frame-level representation timeline.
All filtering rules are fixed before model evaluation.
We exclude tracks that cannot be decoded, are shorter than one evaluation
segment, lack matching annotations, or contain key changes, since key detection
assumes a single global key per song.
Accordingly, GiantSteps entries with multiple key labels and GTZAN-Key entries
marked as modulation or unknown are removed.

For chord recognition, frames labeled as unknown are excluded from all
metrics.
Root WCR evaluates the remaining chord frames using only the root pitch class,
regardless of chord quality.
MajMin WCR follows the \texttt{mir\_eval} vocabulary: reference frames that
cannot be reduced to a major or minor triad are excluded from this metric.

\section{Baselines}
\label{app:baselines}

We compare our method with supervised linear probes and task-specific
music-analysis systems. The probes are trained on the same data and frozen
foundation-model representations as our method. The task-specific systems are
evaluated using their publicly released checkpoints, which were trained on
external datasets.

Throughout this appendix, $n$ indexes songs, $t$ indexes frames, $p$ indexes pitch classes, and $s$ indexs sub-SAEs. We reserve $i$ and $j$ for SAE
feature indices and use $I$ to denote a
temporal segment or interval. Let
$\mathcal{P}=\{0,\ldots,11\}$ denote the pitch-class set. 

\subsection{Chord Recognition}

\paragraph{LVCR.}
LVCR~\cite{jiang2019large} is a supervised large-vocabulary chord
transcription system. It predicts chord components, including root, bass,
triad, seventh, and extensions, using multitask learning and structured
decoding. Its predictions are reduced to the Root and MajMin vocabularies at
evaluation time.

\paragraph{Linear probe.}
Given a frozen frame-level representation
$x_t\in\mathbb{R}^{d}$, we train a single affine layer
\begin{equation}
\label{eq:chord_probe}
h^{\mathrm{ch}}_t
=
W^{\mathrm{ch}}x_t
+
b^{\mathrm{ch}},
\qquad
W^{\mathrm{ch}}
\in
R^{25\times d},
\end{equation}
where the 25 outputs correspond to 12 major chords, 12 minor chords, and the
no-chord class $\mathrm{N}$. Reference labels are reduced to the MajMin
vocabulary before training. Extensions that preserve an underlying major or
minor triad are folded into the corresponding triad, whereas unsupported
chord qualities and unknown annotations are ignored. The no-chord label is
retained as the twenty-fifth class.

Let $\mathcal{T}_{\mathrm{ch}}$ denote the set of valid training-frame
indices and let
$y^{\mathrm{ch}}_t\in\{0,\ldots,24\}$ denote the reduced chord target. The
probe is optimized using frame-level cross-entropy:
\begin{equation}
\label{eq:chord_probe_loss}
\mathcal{L}_{\mathrm{ch}}
=
-\frac{1}{|\mathcal{T}_{\mathrm{ch}}|}
\sum_{t\in\mathcal{T}_{\mathrm{ch}}}
\log
\left[
\operatorname{softmax}
\left(
h^{\mathrm{ch}}_t
\right)
\right]_{y^{\mathrm{ch}}_t}.
\end{equation}

\subsection{Key Detection}

\paragraph{S-KEY.}
S-KEY~\cite{kong2025} is a self-supervised key-estimation model extending
STONE~\cite{kong:hal-04675781} with transposition-invariant chroma pseudo-labels. This
enables major--minor key prediction without human key annotations. We evaluate
its released checkpoint without fine-tuning.

\paragraph{Linear probes.}
We independently train frame-level probes for the tonic, subdominant, and
dominant functions. Each probe has 12 outputs corresponding to pitch classes:
\begin{equation}
\label{eq:key_probe}
h^{(q)}_t
=
W^{(q)}x_t
+
b^{(q)},
\qquad
q\in
\{
\mathrm{ton},
\mathrm{sub},
\mathrm{dom}
\},
\end{equation}
where $q$ denotes the target harmonic function. Let
$r^{\mathrm{rel}}_t\in\mathcal{P}$ denote the relative-major key signature at
frame $t$. The pitch class associated with harmonic function $q$ is
\begin{equation}
\label{eq:key_probe_degree}
\xi^{(q)}_t
=
\left(
r^{\mathrm{rel}}_t+\delta_q
\right)
\bmod 12,
\qquad
\delta_q
=
\begin{cases}
0, & q=\mathrm{ton},\\
5, & q=\mathrm{sub},\\
7, & q=\mathrm{dom}.
\end{cases}
\end{equation}
Given the reference chord root
$r^{\mathrm{ch}}_t\in\mathcal{P}$, the binary target for output pitch class
$p$ is
\begin{equation}
\label{eq:key_probe_target}
y^{(q)}_{t,p}
=
\mathbbm{1}
\left[
r^{\mathrm{ch}}_t=\xi^{(q)}_t
\right]
\mathbbm{1}
\left[
p=\xi^{(q)}_t
\right].
\end{equation}
Thus, matching frames receive a one-hot target, whereas all other frames
receive an all-zero target. Let $\mathcal{T}_{\mathrm{key}}$ denote the set
of valid training frames. Each probe is trained independently using binary
cross-entropy:
\begin{equation}
\label{eq:key_probe_loss}
\mathcal{L}^{(q)}
=
-\frac{1}{
12|\mathcal{T}_{\mathrm{key}}|
}
\sum_{t\in\mathcal{T}_{\mathrm{key}}}
\sum_{p\in\mathcal{P}}
\left[
y^{(q)}_{t,p}
\log
\sigma
\left(
h^{(q)}_{t,p}
\right)
+
\left(
1-y^{(q)}_{t,p}
\right)
\log
\left(
1-
\sigma
\left(
h^{(q)}_{t,p}
\right)
\right)
\right].
\end{equation}
The main comparison uses the subdominant probe, while the tonic and dominant
probes are included in additional analyses.

\section{Downstream-Task Inference}
\label{app:downstream_inference}

We evaluate whether the discovered SAE orbits can be used directly for
downstream prediction, without fitting an additional classifier on their
activations. For controlled comparison, probe outputs are processed using the
same aggregation and decision rules whenever possible.

\begin{figure}[t]
  \centering

  \begin{subfigure}[t]{0.49\textwidth}
    \centering
    \includegraphics[width=\linewidth]{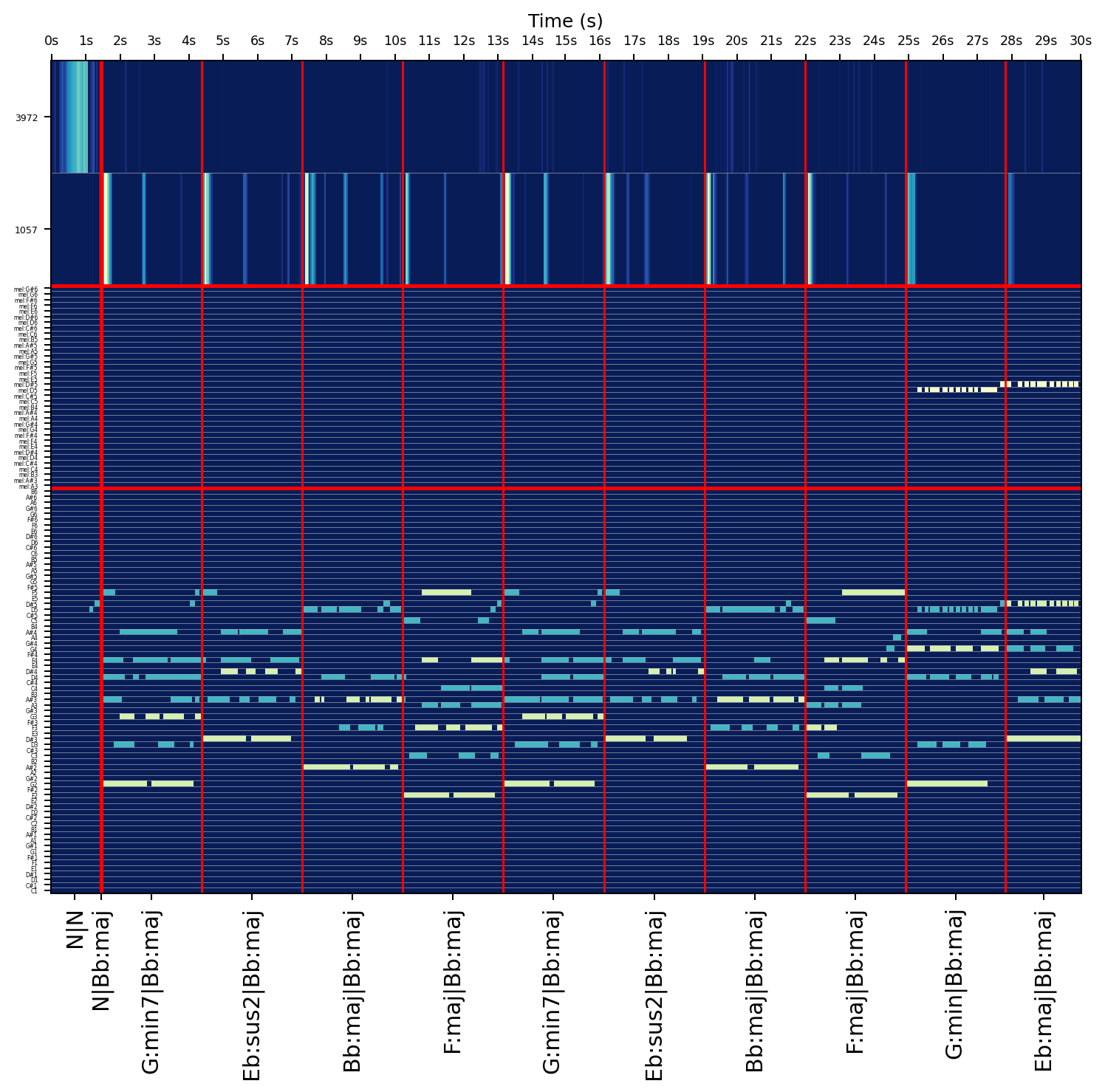}
  \end{subfigure}
  \hfill
  \begin{subfigure}[t]{0.49\textwidth}
    \centering
    \includegraphics[width=\linewidth]{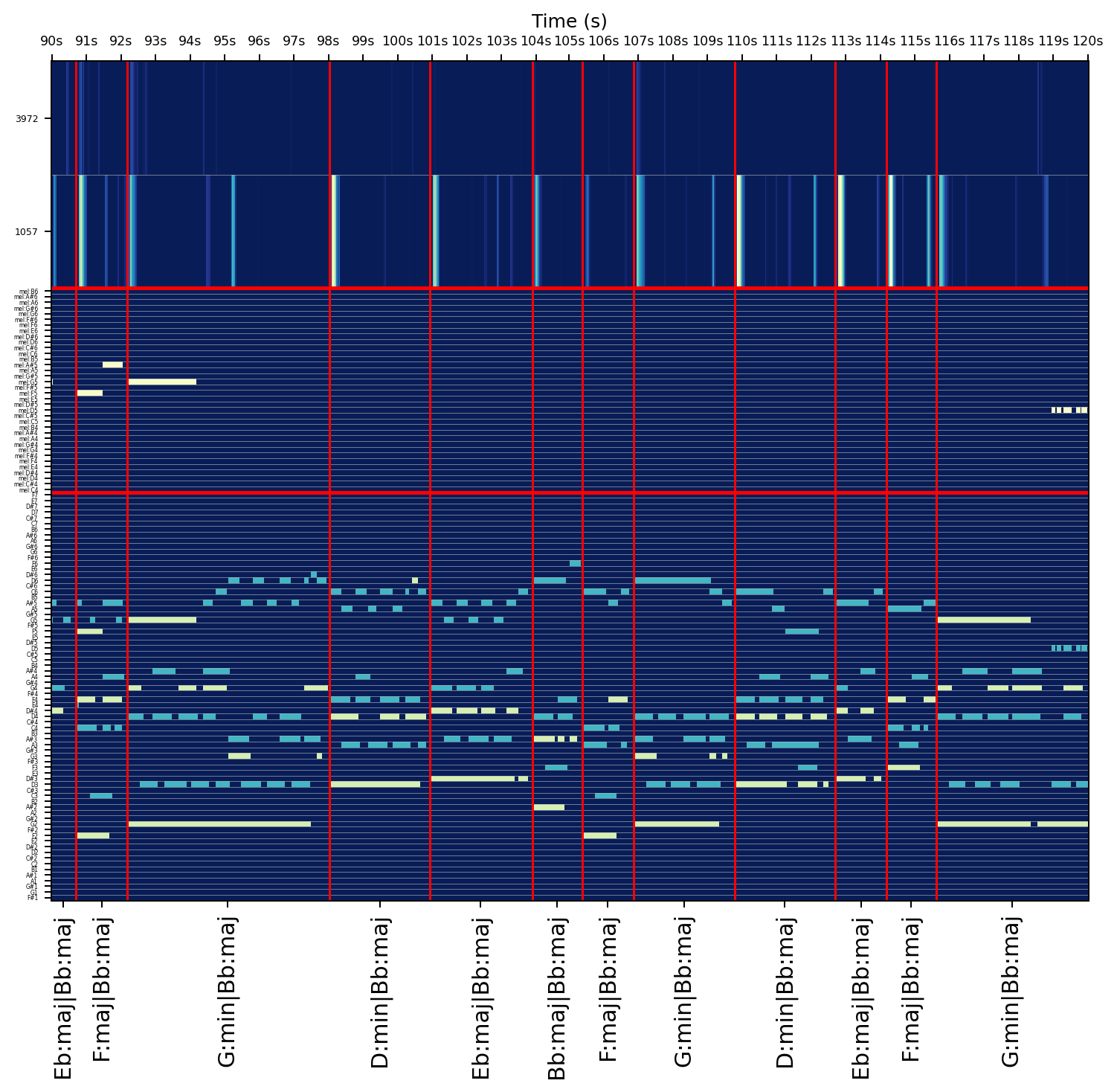}

  \end{subfigure}

  \caption{
Activations of the selected silence and chord-boundary features.
Both features are identified from a few music samples;
in batched experiments, we approximate the same manual procedure using a
small set of labeled anchors.
}
\label{fig:bd_sil}
 
\end{figure}

\subsection{Chord Recognition}
\label{app:chord_inference}

The major- and minor-chord orbits each contain 12 features ordered by chord
root. We smooth their frame-level activations using a nine-frame moving
average. We consider two chord boundary settings. Under \emph{Oracle Bd}, the reference chord annotations gives the boundaries. Under
\emph{Est. Bd}, we use local peaks in the activation
of a discovered chord-boundary feature. The chord boundaries divide the song into segments, each segment $I$ containing some contiguous frames $t \in I$.

Let $u^{m}_{t,p}$ denote the smoothed activation at frame $t$ for root pitch
class $p$ and quality $m\in\{\mathrm{maj},\mathrm{min}\}$. We first average the activation of each chord feature over time within each segment $I$:
\begin{equation}
\bar{u}^{m}_{I,p} = \frac{1}{|I|} \sum_{t\in I} u^{m}_{t,p}.
\label{eq:segment_avg_activation}
\end{equation}

The estimated chord quality $\hat{m}_I$ and root pitch $\hat{p}_I$ for the segment are jointly selected by maximizing the segment-averaged activation:
\begin{equation}
(\hat{m}_I, \hat{p}_I) = \arg\max_{\substack{m \in \{\mathrm{maj}, \mathrm{min}\} \\ p \in \mathcal{P}}} \bar{u}^{m}_{I,p}.
\label{eq:chord_quality_and_root}
\end{equation}

Under \emph{Est. Bd}, the no-chord evidence is obtained from the discovered
silence features. A segment is labeled as no-chord when
its accumulated silence evidence exceeds both its accumulated major- and
minor-chord evidence. Under \emph{Oracle Bd}, it is obtained from reference frames
labeled as no-chord. 

For the supervised chord probe, we apply softmax to its 25 outputs. The first
12 probabilities represent major chords, the next 12 represent minor chords,
and the final probability represents no-chord. We apply the same smoothing as above, use the oracle boundaries, and follow the same decoding procedure as above.

\paragraph{Chord-boundary and silence features.}
As shown in Figure~\ref{fig:bd_sil}, We identify silence features and chord boundary features using a small number of anchor examples.

\subsection{Key-Signature Detection from Scale-Degree Features}
\label{app:relative_key_inference}

The tonic, subdominant, and dominant methods each produce a ring of 12 pitch-class activations. For a 30-second segment, we average the ring over time and obtain a 12-size vector. Then we teat the most active pitch as the the detected harmonic-function root and infer the corresponding tonic. This method predicts only the 12-class key signature and does not distinguish relative major and minor keys. For the corresponding linear probes, again we use the same method but treating probe outputs as feature activations. At the song level, the 12-dimensional signature-score vectors are averaged across segments before selecting the final prediction.

\subsection{Key Detection from Chord Orbits}
\label{app:chord_key_inference}

To investigate whether key information is implied by chord-feature
activations, we design a \textbf{music-theory-driven, template-based} method
for estimating musical key from chord activations.

For each segment $I$, we aggregate the activations over time, apply
square-root compression, and normalize the two chord rings separately:
\begin{equation}
b^{m}_{I,p}
=
\frac{
\left(\bar{u}^{m}_{I,p}\right)^{1/2}
}{
\sum_{q\in\mathcal{P}}
\left(\bar{u}^{m}_{I,q}\right)^{1/2}
},
\qquad
m\in\{\mathrm{maj},\mathrm{min}\}.
\label{eq:chord_ring_normalization}
\end{equation}
This normalization prevents the overall activation strength of one chord
quality from dominating the key score. For a candidate tonic $k$, let
\begin{equation}
M_d=b^{\mathrm{maj}}_{I,(k+d)\bmod 12},
\qquad
m_d=b^{\mathrm{min}}_{I,(k+d)\bmod 12},
\label{eq:relative_chord_activation}
\end{equation}
where $d$ denotes the pitch offset in semitones relative to the candidate
tonic. Following common-practice tonal harmony, we compute a weighted sum of
these normalized activations. For a major key with tonic $k$, we emphasize the
tonic ($M_0$), subdominant ($M_5$), and dominant ($M_7$) chords, with
additional support from either the minor supertonic ($m_2$) or minor
submediant ($m_9$):
\begin{equation}
\mathrm{ScaleScore}_{\mathrm{maj}}(k)
=
M_0+0.4M_5+0.4M_7
+0.1\max(m_2,m_9).
\label{eq:major_key_template}
\end{equation}

For minor keys, we take the maximum over three templates corresponding to
common natural-, harmonic-, and melodic-minor chord patterns:
\begin{align}
\mathrm{ScaleScore}_{\mathrm{min}}(k)
=
\max\Big\{&
m_0+0.4m_5+0.4m_7+0.1M_8,\nonumber\\
&
m_0+0.4m_5+0.4M_7+0.1M_8,\nonumber\\
&
m_0+0.1m_2+0.4M_5+0.4M_7
\Big\}.
\label{eq:minor_key_template}
\end{align}

The predicted tonic and mode are selected jointly from the 24 candidates:
\begin{equation}
(\hat{k}_I,\hat{m}_I)
=
\arg\max_{\substack{k\in\mathcal{P}\\
m\in\{\mathrm{maj},\mathrm{min}\}}}
\mathrm{ScaleScore}_{m}(k).
\label{eq:joint_key_decision}
\end{equation}
At the song level, the 24 candidate scores are averaged across segments before
the final decision is made. For the supervised chord probe, we use the same scoring and decision procedure except treating the rectified major- and minor-chord logits as feature activations.

\section{Evaluation Metrics}
\label{app:evaluation_metrics}

\subsection{Chord Recognition}

We evaluate chord recognition using Root and MajMin Weighted Chord Symbol
Recall (WCSR), following the vocabularies implemented in
\texttt{mir\_eval}~\cite{raffel2014mir_eval}. Root WCSR evaluates only whether the predicted chord root is correct and
ignores chord quality. MajMin WCSR requires both the root and the major or
minor quality to be correct. Chord extensions that preserve an underlying
major or minor triad are mapped to the corresponding major or minor class. Reference chords outside the MajMin vocabulary are excluded from the MajMin
denominator. Chords with
unsupported qualities may still be evaluated under Root when their root pitch
class is defined. Unknown reference annotations are excluded from all chord
metrics, while the no-chord label is retained as an evaluated
class.

\subsection{Key Detection}

The tonic, subdominant, and dominant methods predict one of 12 key signatures,
where relative major and minor keys are treated as equivalent. Major keys are
represented by their tonic, while minor keys are mapped to their relative-major
tonic, three semitones above. We report exact key-signature accuracy, with no
partial credit for harmonically related errors. The metrics are computed and reported at both the segment and song levels.

For methods that predict both tonic and mode, we additionally report exact
24-class key accuracy and the standard MIREX weighted score:
\begin{equation}
\label{eq:mirex_key_weight}
\mathrm{MIREX Score Weight}
=
\begin{cases}
1.0, & \text{exact key},\\
0.5, & \text{fifth-related key in the same mode},\\
0.3, & \text{relative major--minor key},\\
0.2, & \text{parallel major--minor key},\\
0,   & \text{otherwise}.
\end{cases}
\end{equation}

\section{Batched Post-hoc Evaluation Protocol}
\label{app:automated_protocol}

Our qualitative analyses assign semantic labels to recovered orbits by
inspecting their activations on one or a few annotated music excerpts. To
evaluate many model layers, sparsity settings, and random seeds efficiently
and consistently, we batch the same post-hoc procedure. Using a small set of
labeled anchors, the scripts match recovered structures to musical concepts,
resolve their cyclic offsets, and select auxiliary features such as
chord-boundary and silence features.

Importantly, this procedure is applied only after SAE training and orbit
recovery. It does not introduce an additional training objective, modify the
recovered graph, or affect orbit membership. It therefore approximates manual
inspection without influencing how the orbit structures emerge.

The supervised probe baselines use validation performance for checkpoint
selection. To provide a similarly stable criterion for SAE checkpoints, we
use a fixed trained chord probe only as a post-hoc semantic reference. At each
checkpoint, the SAE features with the highest cosine similarity to the 12
major- and 12 minor-chord probe weight vectors are selected and evaluated on
the validation set. The resulting chord-recognition accuracy is used solely
to rank checkpoints; neither gradients nor probe supervision are introduced
during SAE training. The chord orbits themselves emerge independently of this
selection procedure.

\begin{figure}[t]
  \centering
\includegraphics[width=\linewidth]{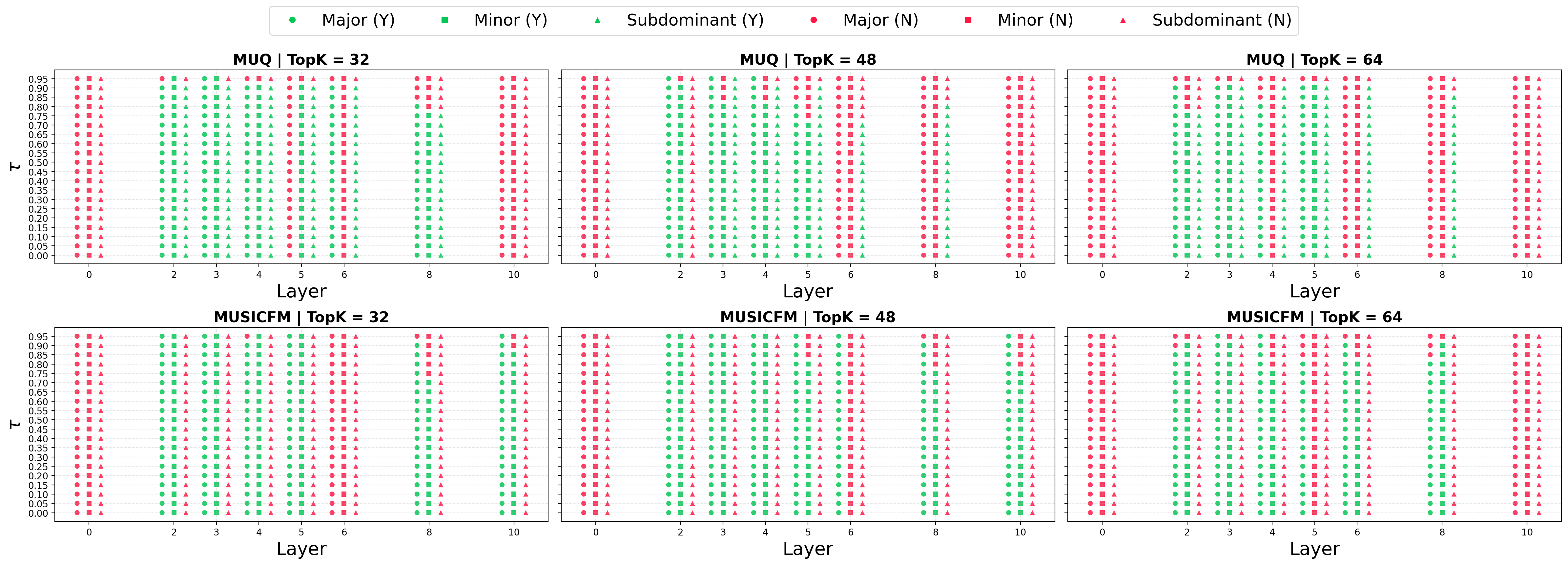}

  \caption{
  The recovery of three core orbits across different topK settings and different-layer representations under different $\tau$.
  }
  \label{fig:tau_trhee_rings}
\end{figure}

\begin{figure}[p]
  \centering

  \begin{subfigure}[t]{0.47\textwidth}
    \centering
    \includegraphics[width=0.8\linewidth]{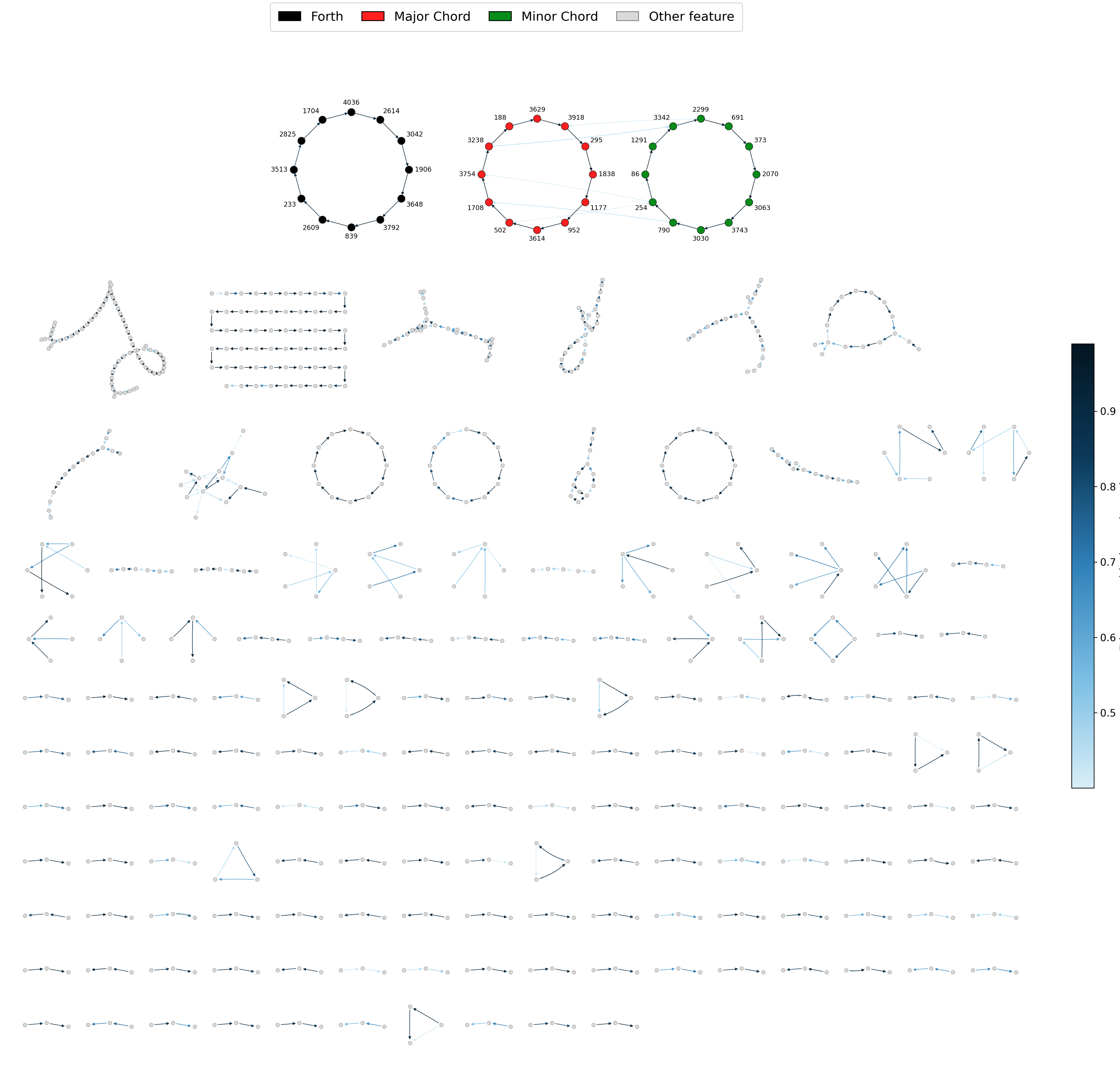}
    \caption{MuQ (2), $\tau=0.4$}

  \end{subfigure}
\hfill
  \begin{subfigure}[t]{0.47\textwidth}
    \centering
    \includegraphics[width=0.8\linewidth]{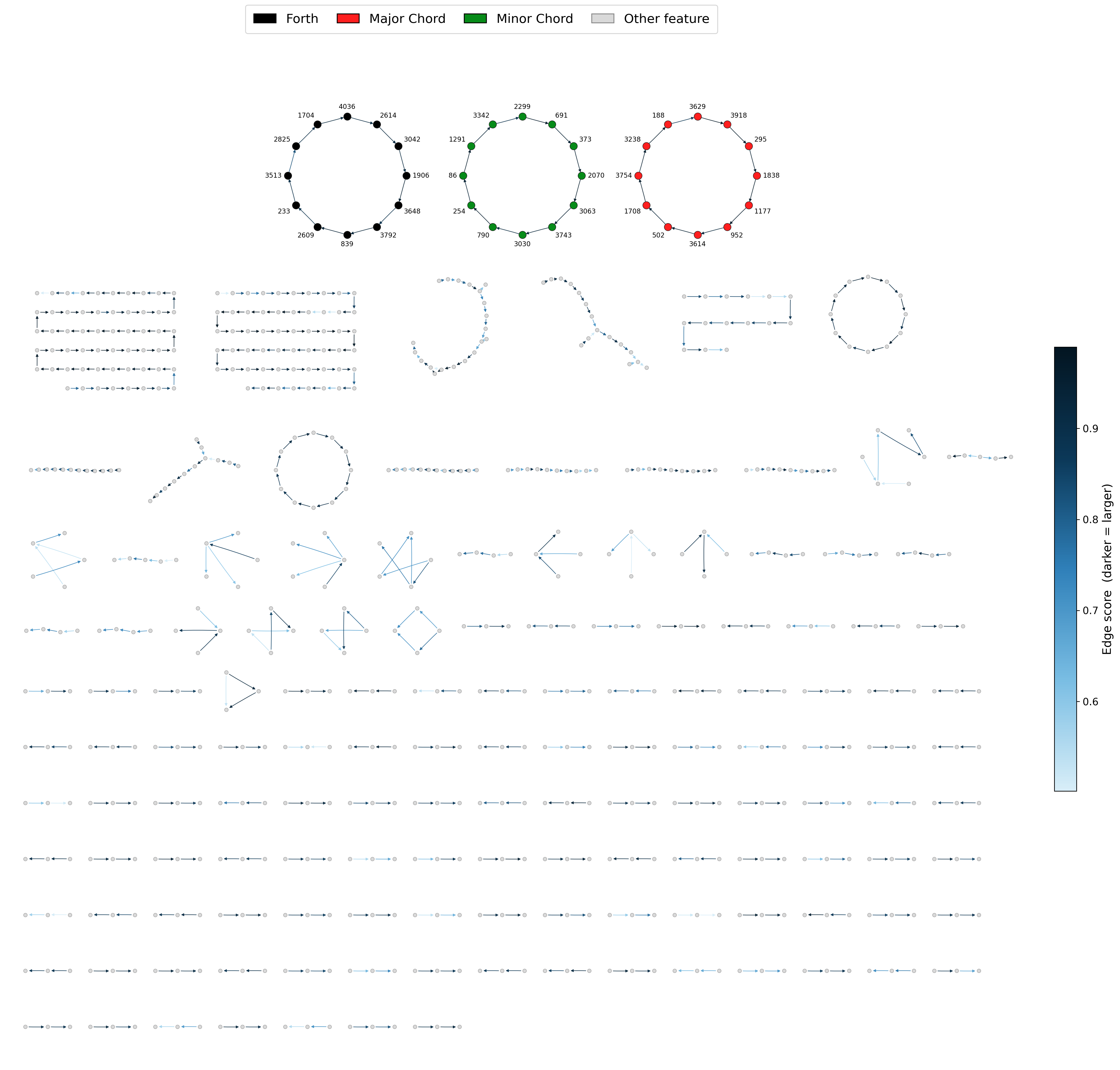}
    \caption{MuQ (2), $\tau=0.5$}\label{fig:orbit_muq_2_05}
  \end{subfigure}

  \begin{subfigure}[t]{0.47\textwidth}
    \centering
    \includegraphics[width=0.8\linewidth]{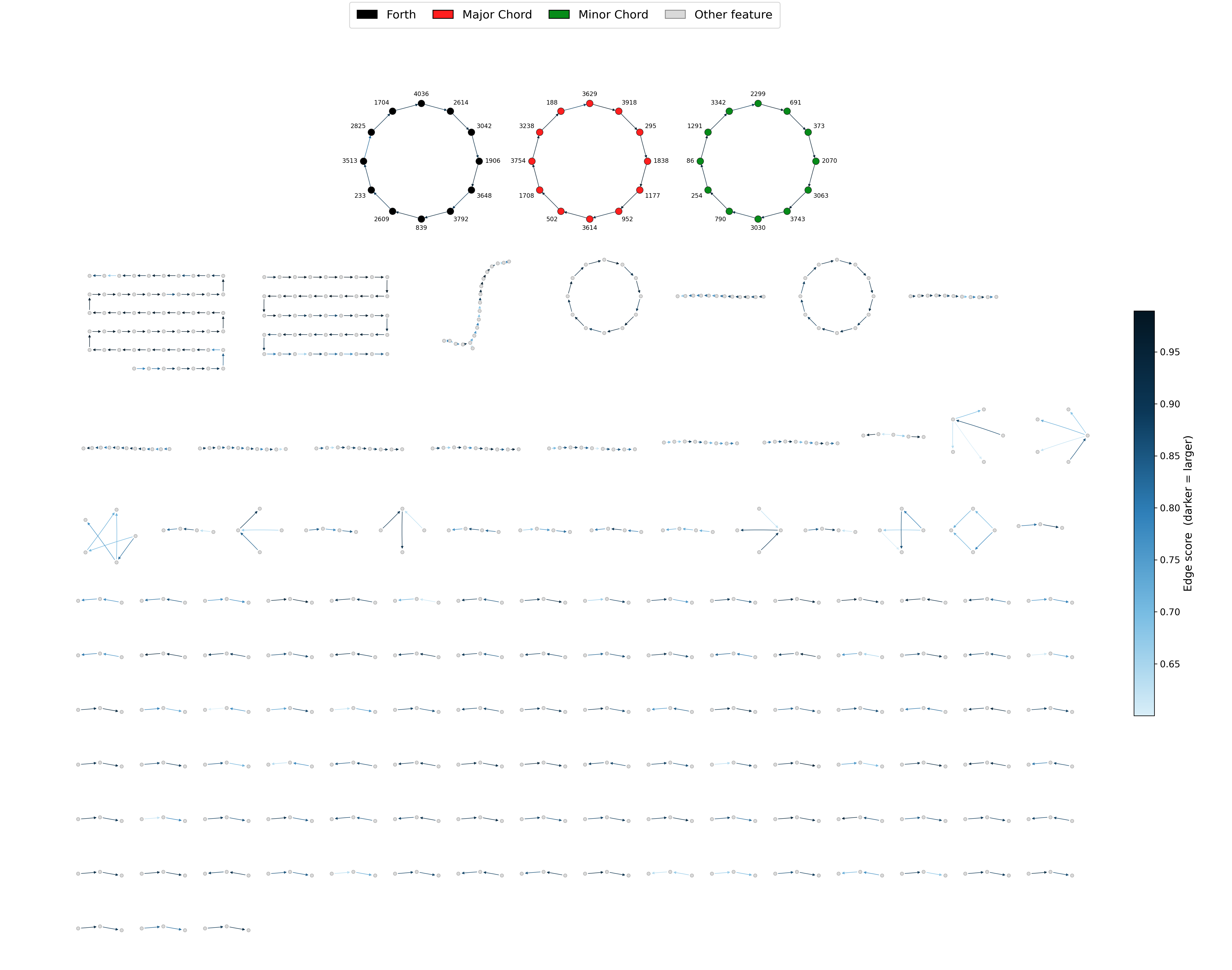}
    \caption{MuQ (2), $\tau=0.6$}
  \end{subfigure}
\hfill
  \begin{subfigure}[t]{0.47\textwidth}
    \centering
    \includegraphics[width=0.8\linewidth]{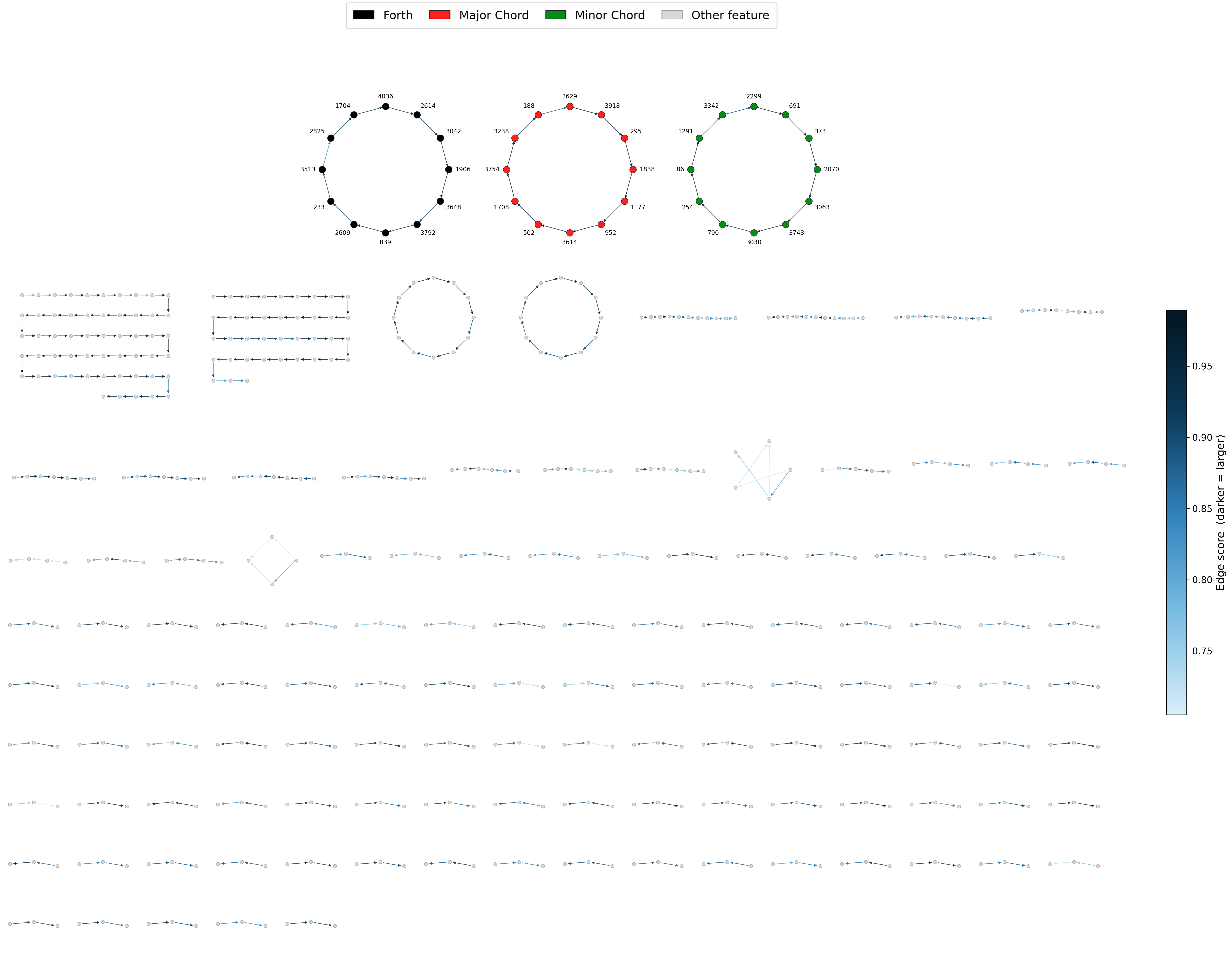}
    \caption{MuQ (2), $\tau=0.7$}
  \end{subfigure}
  
  \begin{subfigure}[t]{0.47\textwidth}
    \centering
    \includegraphics[width=0.8\linewidth]{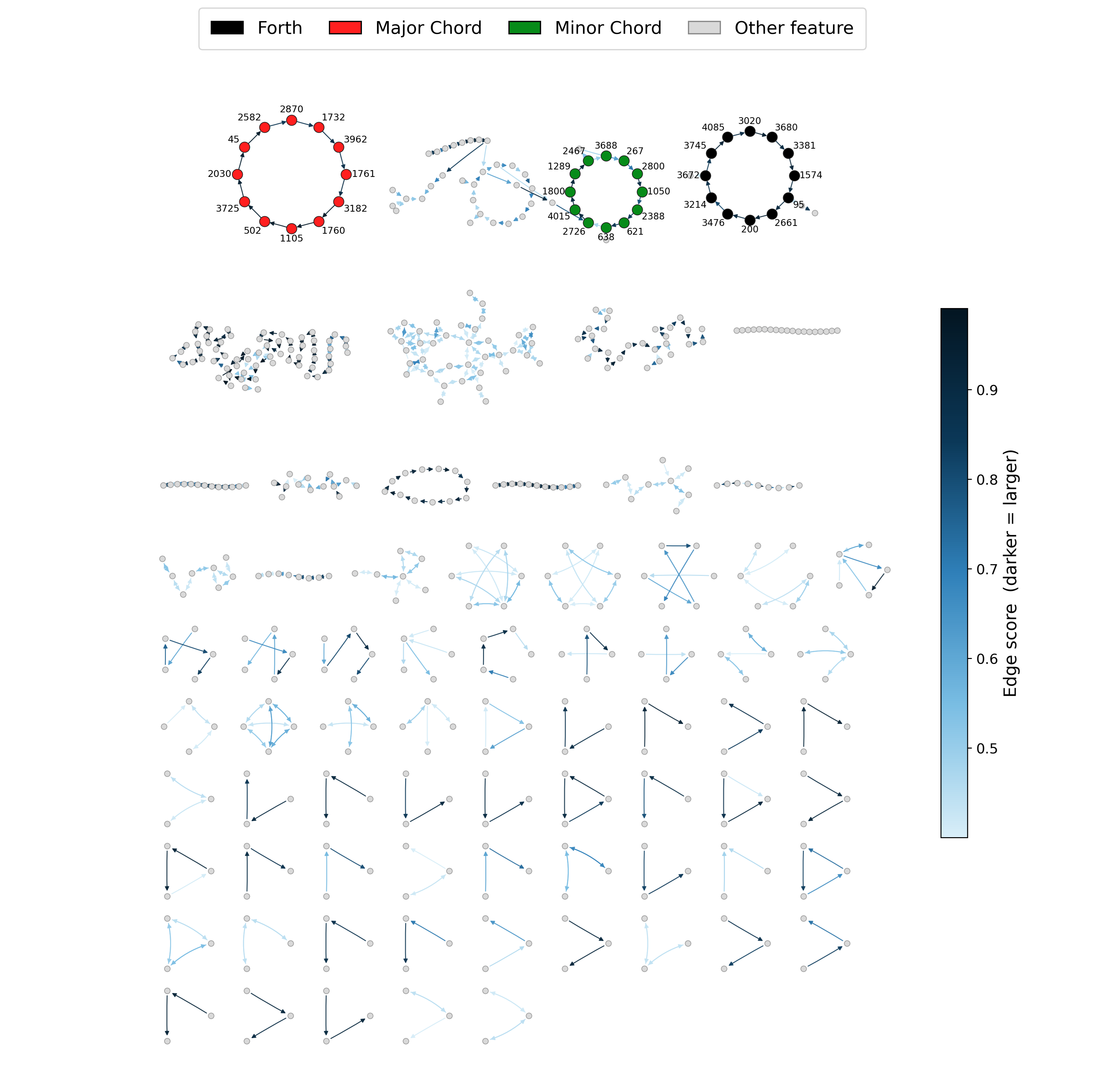}
    \caption{MuQ (6), $\tau=0.4$}

  \end{subfigure}
\hfill
  \begin{subfigure}[t]{0.47\textwidth}
    \centering
    \includegraphics[width=0.8\linewidth]{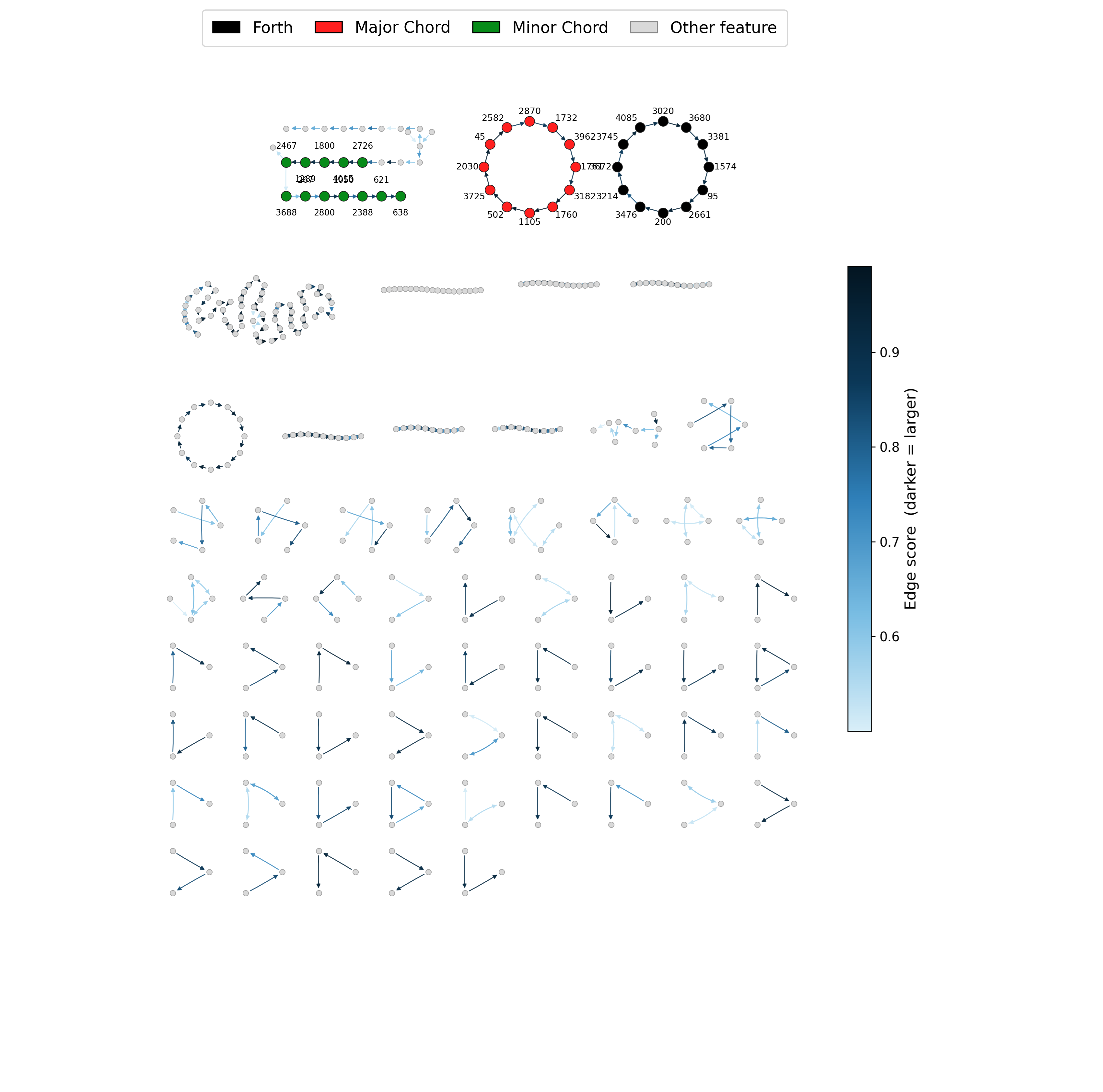}
    \caption{MuQ (6), $\tau=0.5$}
  \end{subfigure}

  \begin{subfigure}[t]{0.47\textwidth}
    \centering
    \includegraphics[width=0.8\linewidth]{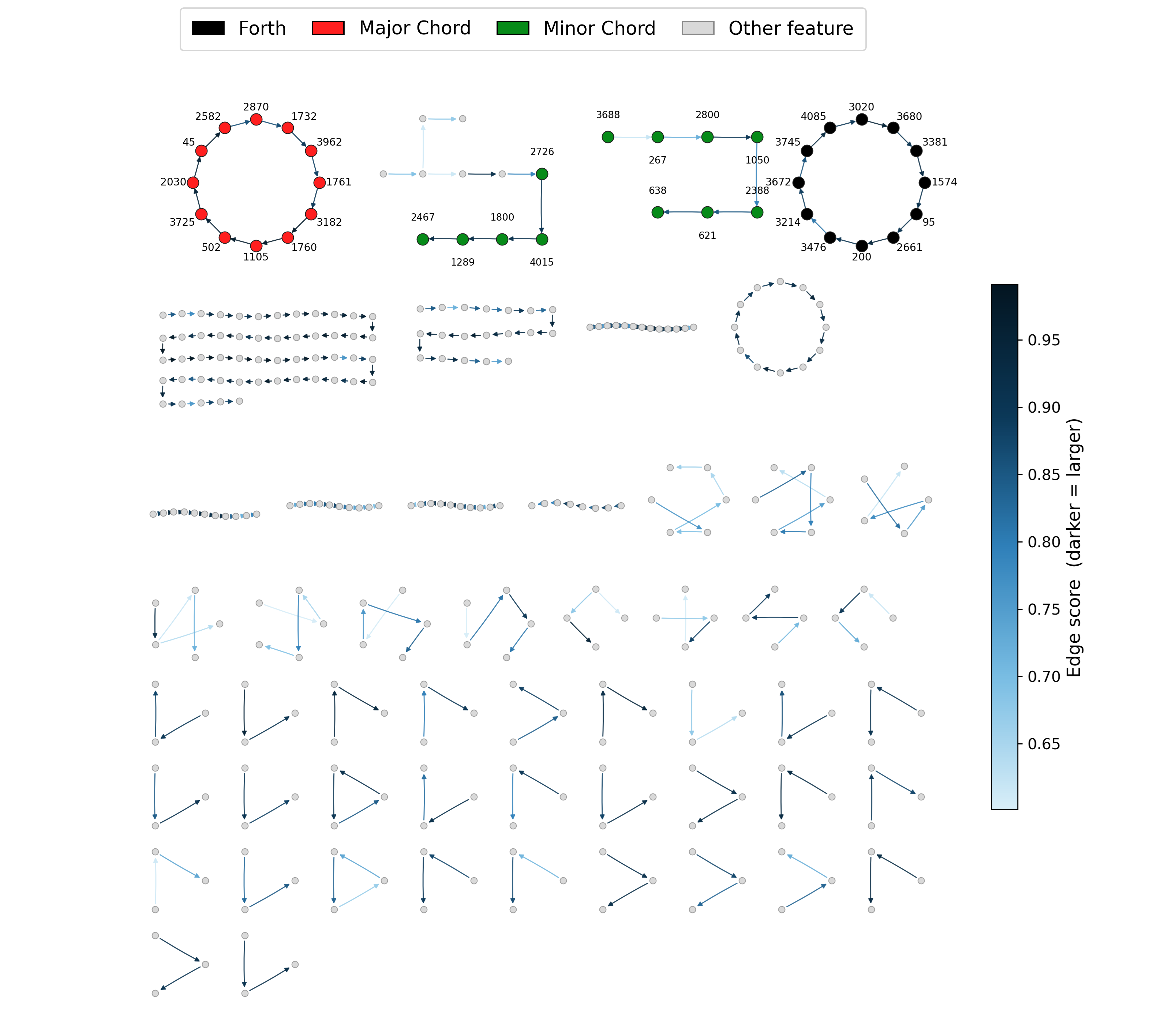}
    \caption{MuQ (6), $\tau=0.6$}
  \end{subfigure}
\hfill
  \begin{subfigure}[t]{0.47\textwidth}
    \centering
    \includegraphics[width=0.8\linewidth]{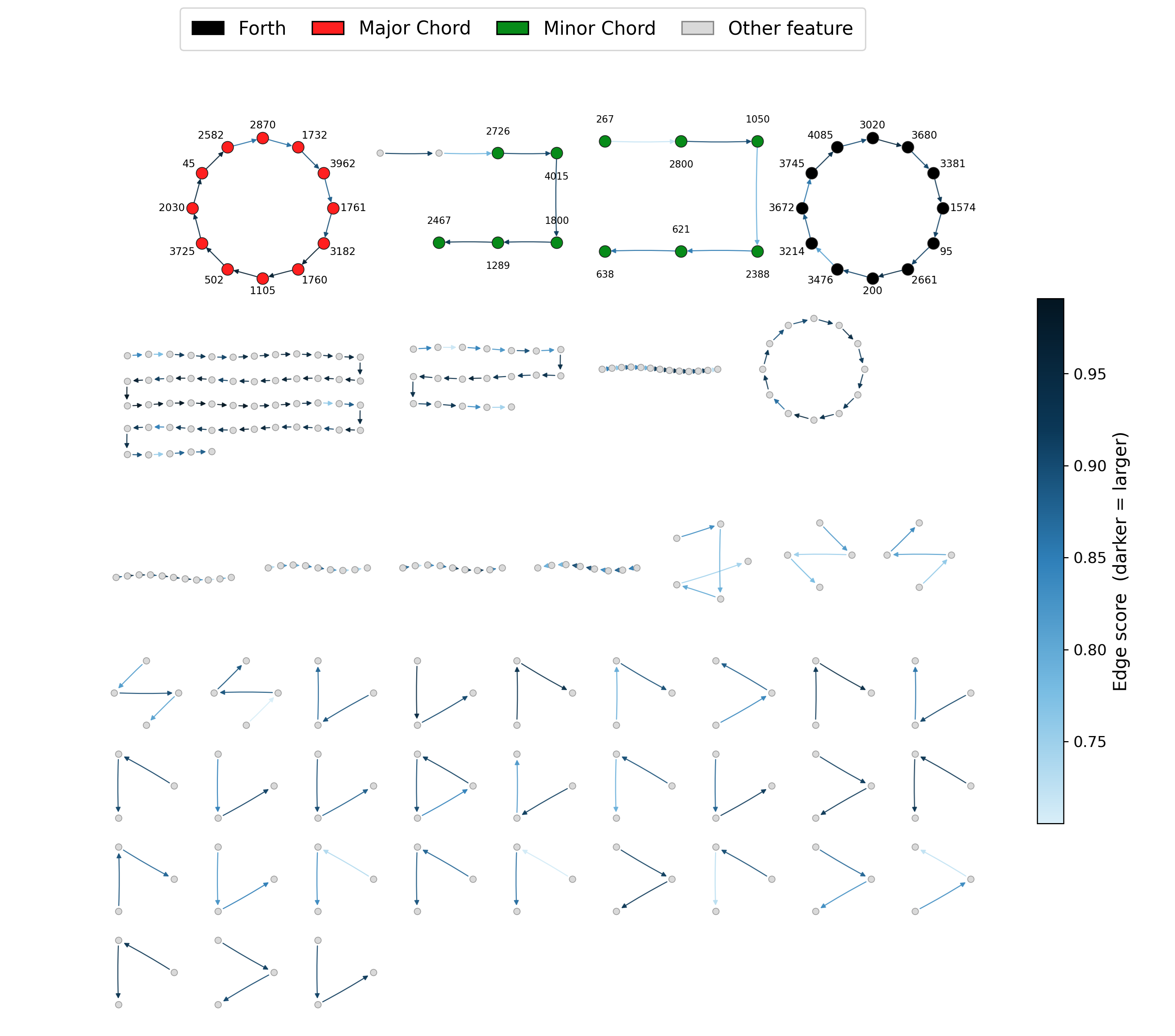}
    \caption{MuQ (6), $\tau=0.7$}
  \end{subfigure}

  \caption{Directed orbit graphs recovered from MuQ at $K=32$ with $n_e=10$ and varying $\tau$. Fixed points and components with fewer than three nodes are omitted, and ``forth'' refers to the subdominant.}\label{fig:orbit_graphs}
 
\end{figure}

\begin{figure}[t]
  \centering

  \begin{subfigure}[t]{0.48\textwidth}
    \centering
    \includegraphics[width=0.8\linewidth]{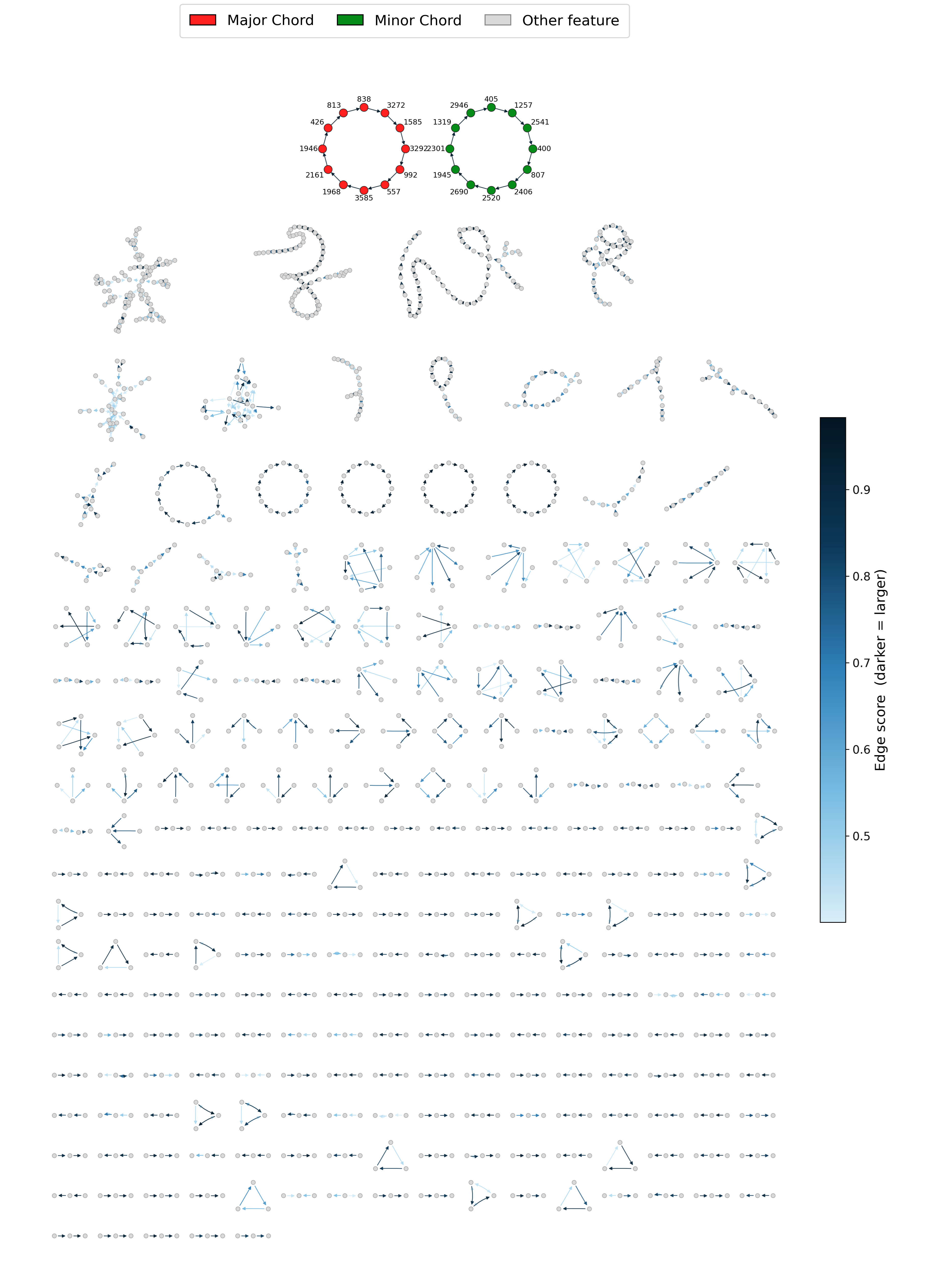}
    \caption{MusicFM (2), $\tau=0.4$}

  \end{subfigure}
\hfill
  \begin{subfigure}[t]{0.48\textwidth}
    \centering
    \includegraphics[width=0.8\linewidth]{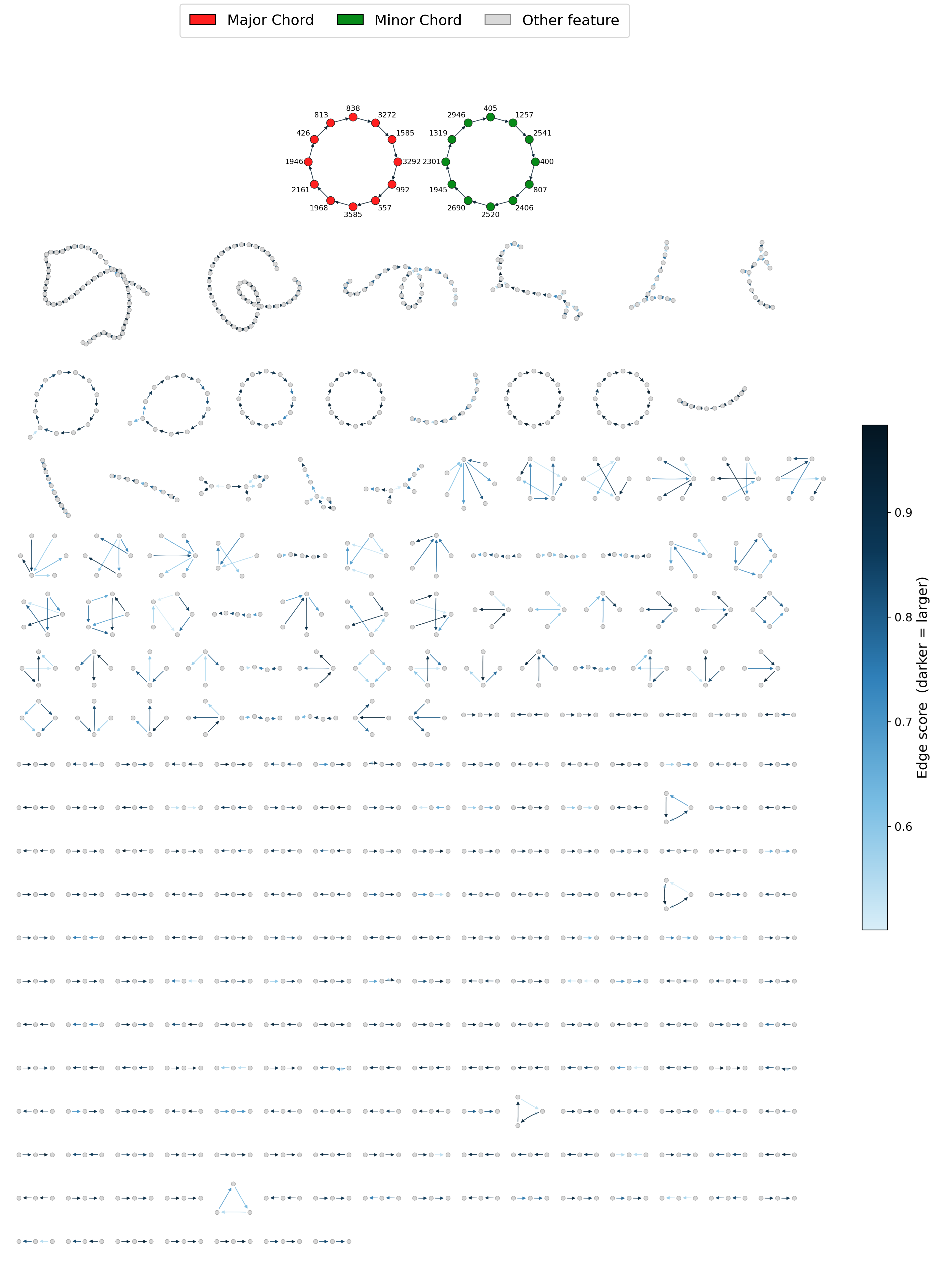}
    \caption{MusicFM (2), $\tau=0.5$}
  \end{subfigure}

  \begin{subfigure}[t]{0.48\textwidth}
    \centering
    \includegraphics[width=0.8\linewidth]{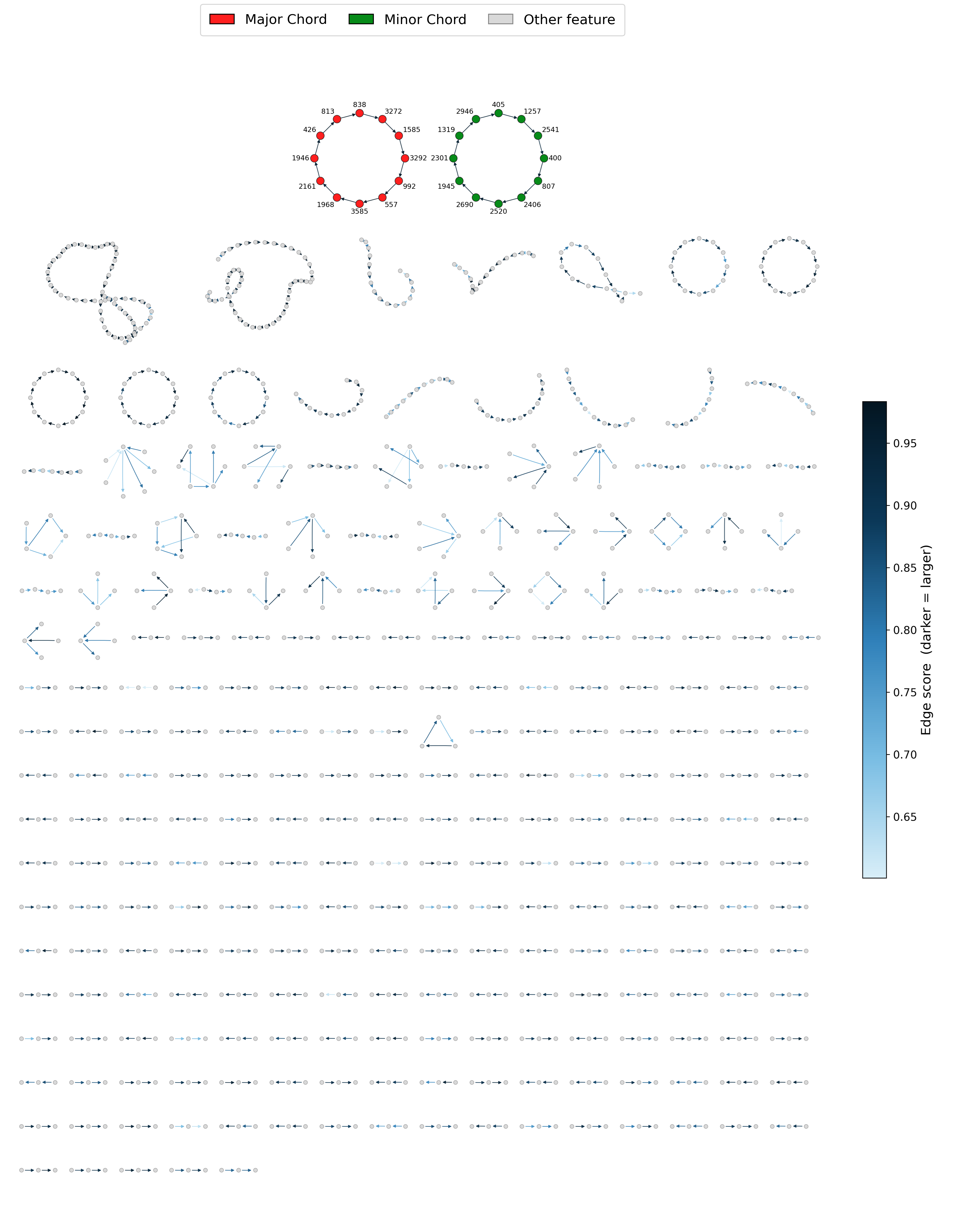}
    \caption{MusicFM (2), $\tau=0.6$}
  \end{subfigure}
\hfill
  \begin{subfigure}[t]{0.48\textwidth}
    \centering
    \includegraphics[width=0.8\linewidth]{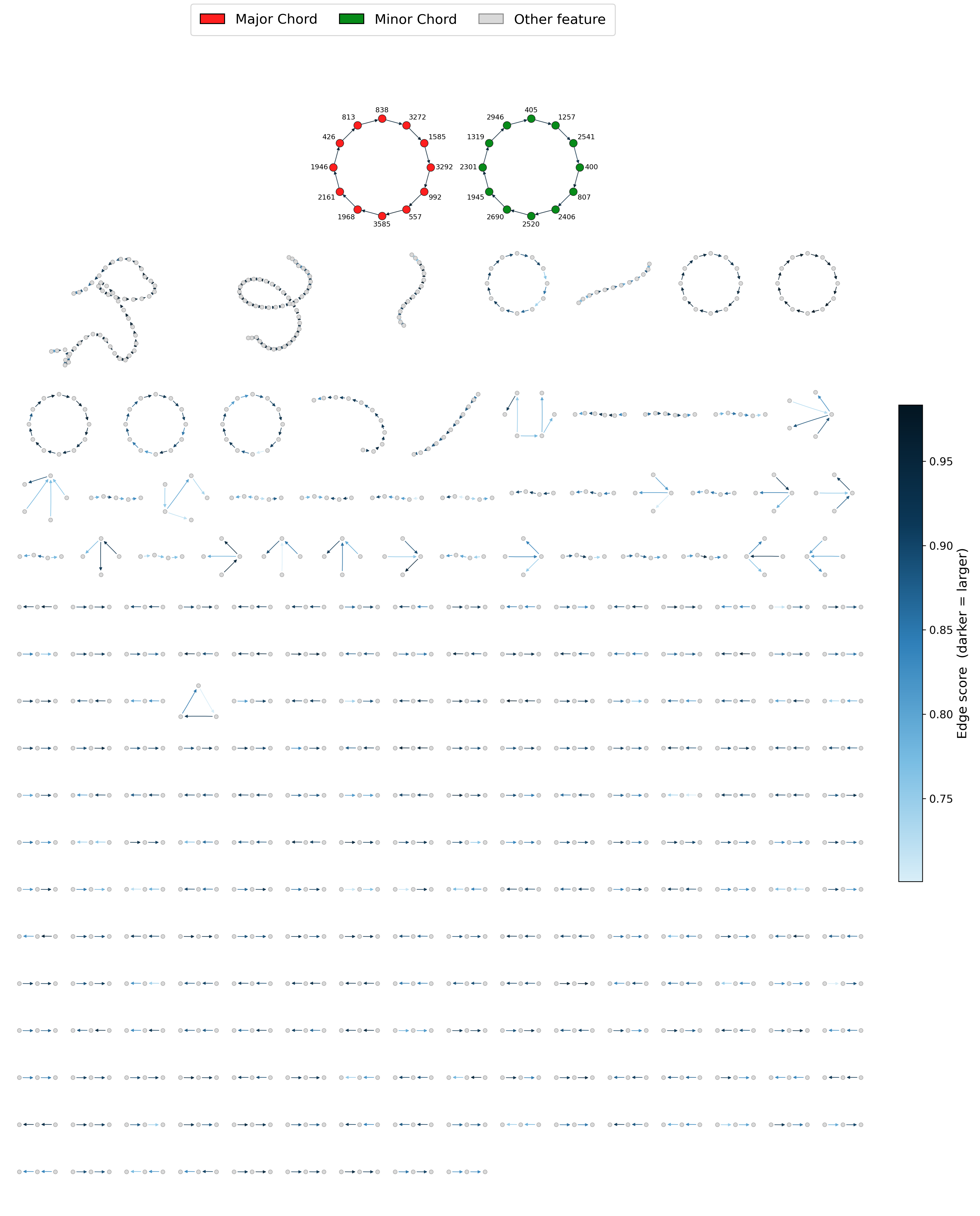}
    \caption{MusicFM (2), $\tau=0.7$}
  \end{subfigure}

\caption{Directed orbit graphs recovered from MusicFM at $K=32$ with $n_e=10$ and varying $\tau$. Fixed points and components with fewer than three nodes are omitted.}\label{fig:orbit_graphs_add}
 
\end{figure}

\section{Visualization of Recovered Orbits}
\label{app:orbit_visualization}

We further analyze the recovered feature orbits from three perspectives:
their sensitivity to the recovery threshold, their structures in a less
restrictive graph, and their activation patterns across music samples.

\subsection{Sensitivity to the Recovery Threshold}

The confidence threshold $\tau$ controls which feature transitions are retained
during orbit recovery. We vary $\tau$ and record whether the major-chord,
minor-chord, and subdominant orbits are recovered across different model
layers and sparsity settings. Figure~\ref{fig:tau_trhee_rings} summarizes the
results and shows how consistently each core orbit emerges across the range of
thresholds. Interestingly, no minor-chord orbit is recovered from MuQ layer~6
at any tested threshold. We investigate this result further in the section below.

\begin{figure}[t]
  \centering
   \begin{subfigure}[t]{0.19\textwidth}
    \centering
\includegraphics[width=\linewidth]{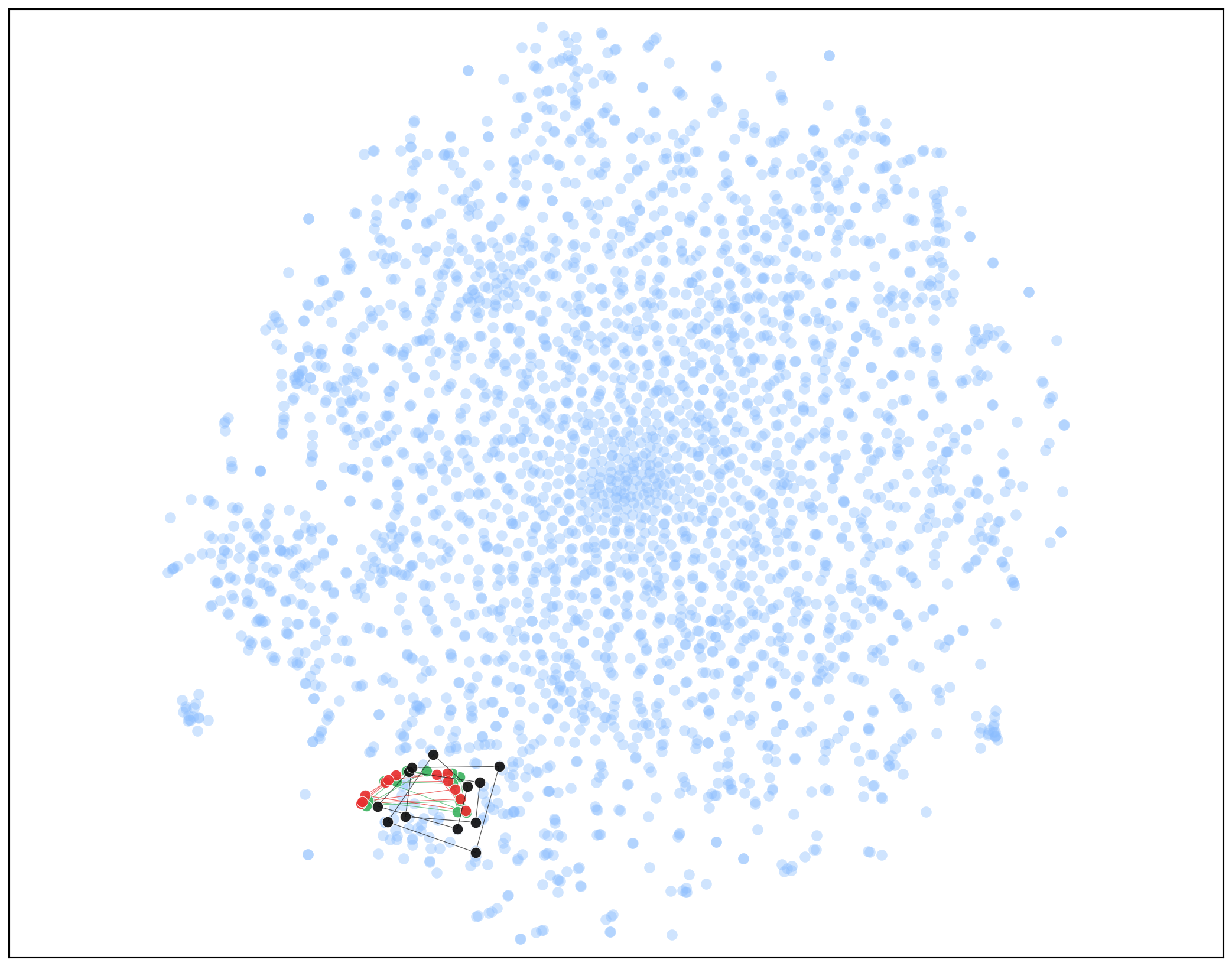}

  \end{subfigure}
   \hfill
\begin{subfigure}[t]{0.19\textwidth}
    \centering
\includegraphics[width=\linewidth]{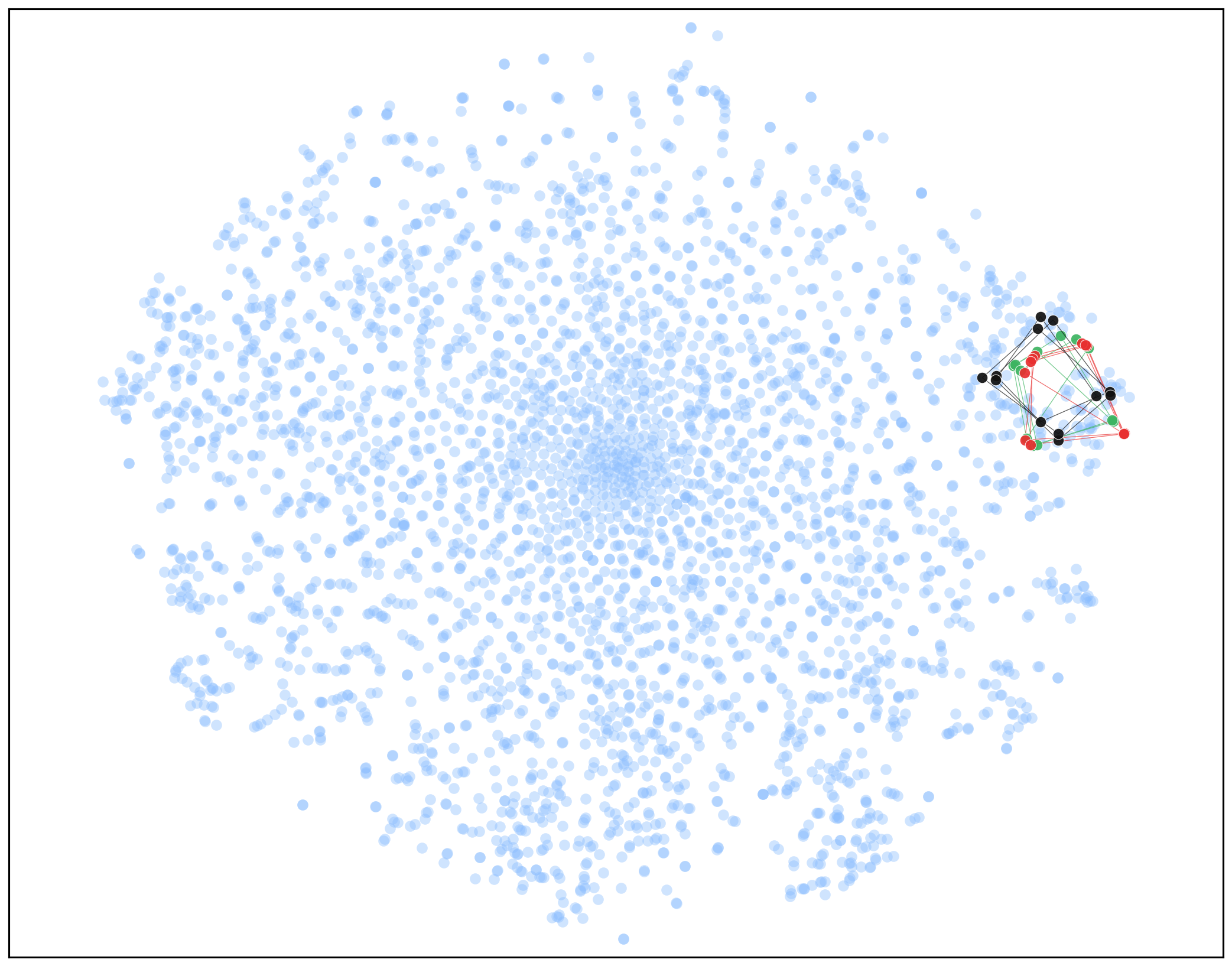}

  \end{subfigure}
   \hfill
\begin{subfigure}[t]{0.19\textwidth}
    \centering
\includegraphics[width=\linewidth]{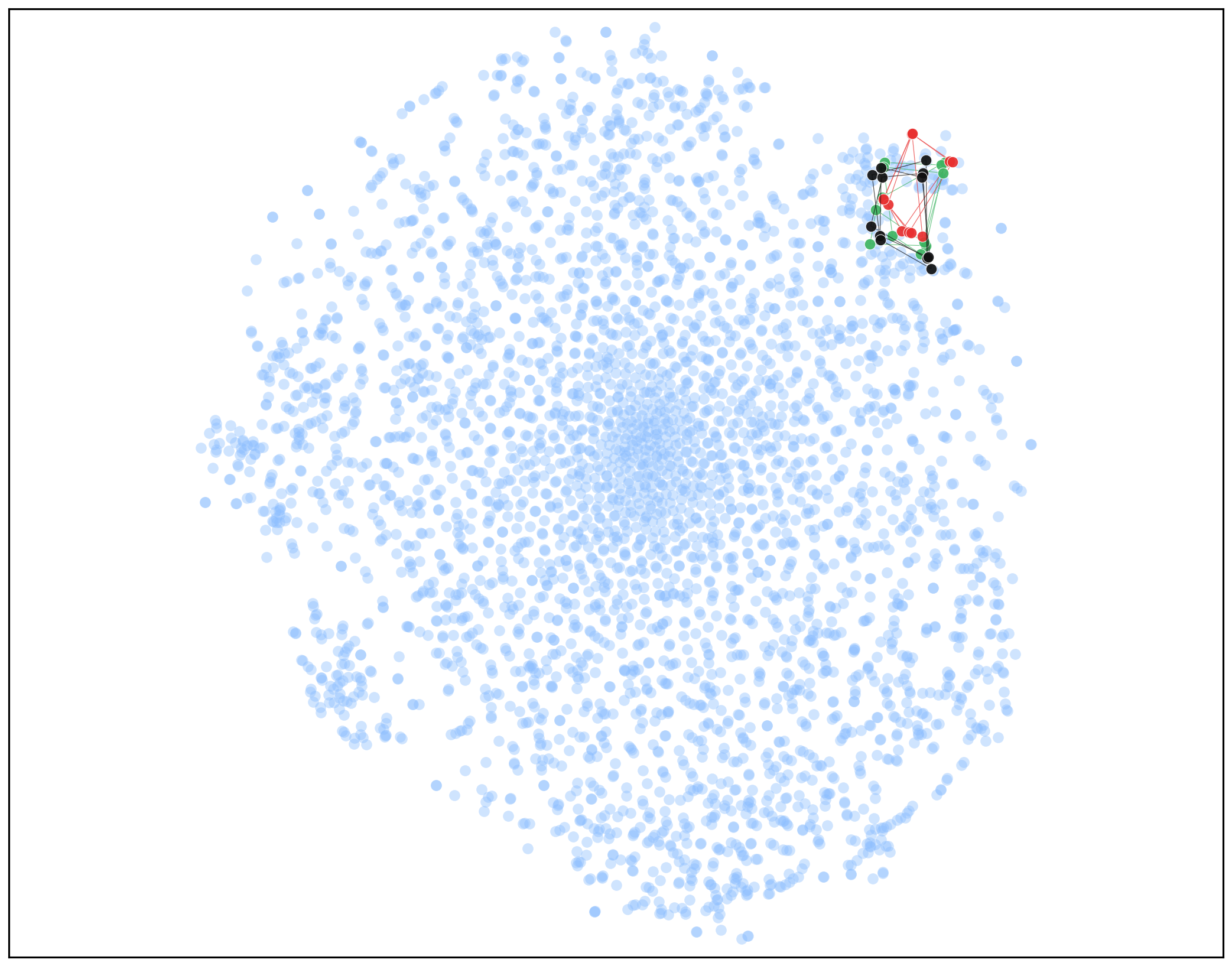}

  \end{subfigure}
   \hfill
\begin{subfigure}[t]{0.19\textwidth}
    \centering
\includegraphics[width=\linewidth]{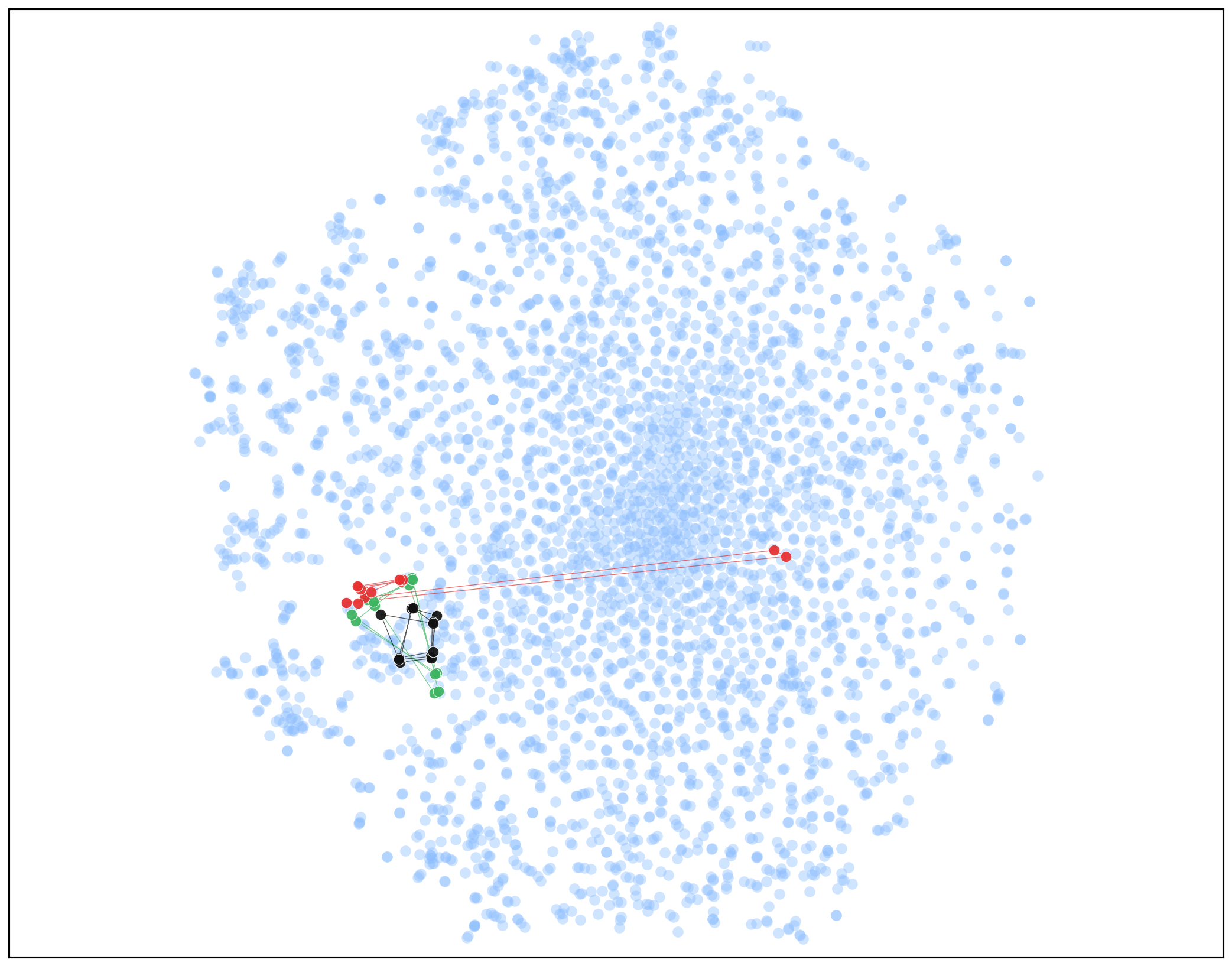}

  \end{subfigure}
   \hfill
\begin{subfigure}[t]{0.19\textwidth}
    \centering
\includegraphics[width=\linewidth]{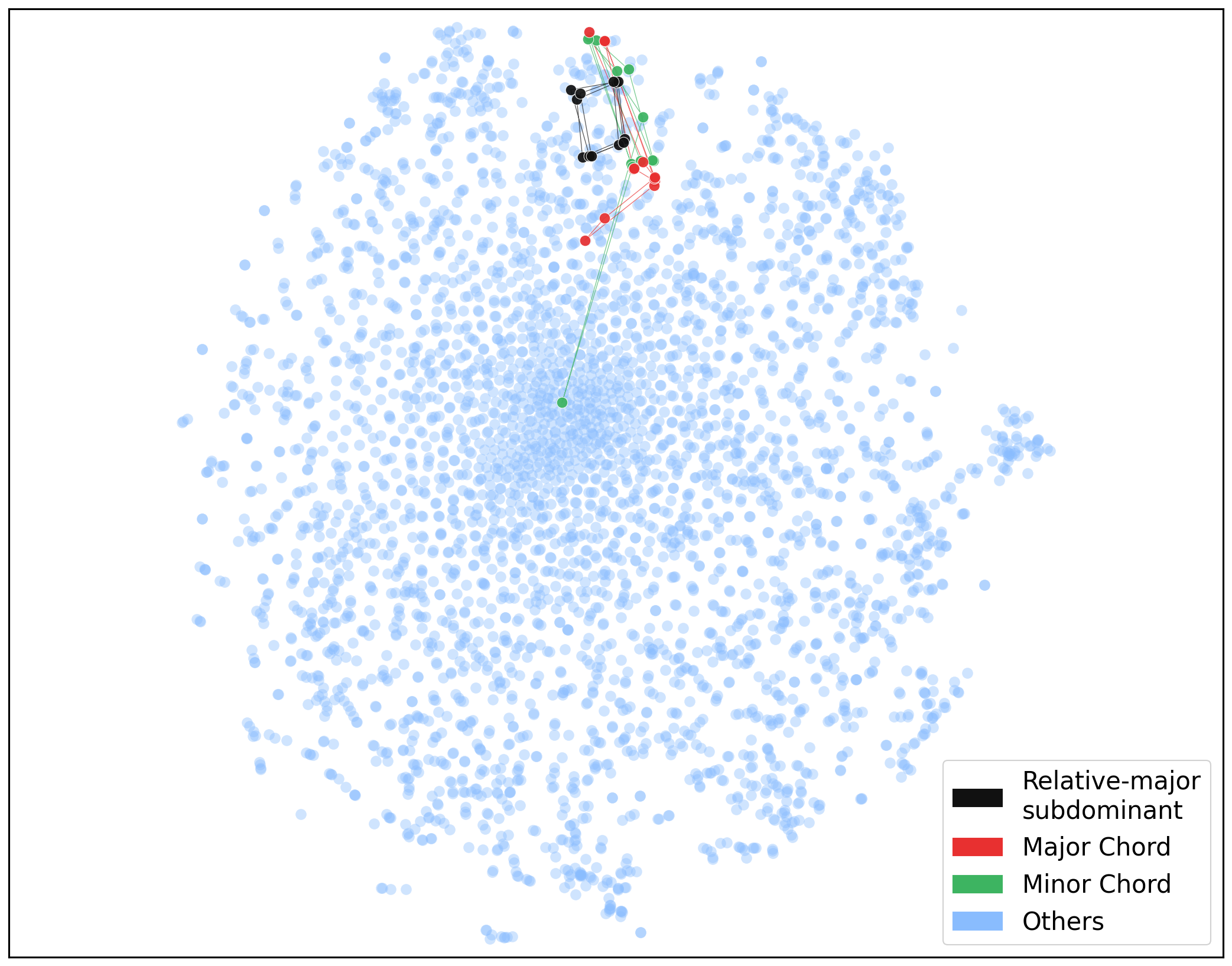}

  \end{subfigure}
   \hfill
 
  \centering
   \begin{subfigure}[t]{0.19\textwidth}
    \centering
\includegraphics[width=\linewidth]{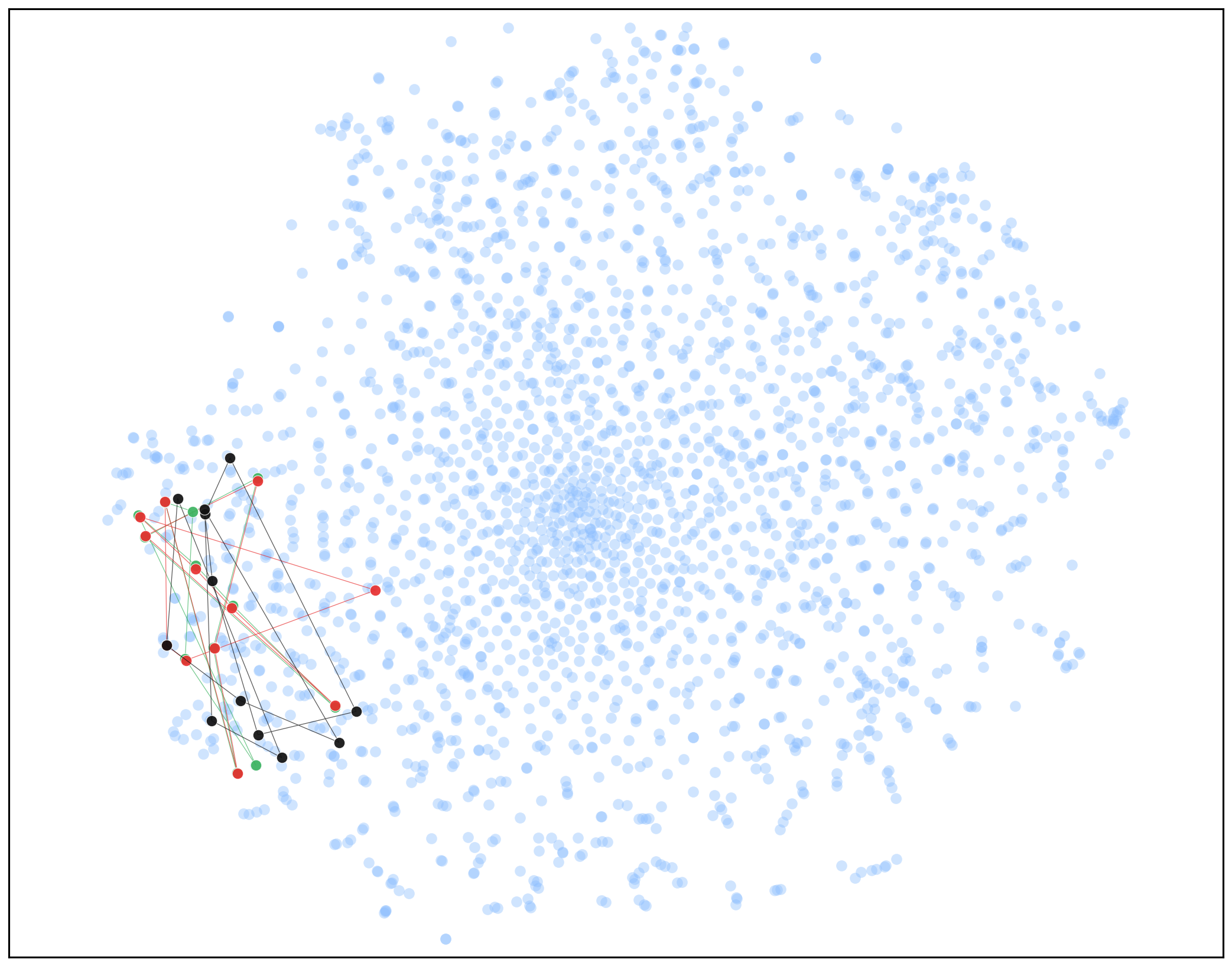}
    \caption{MuQ (2)}
   
  \end{subfigure}
   \hfill
\begin{subfigure}[t]{0.19\textwidth}
    \centering
\includegraphics[width=\linewidth]{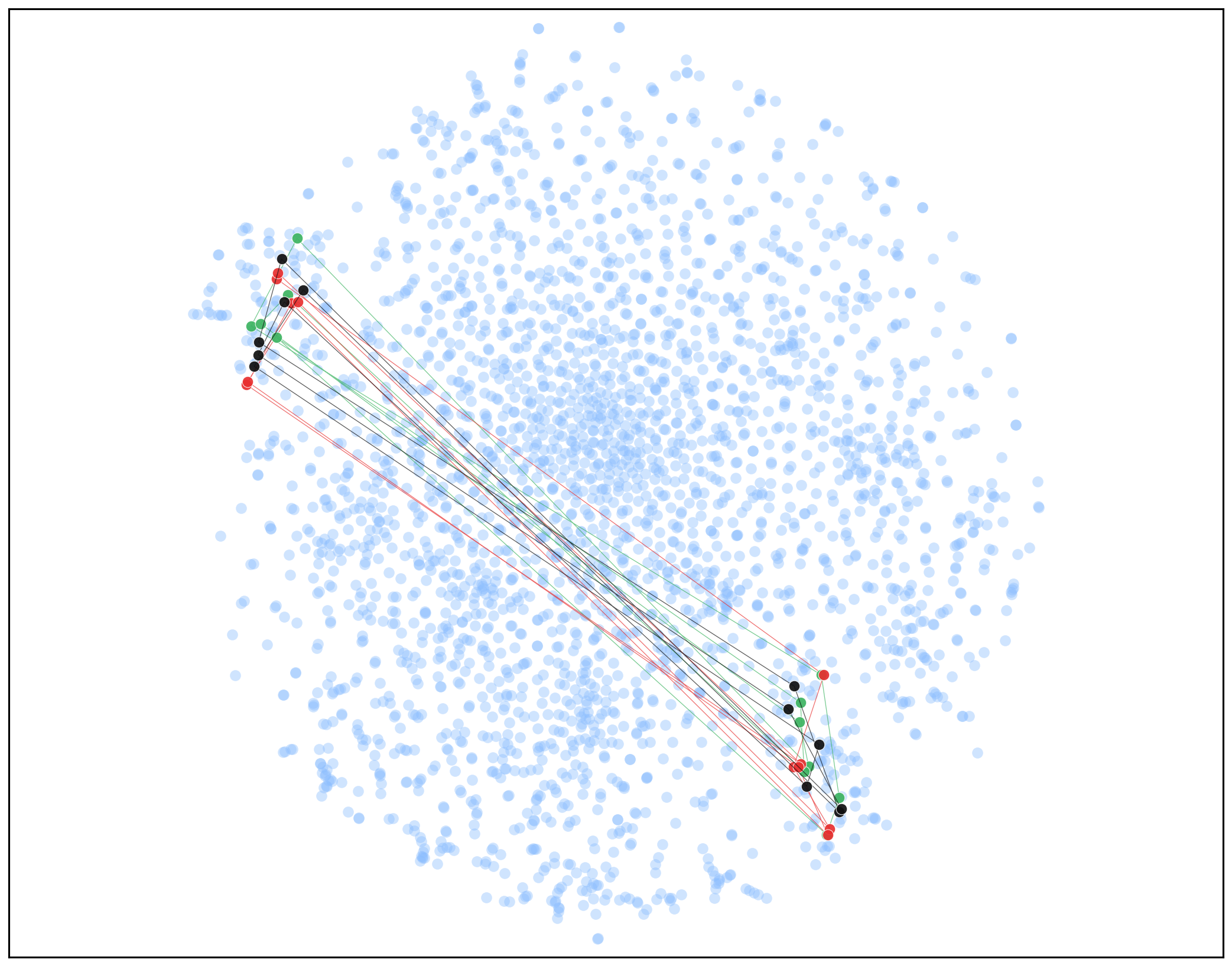}
    \caption{MuQ (3)}
   
  \end{subfigure}
   \hfill
\begin{subfigure}[t]{0.19\textwidth}
    \centering
\includegraphics[width=\linewidth]{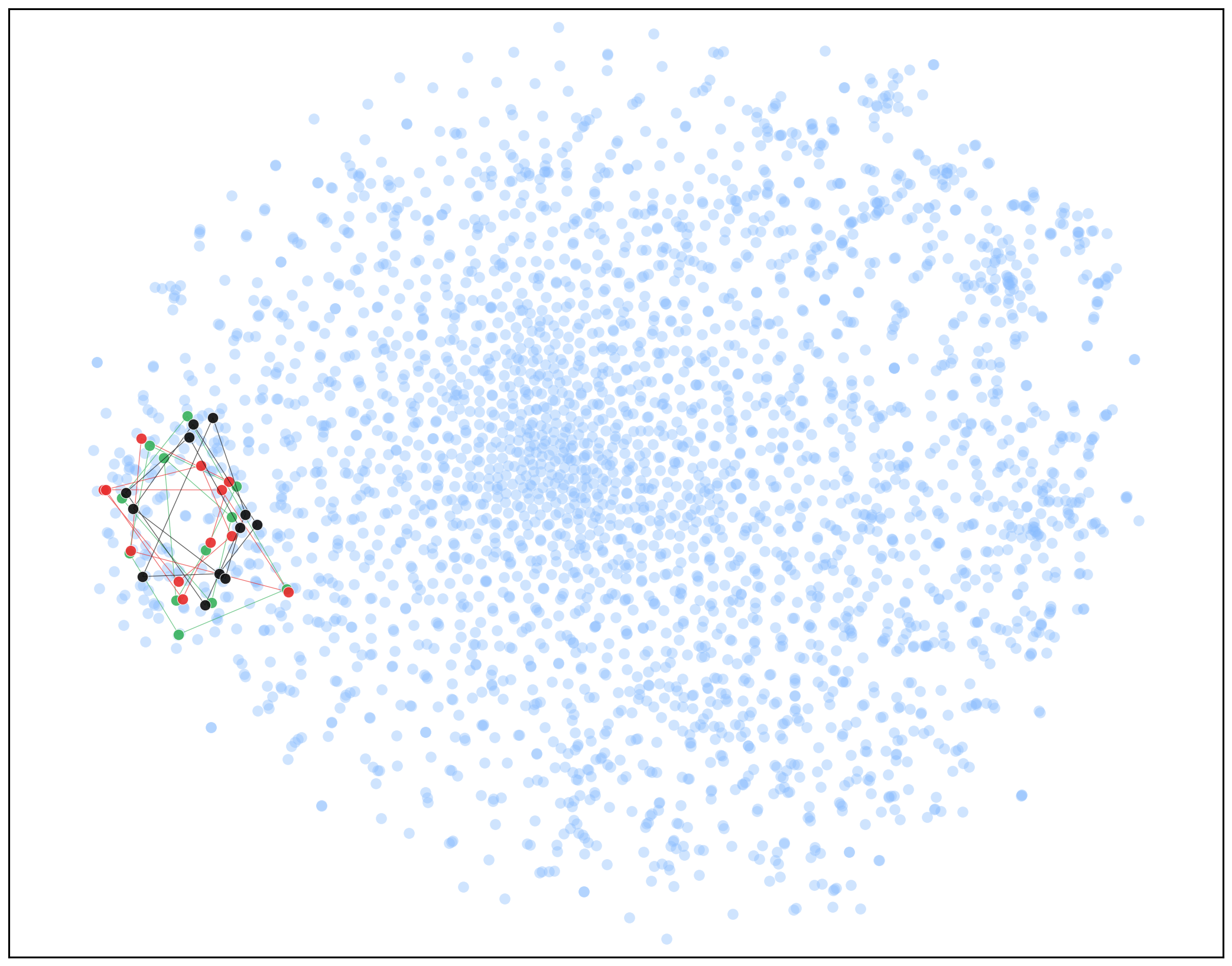}
    \caption{MuQ (4)}
   
  \end{subfigure}
   \hfill
\begin{subfigure}[t]{0.19\textwidth}
    \centering
\includegraphics[width=\linewidth]{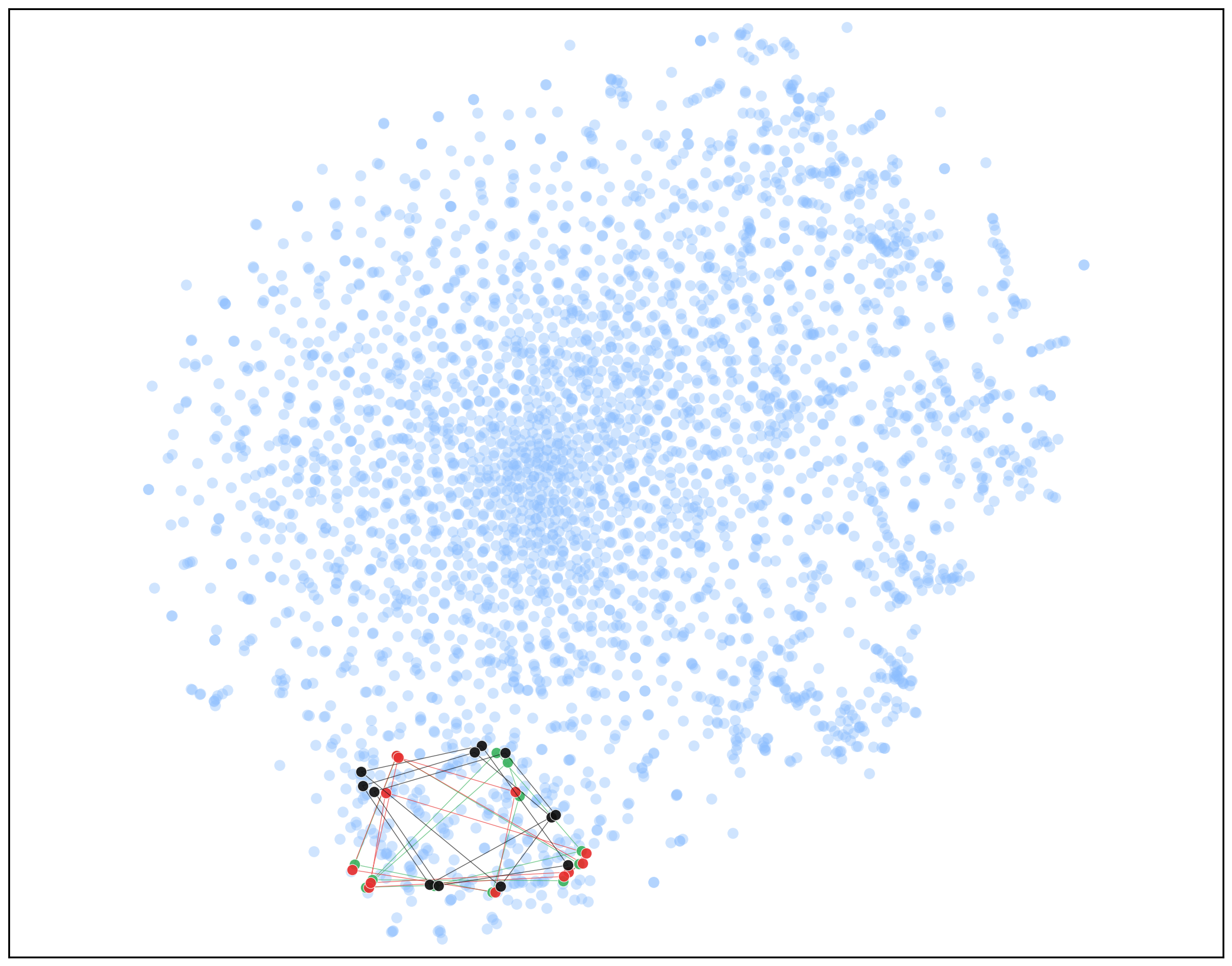}
    \caption{MuQ (5)}
   
  \end{subfigure}
   \hfill
\begin{subfigure}[t]{0.19\textwidth}
    \centering
\includegraphics[width=\linewidth]{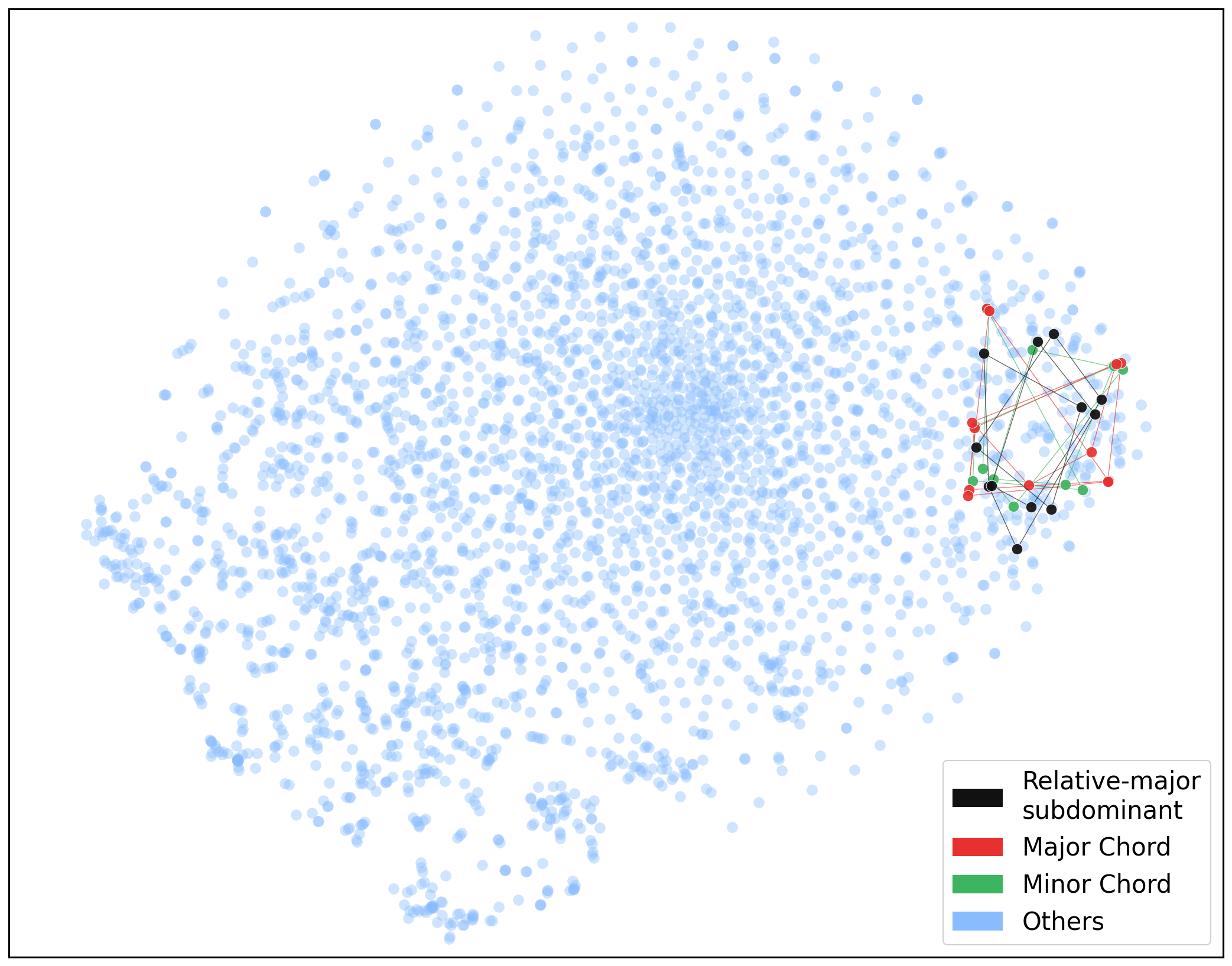}
    \caption{MuQ (6)}
   
  \end{subfigure}
   \hfill
 
  \caption{t-SNE projections of normalized SAE decoder vectors across different MuQ
layers at $K=32$. The columns correspond to different layers; the top row
shows o-SAE, and the bottom row shows s-SAE. Black, red, and green points
denote features in the subdominant, major-chord, and minor-chord orbits,
respectively, while blue points denote the remaining active features.}\label{fig:feature_tsne}
  
\end{figure}

\begin{figure}[t]
  \centering
   \begin{subfigure}[t]{0.19\textwidth}
    \centering
\includegraphics[width=\linewidth]{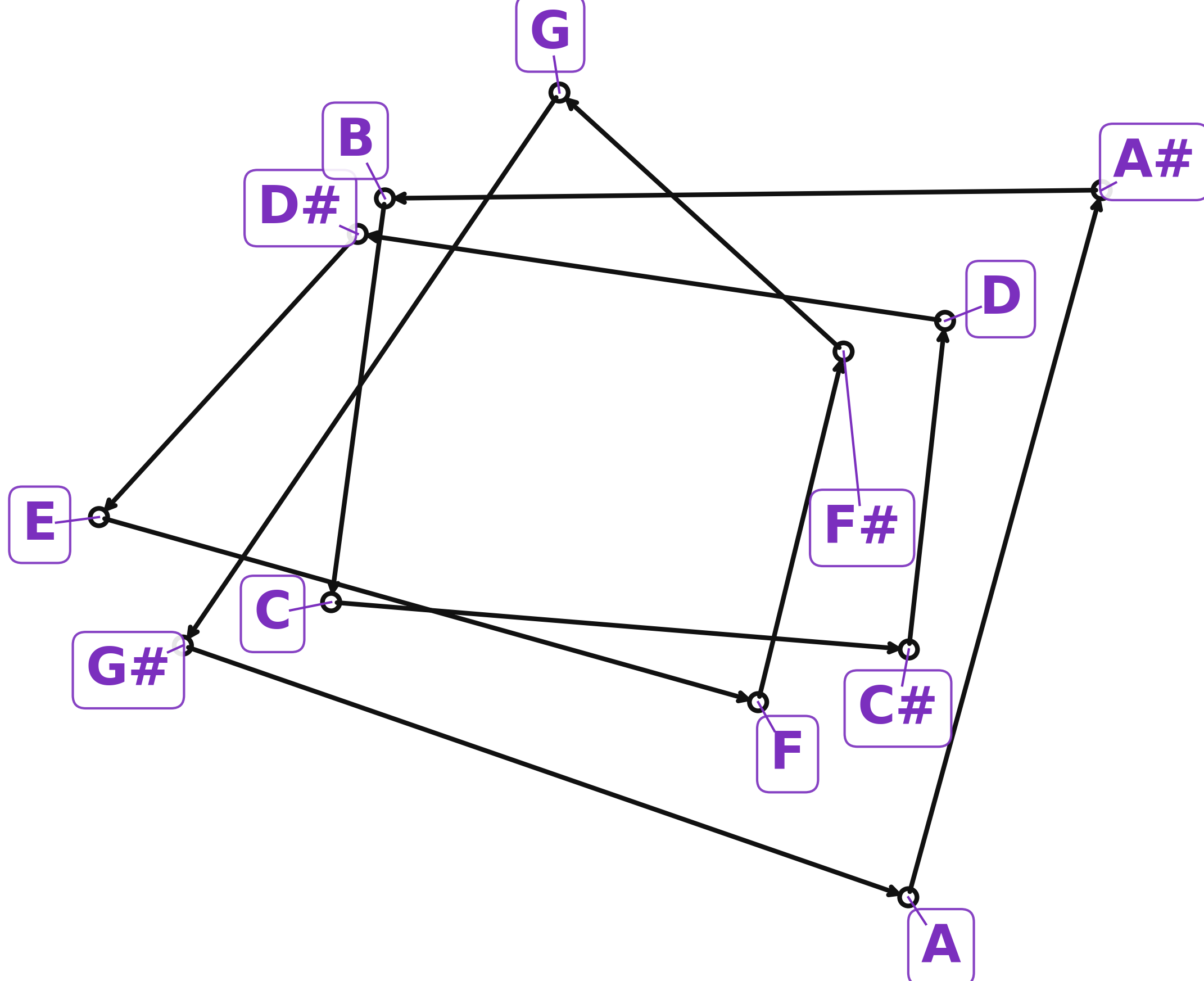}

  \end{subfigure}
   \hfill
\begin{subfigure}[t]{0.19\textwidth}
    \centering
\includegraphics[width=\linewidth]{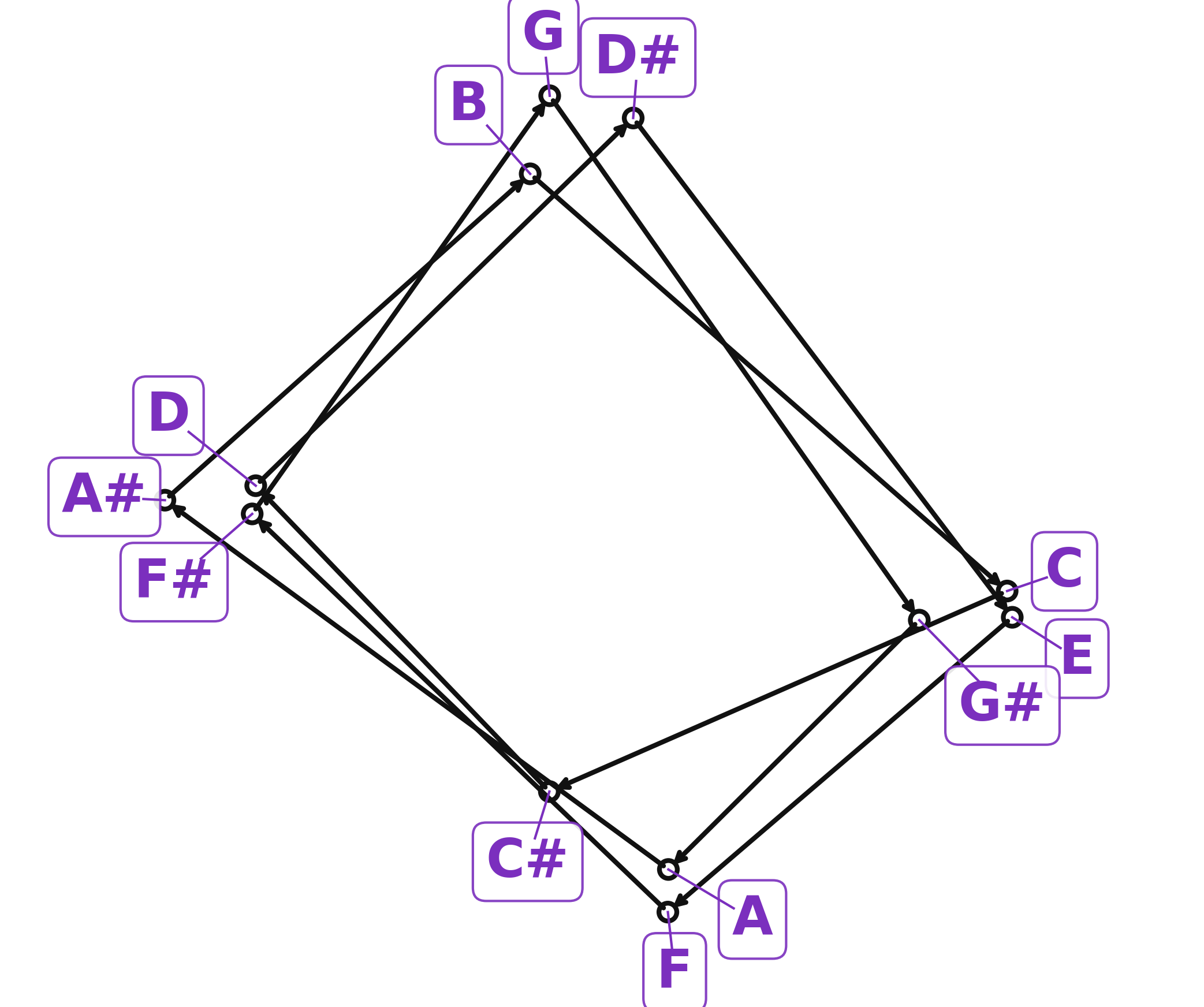}

  \end{subfigure}
   \hfill
\begin{subfigure}[t]{0.19\textwidth}
    \centering
\includegraphics[width=\linewidth]{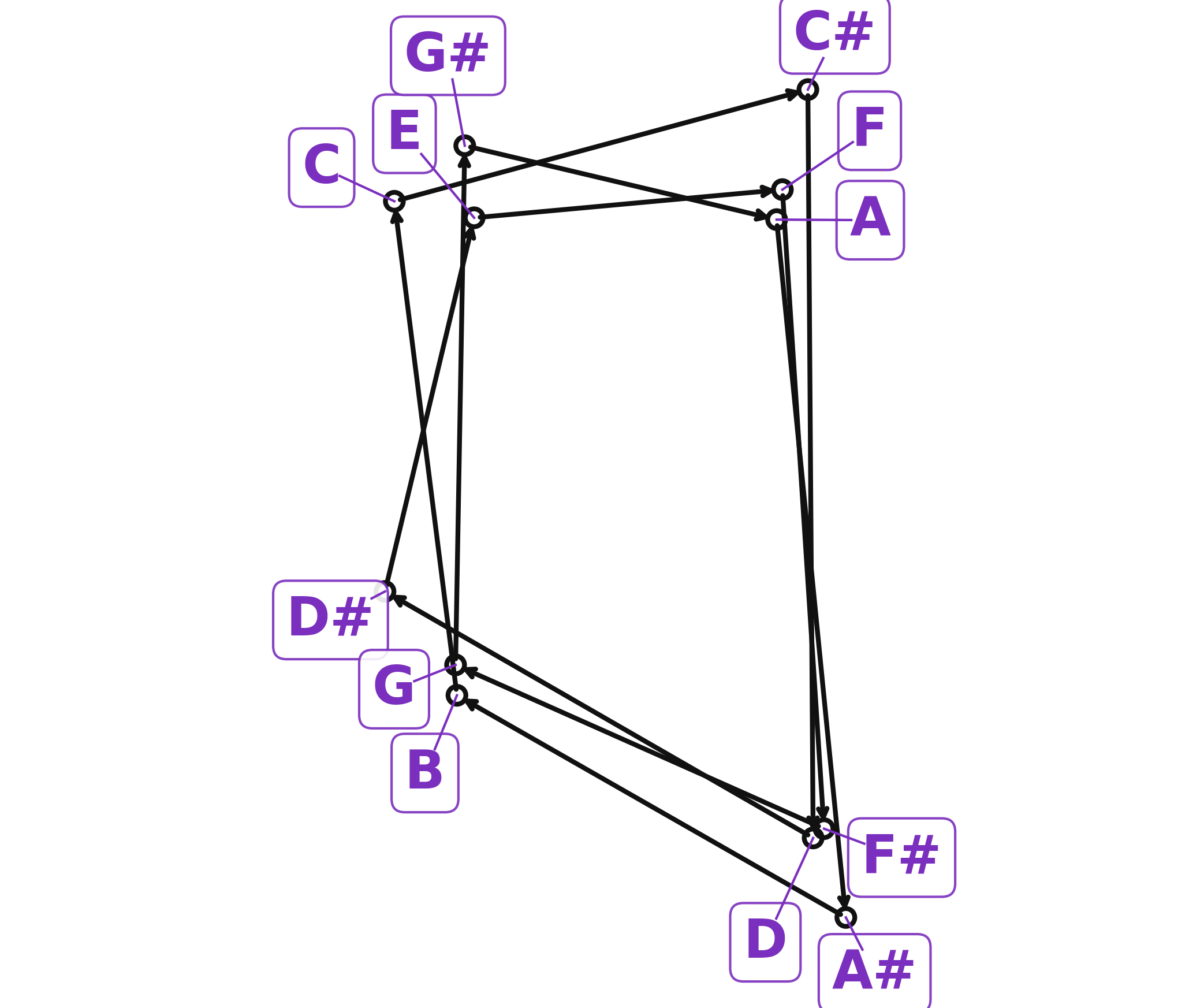}

  \end{subfigure}
   \hfill
\begin{subfigure}[t]{0.19\textwidth}
    \centering
\includegraphics[width=\linewidth]{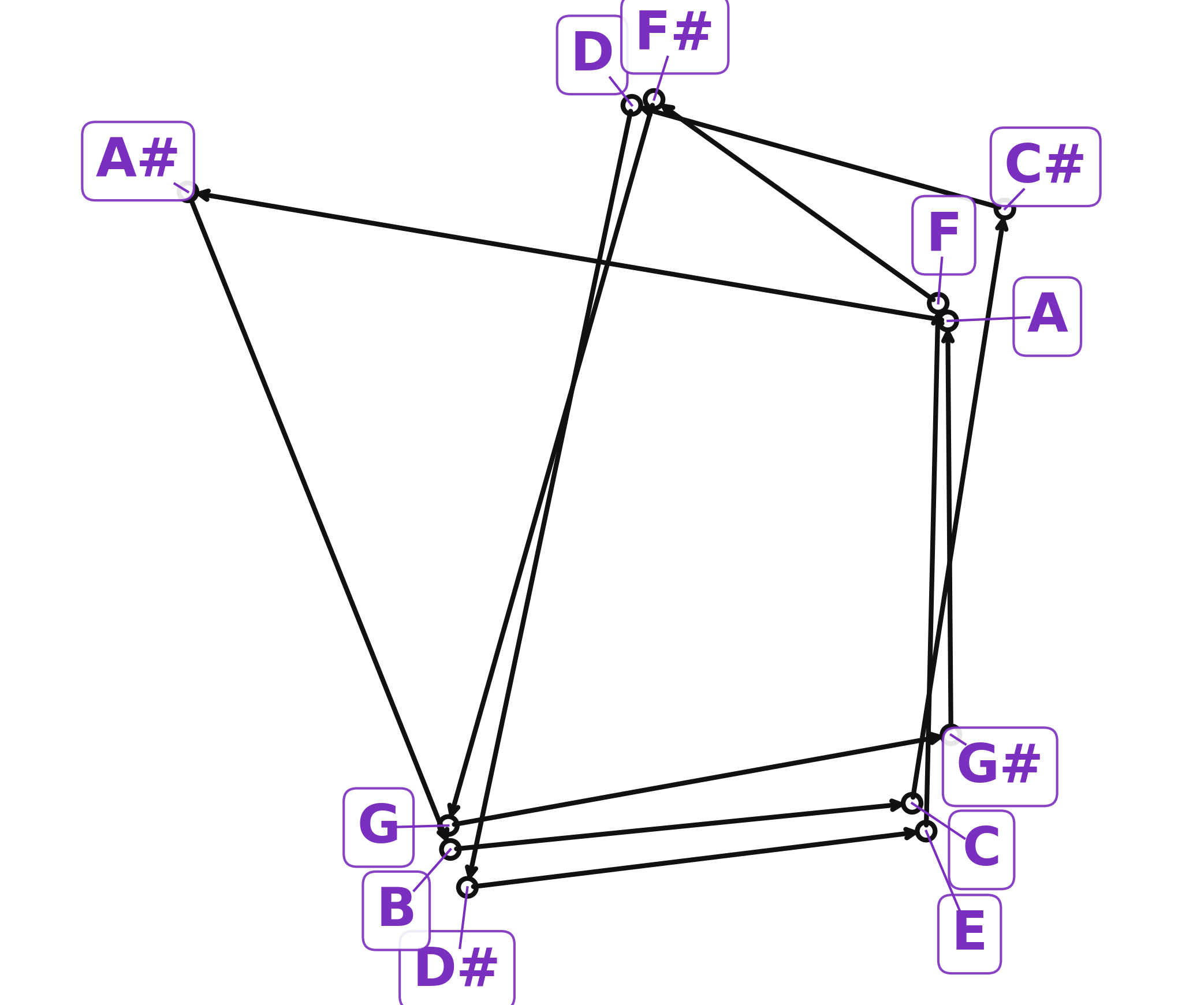}

  \end{subfigure}
   \hfill
\begin{subfigure}[t]{0.19\textwidth}
    \centering
\includegraphics[width=\linewidth]{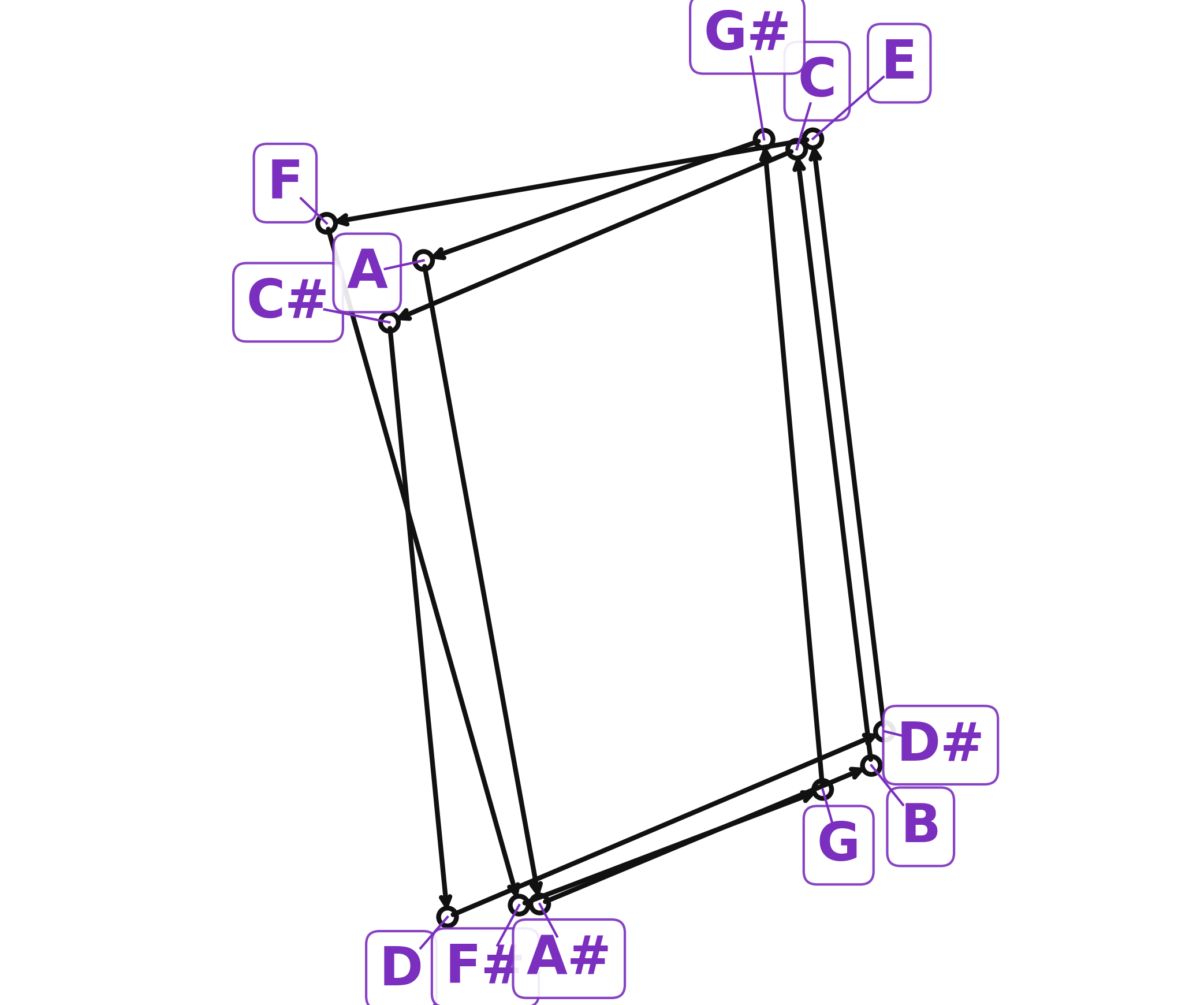}

  \end{subfigure}
   \hfill

  \centering
   \begin{subfigure}[t]{0.19\textwidth}
    \centering
\includegraphics[width=\linewidth]{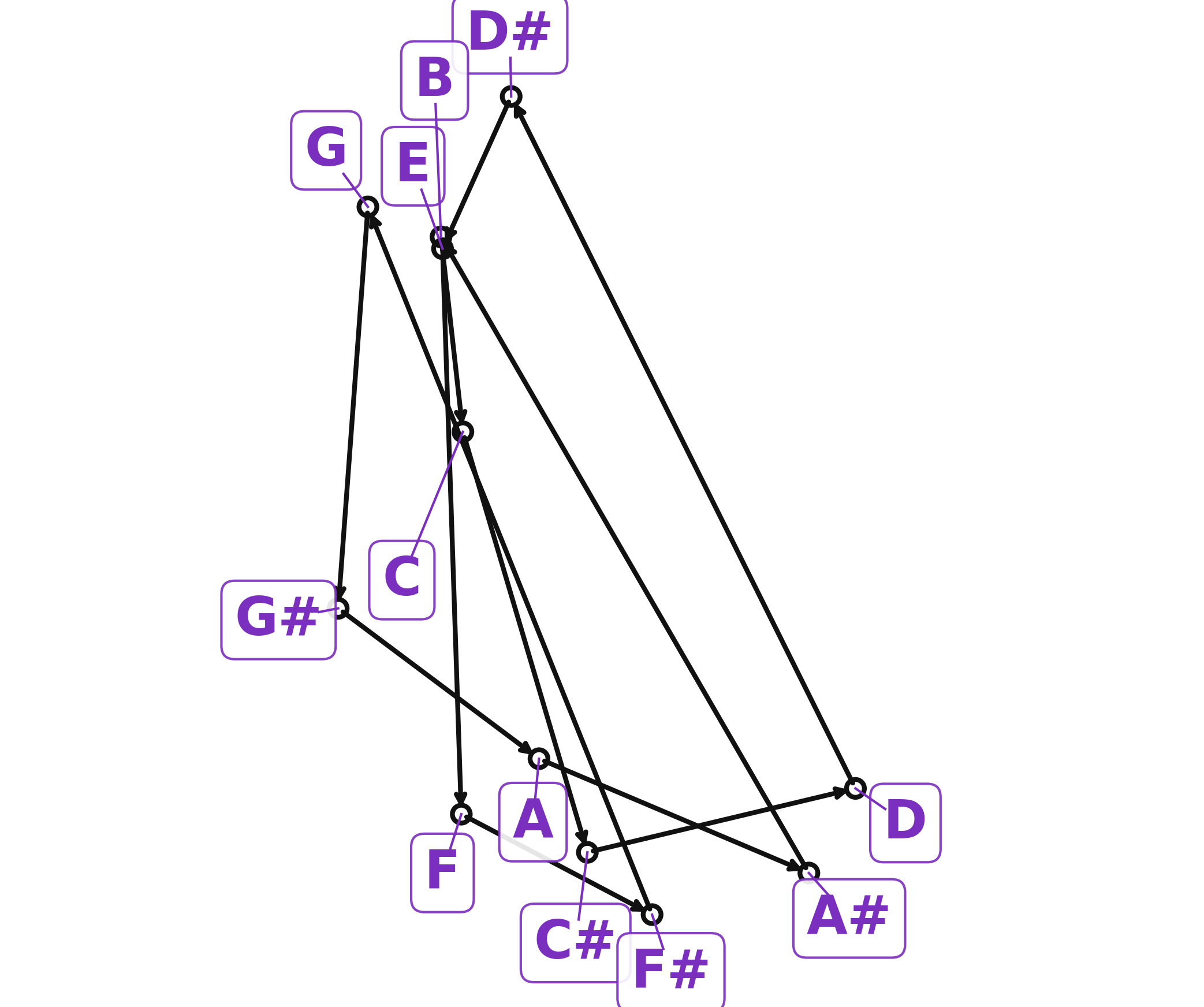}
    \caption{MuQ (2)}
   
  \end{subfigure}
   \hfill
\begin{subfigure}[t]{0.19\textwidth}
    \centering
\includegraphics[width=\linewidth]{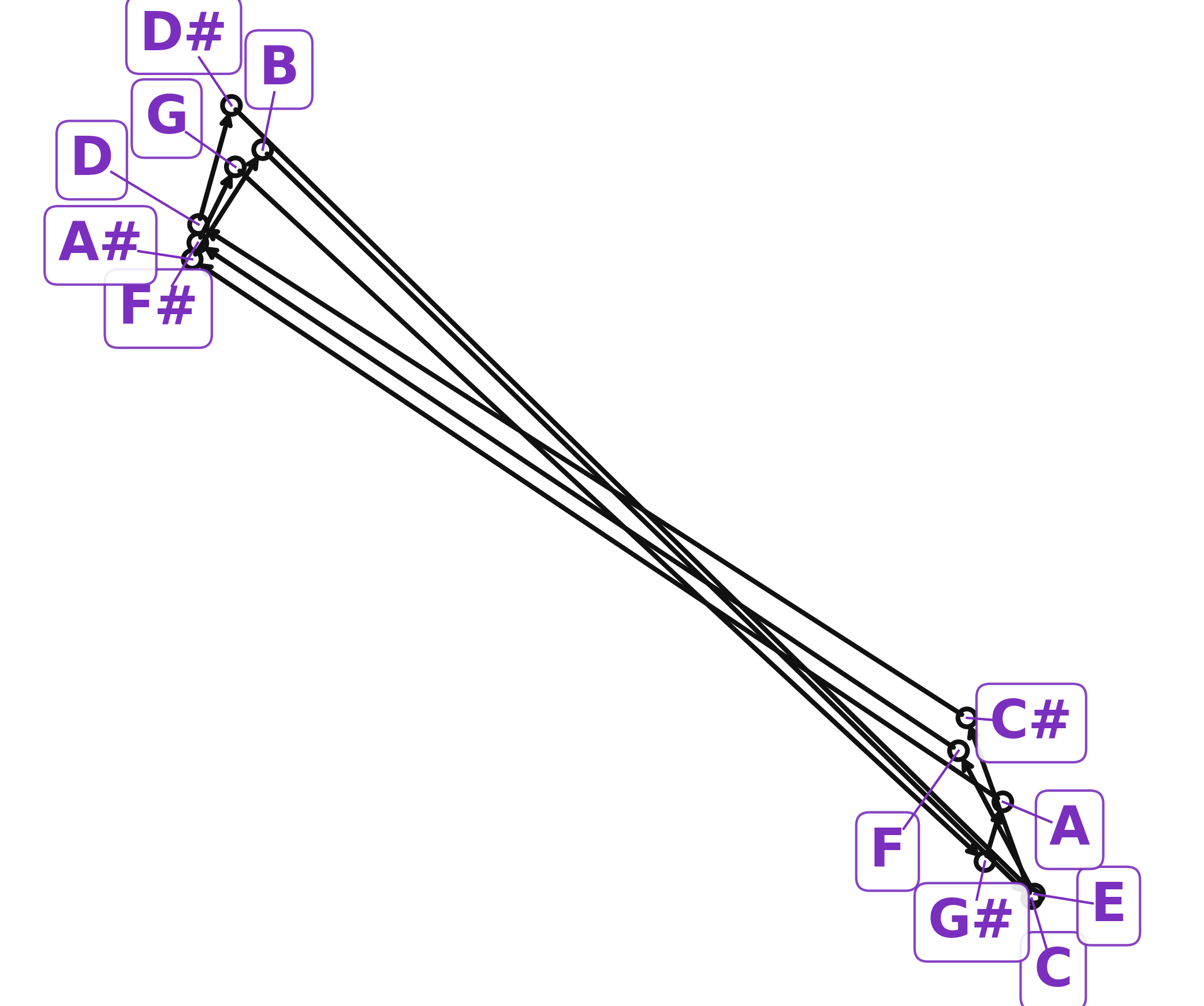}
    \caption{MuQ (3)}
   
  \end{subfigure}
   \hfill
\begin{subfigure}[t]{0.19\textwidth}
    \centering
\includegraphics[width=\linewidth]{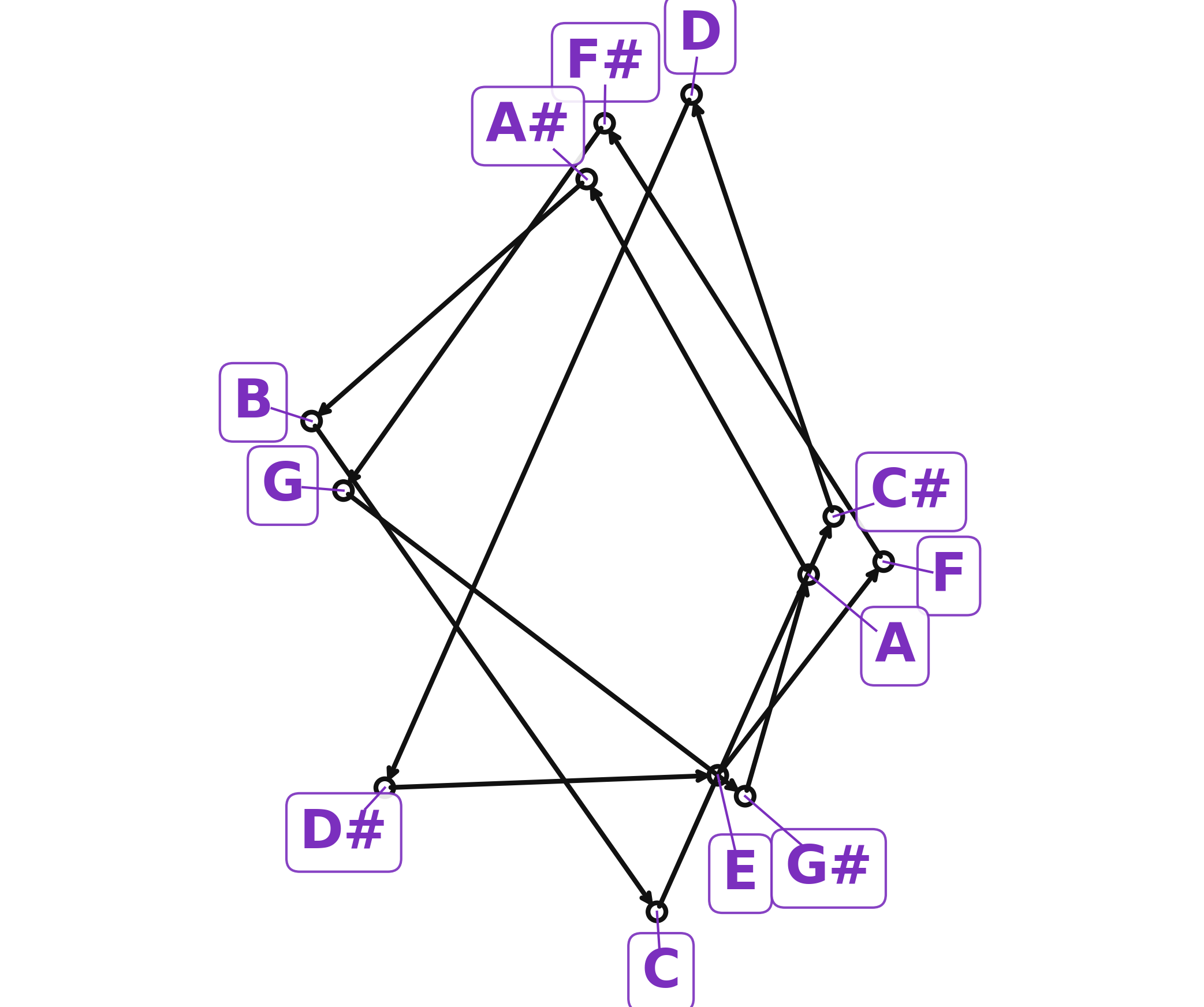}
    \caption{MuQ (4)}
   
  \end{subfigure}
   \hfill
\begin{subfigure}[t]{0.19\textwidth}
    \centering
\includegraphics[width=\linewidth]{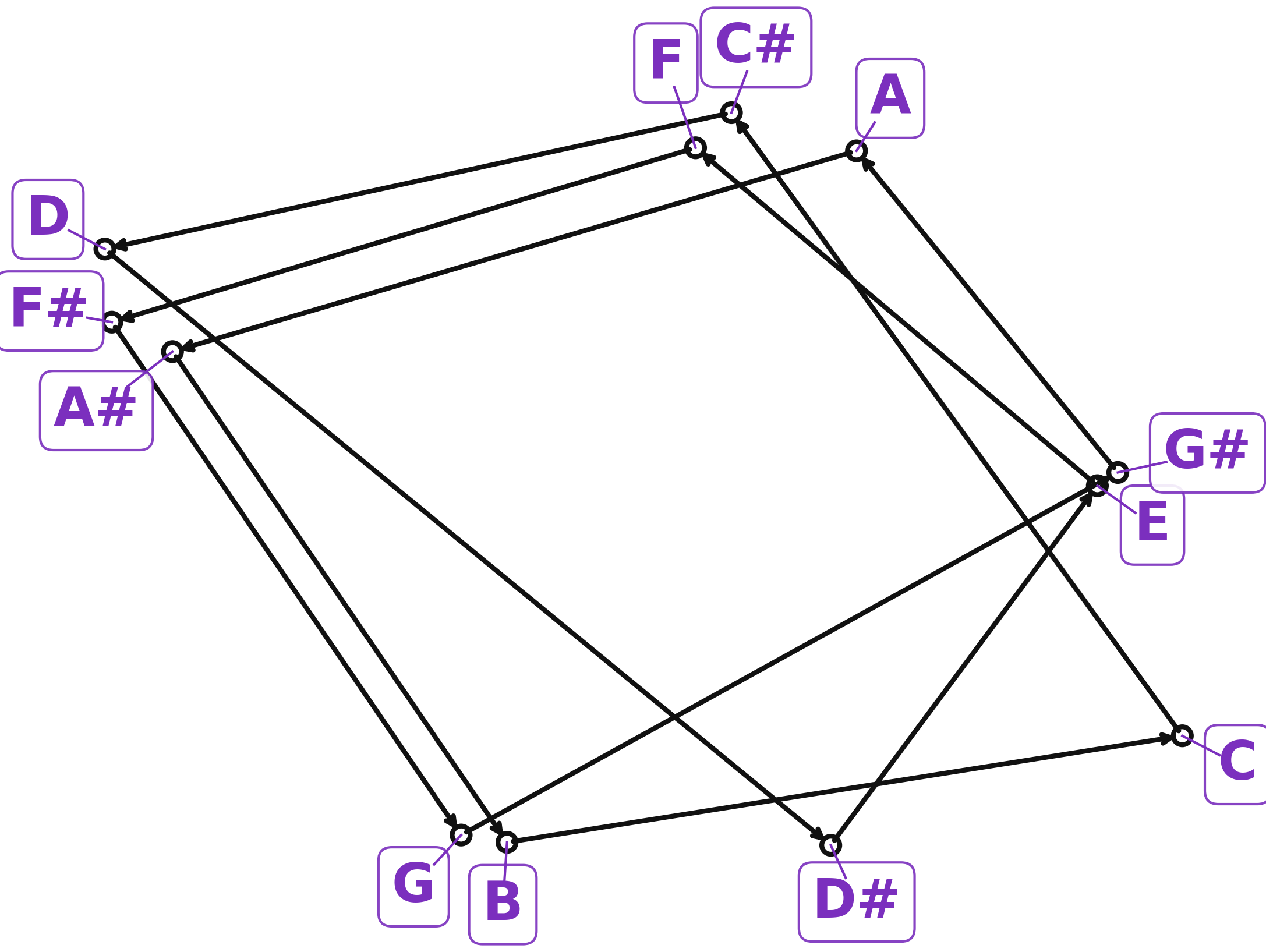}
    \caption{MuQ (5)}
   
  \end{subfigure}
   \hfill
\begin{subfigure}[t]{0.19\textwidth}
    \centering
\includegraphics[width=\linewidth]{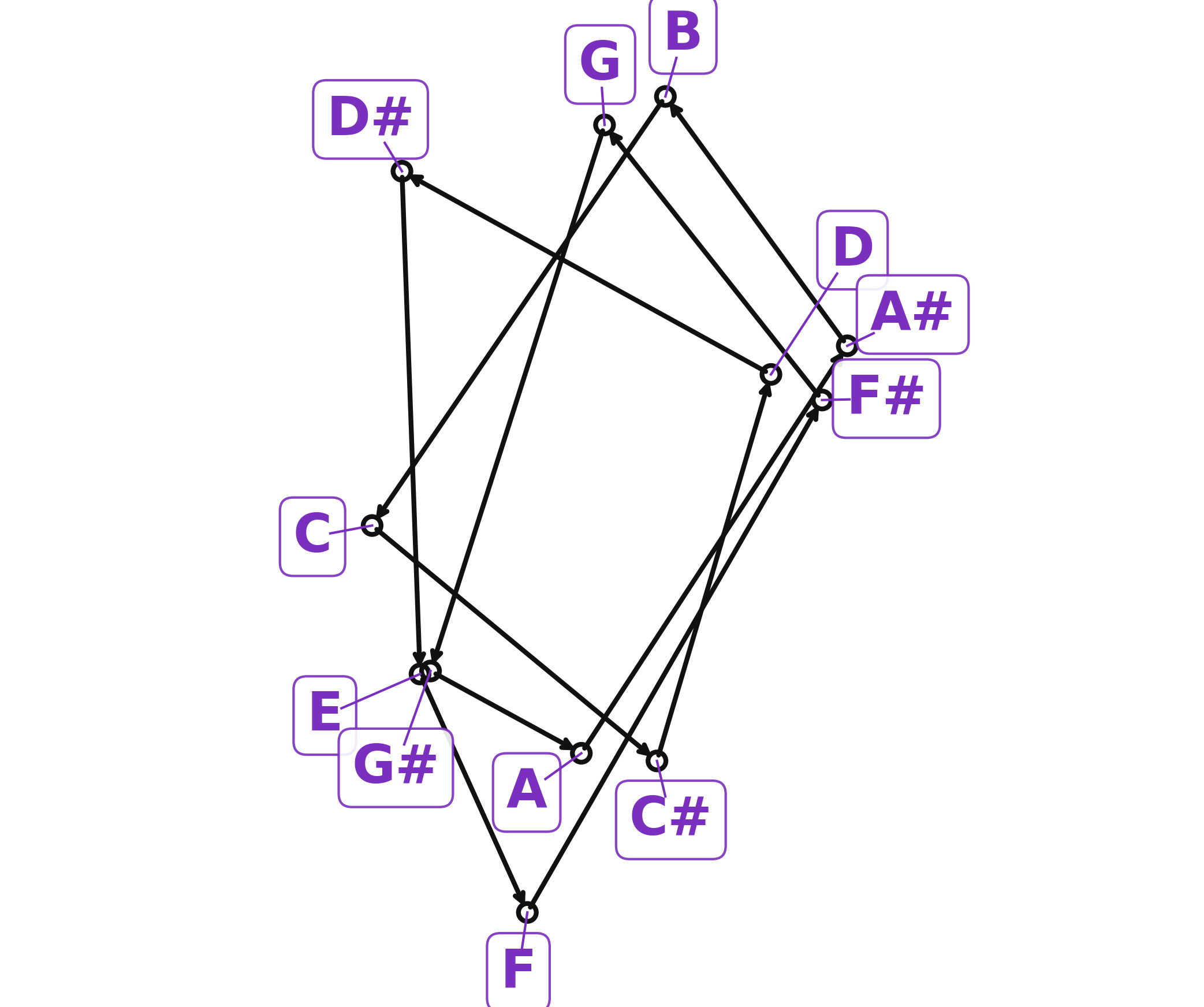}
    \caption{MuQ (6)}
   
  \end{subfigure}
   \hfill
 
  \caption{
  t-SNE visualization of the subdominant features across MuQ layers at $K=32$.
The top row shows o-SAE and the bottom row shows s-SAE. Directed edges follow
the orbit order, and labels denote pitch classes.
}\label{fig:feature_tsne_subdom}
  
\end{figure}

\subsection{Sparse Directed Orbit Graphs}

The orbit-recovery procedure in the main text retains at most one incoming and
one outgoing edge for each feature. To inspect the broader feature structure,
we generalize this constraint using $n_e$, the maximum number of retained
incoming and outgoing edges per node. The original procedure is the special
case $n_e=1$.

Figures~\ref{fig:orbit_graphs} and~\ref{fig:orbit_graphs_add} visualize the
resulting graphs with $n_e=10$ and
$\tau\in\{0.4,0.5,0.6,0.7\}$. Some core orbits that do not form exact rings
under $n_e=1$ still exhibit nearly cyclic structures when more candidate edges
are retained. This suggests that some orbit structures learned by the SAEs may
be imperfect and therefore cannot be recovered directly by our strict
algorithm, yet remain detectable through a simple and intuitive post-hoc graph
analysis. We do so in the downstream task evaluation so that we can have a comprehensive evaluation of every potential core ring for each foundation model layer.

The graphs also contain many small connected components, commonly consisting
of only three or four features. These components may reflect fragmented or
local feature relations and are left for future investigation.

\subsection{Geometric Structure of the Subdominant Ring}
\label{app:subdominant_geometry}

We visualize the SAE feature dictionary using t-SNE and examine the internal geometry of the recovered subdominant ring. We first normalize the
decoder vector of each feature, reduce the vectors to 50 dimensions using PCA,
and then project them into two dimensions using t-SNE. To reduce outliers, we include features that are activated in at least
2,000 segments in noth training and validation samples. 

Figure~\ref{fig:feature_tsne}
shows the three core orbits together with the remaining active o-SAE and s-SAE features across different MuQ layers at top-K $K=32$. We
add directed edges between successive subdominant features according to the
recovered orbit order and annotate them by pitch class. The visualization as shown in Figure~\ref{fig:feature_tsne_subdom} reveals the relative arrangement of the subdominant
features within the broader SAE dictionary. Because t-SNE primarily preserves
local neighborhoods, we use this plot as qualitative evidence of feature
organization rather than interpreting global distances quantitatively.

\begin{figure}[!t]
  \centering
\begin{subfigure}[t]{0.23\textwidth}
    \centering
\includegraphics[width=\linewidth]{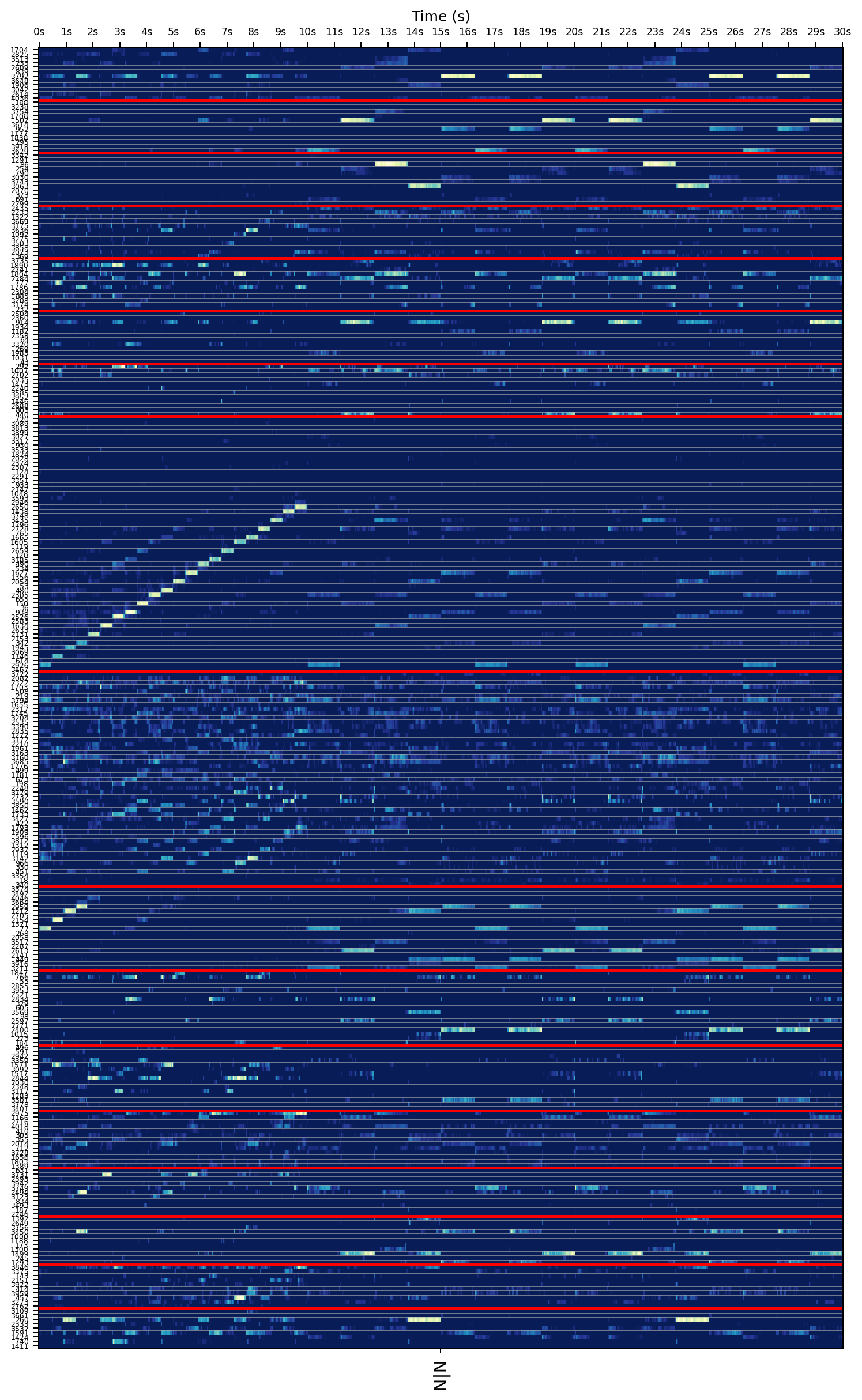}
    \caption{Piano in C major.}
  \end{subfigure}
\hfill
\begin{subfigure}[t]{0.23\textwidth}
    \centering
\includegraphics[width=\linewidth]{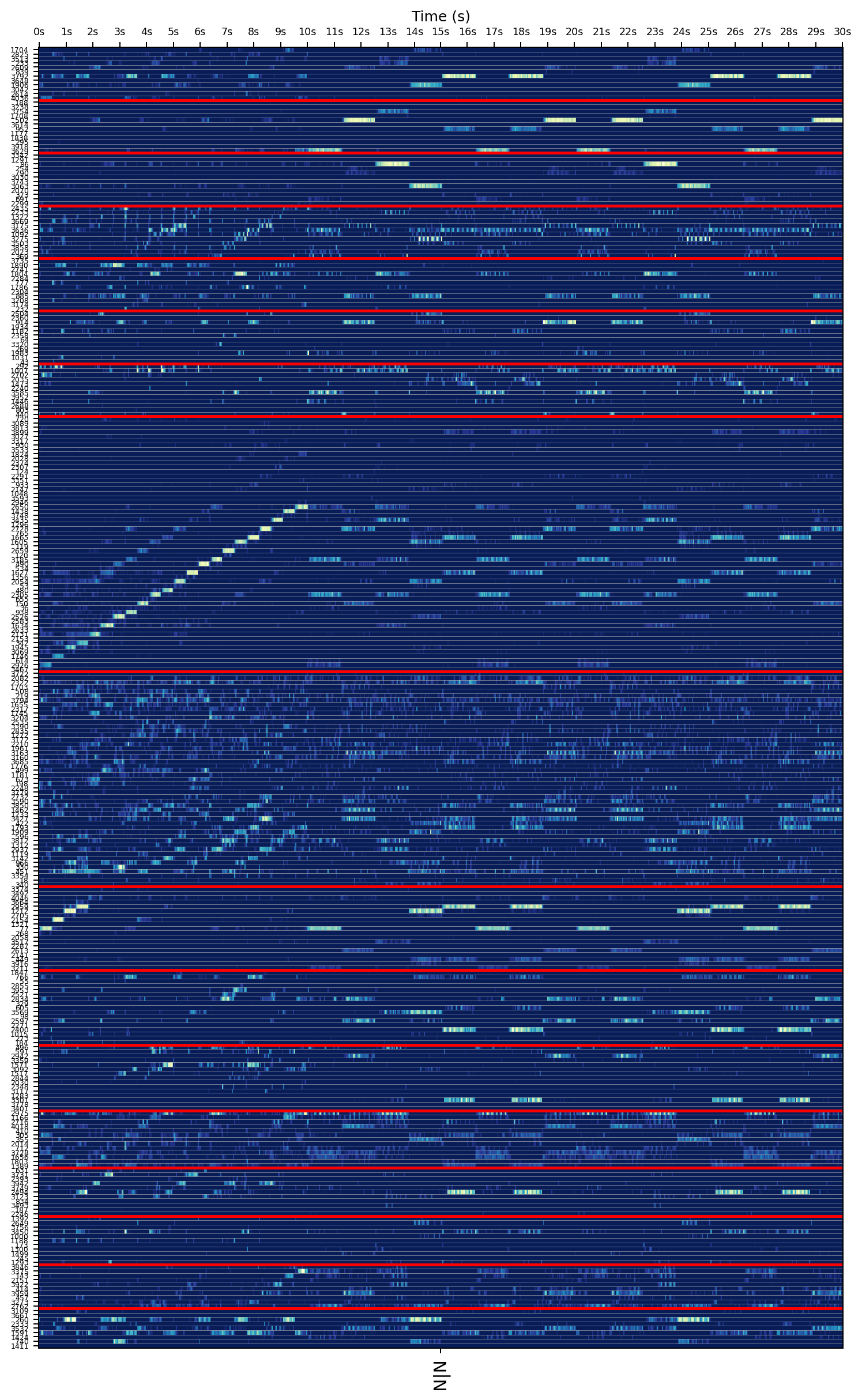}
    \caption{Strings in C major.}
  \end{subfigure}
\hfill
\begin{subfigure}[t]{0.23\textwidth}
    \centering
\includegraphics[width=\linewidth]{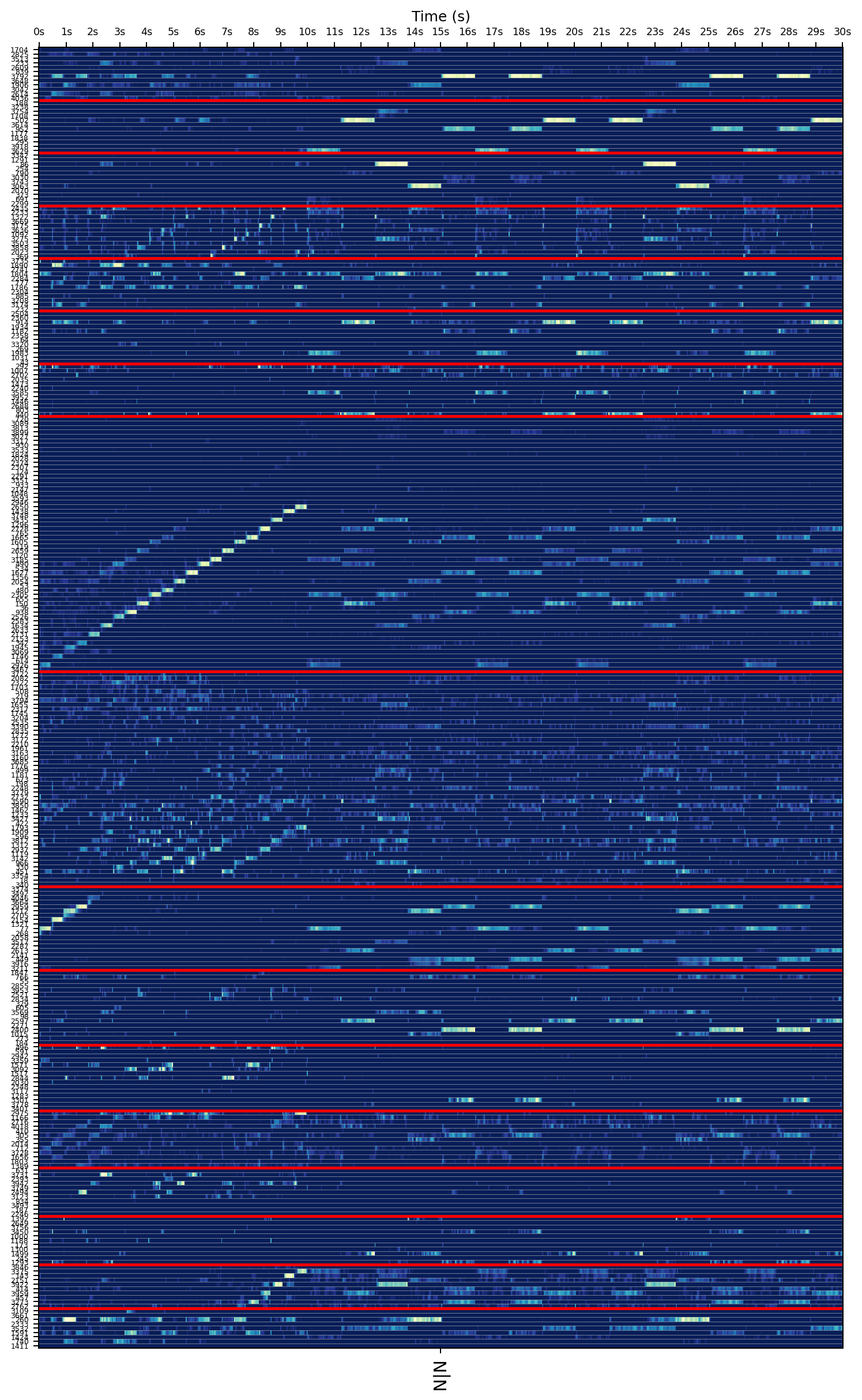}
    \caption{Brass in C major.}
  \end{subfigure}
\hfill
  \begin{subfigure}[t]{0.23\textwidth}
    \centering
\includegraphics[width=\linewidth]{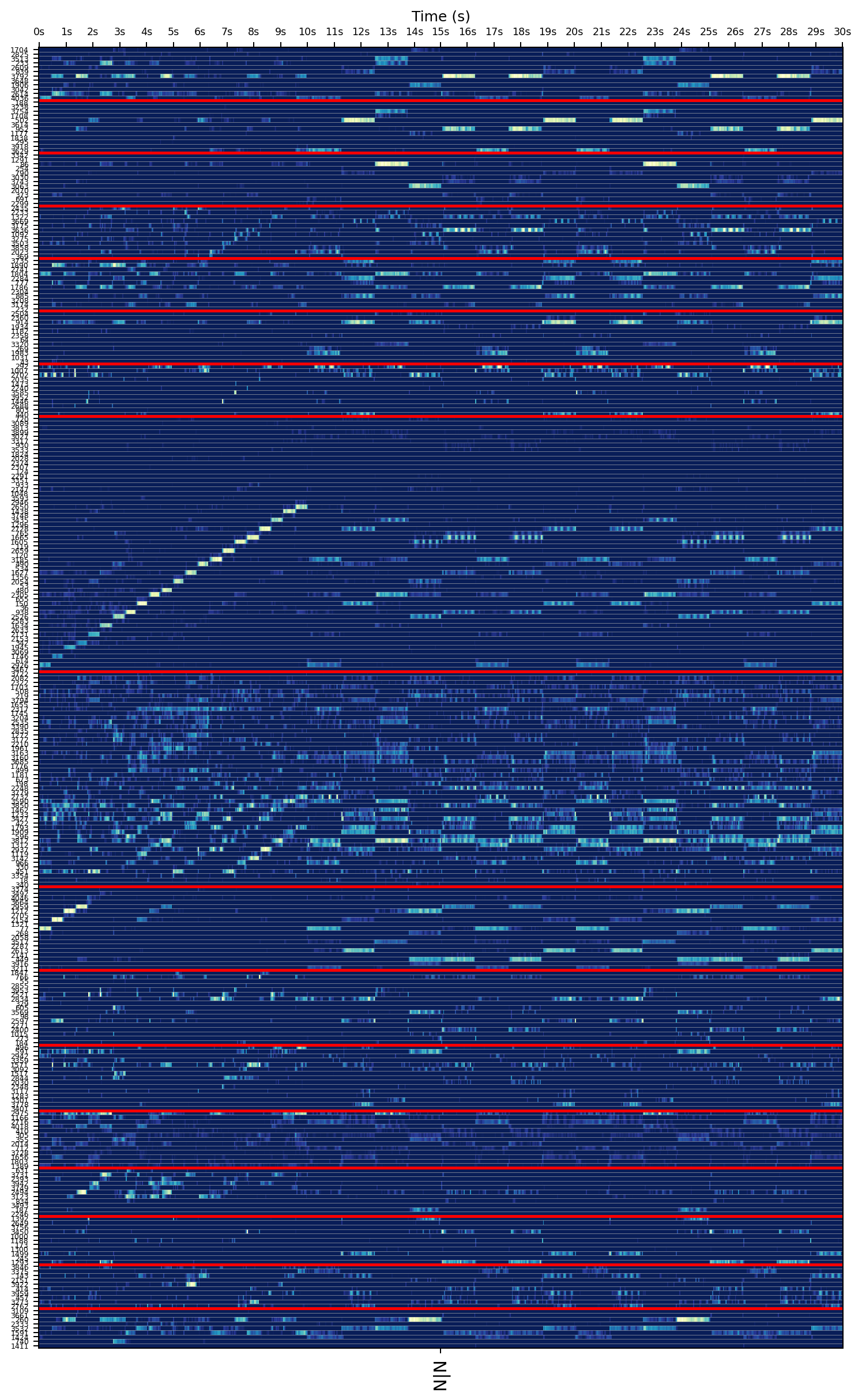}
    \caption{Woodwinds in C major.}
  \end{subfigure}

\begin{subfigure}[t]{0.23\textwidth}
    \centering
\includegraphics[width=\linewidth]{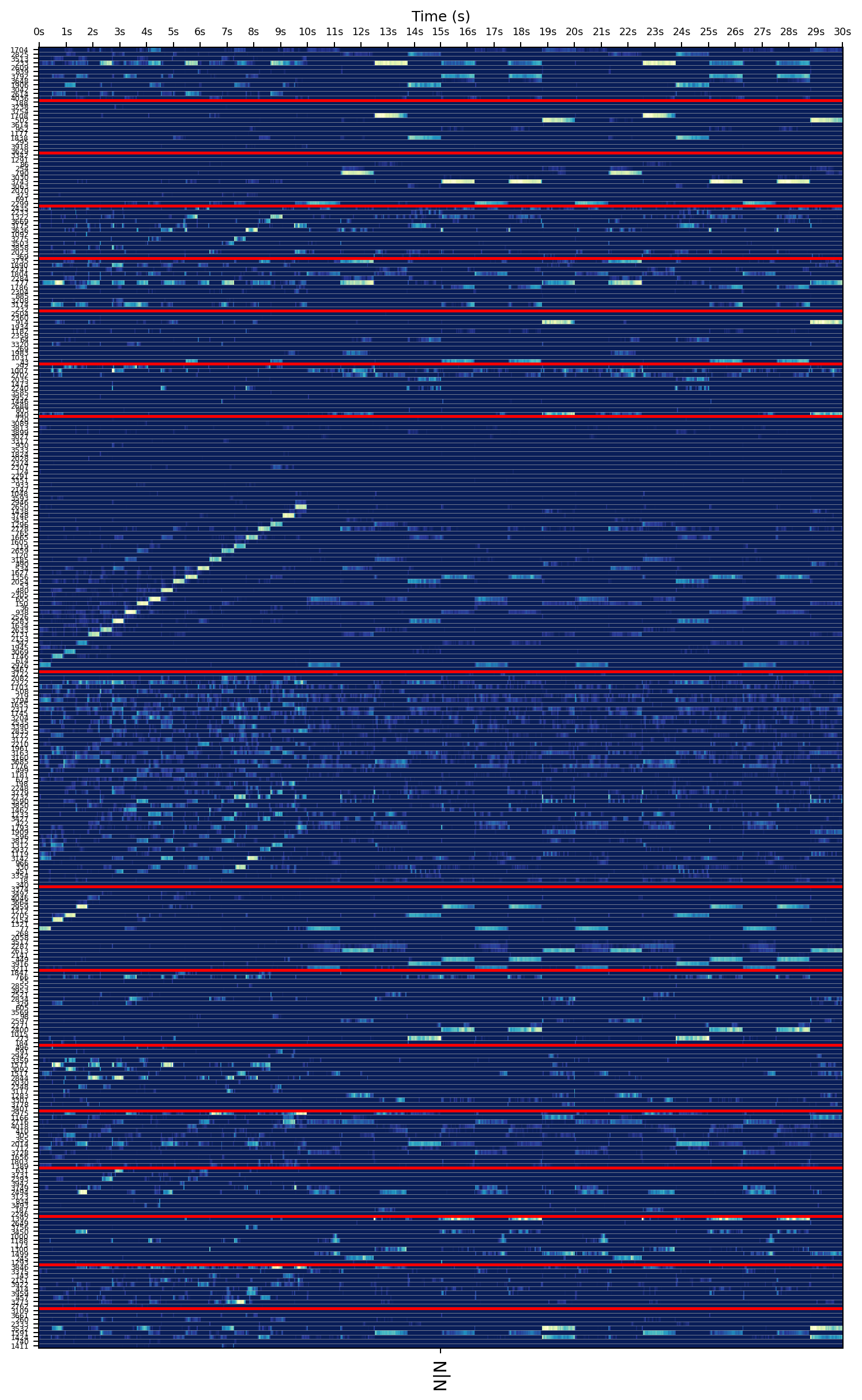}
    \caption{Piano in C minor.}
  \end{subfigure}
\hfill
\begin{subfigure}[t]{0.23\textwidth}
    \centering
\includegraphics[width=\linewidth]{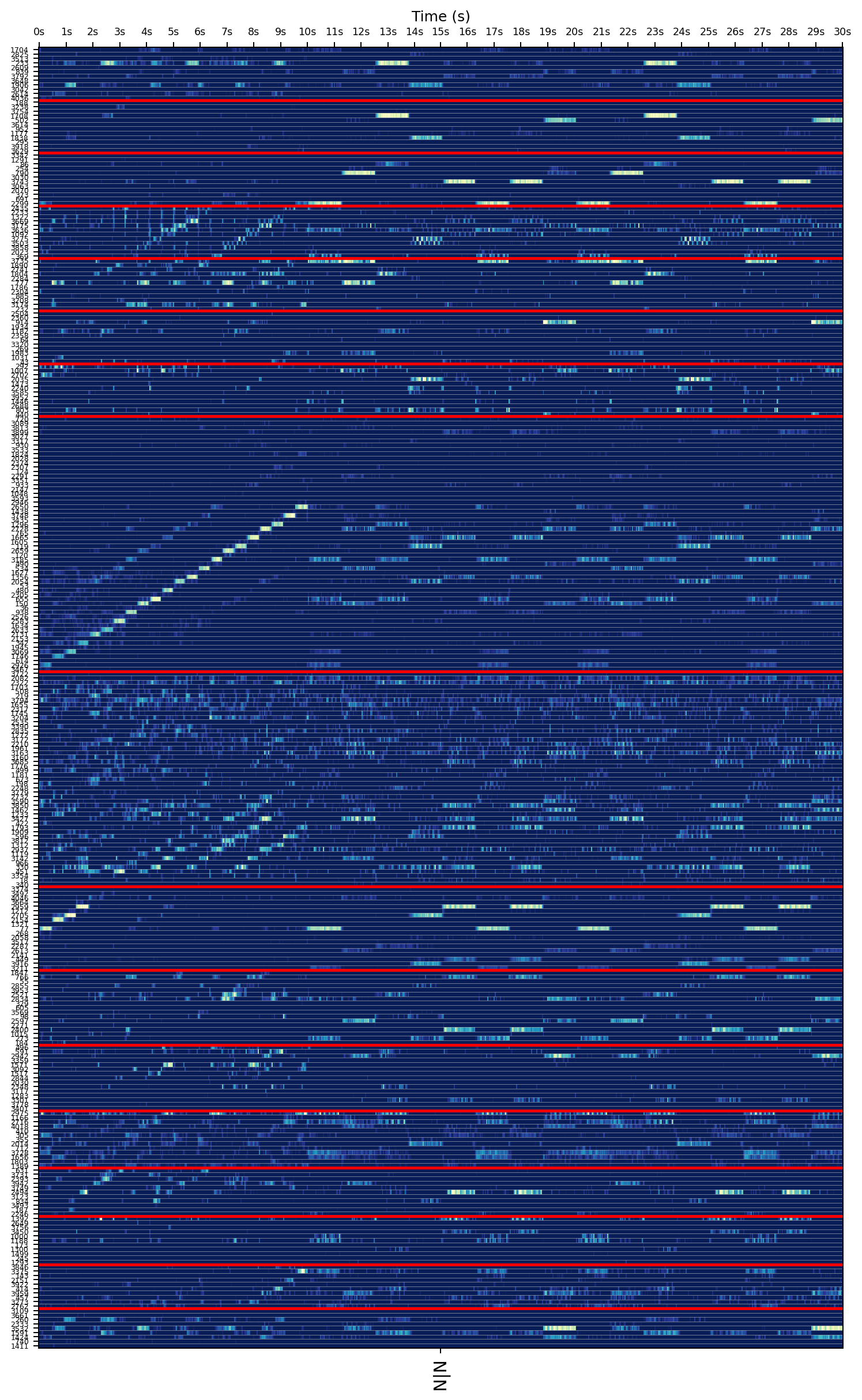}
    \caption{Strings in C minor.}
  \end{subfigure}
\hfill
\begin{subfigure}[t]{0.23\textwidth}
    \centering
\includegraphics[width=\linewidth]{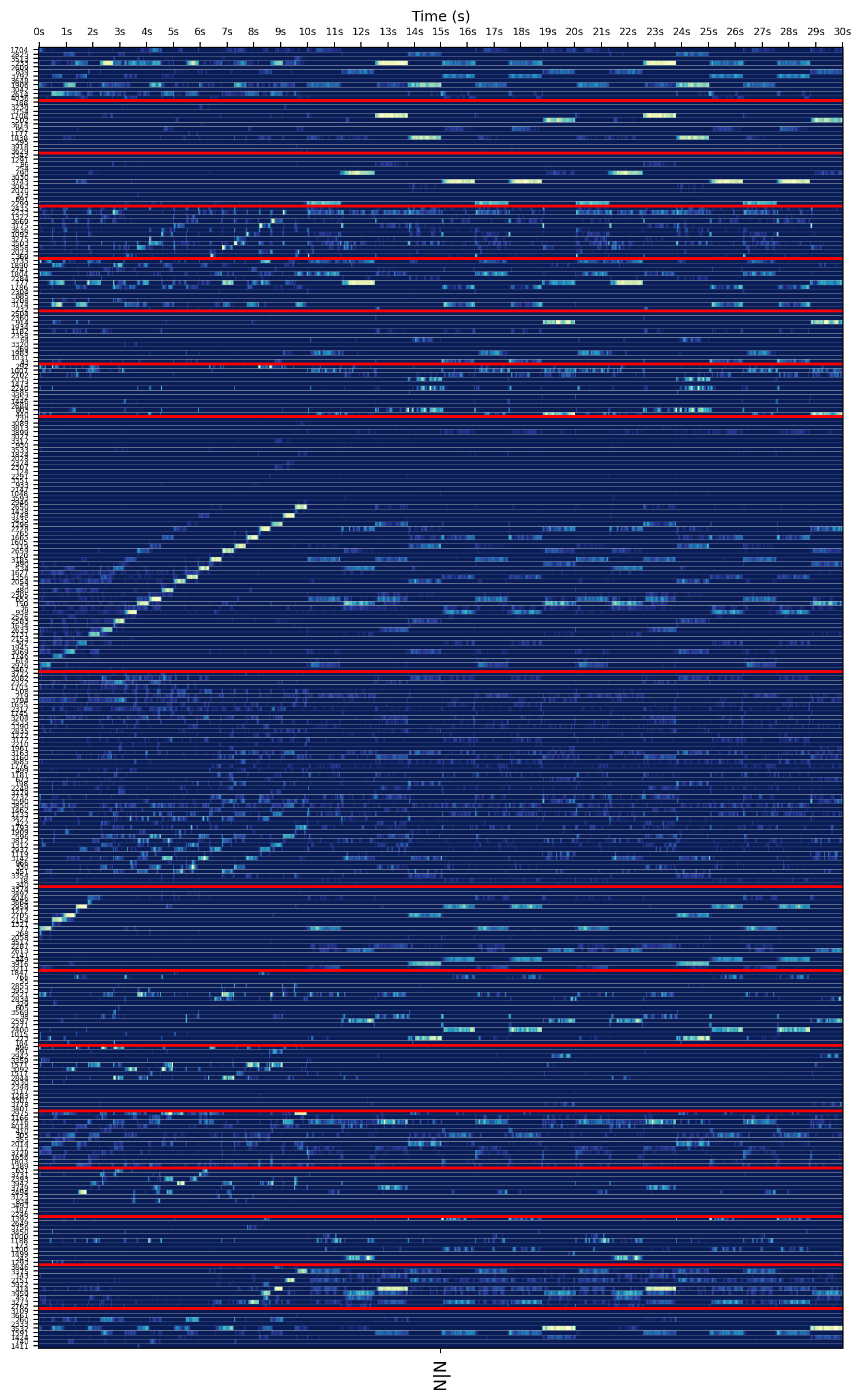}
    \caption{Brass in C minor.}
  \end{subfigure}
\hfill
  \begin{subfigure}[t]{0.23\textwidth}
    \centering
\includegraphics[width=\linewidth]{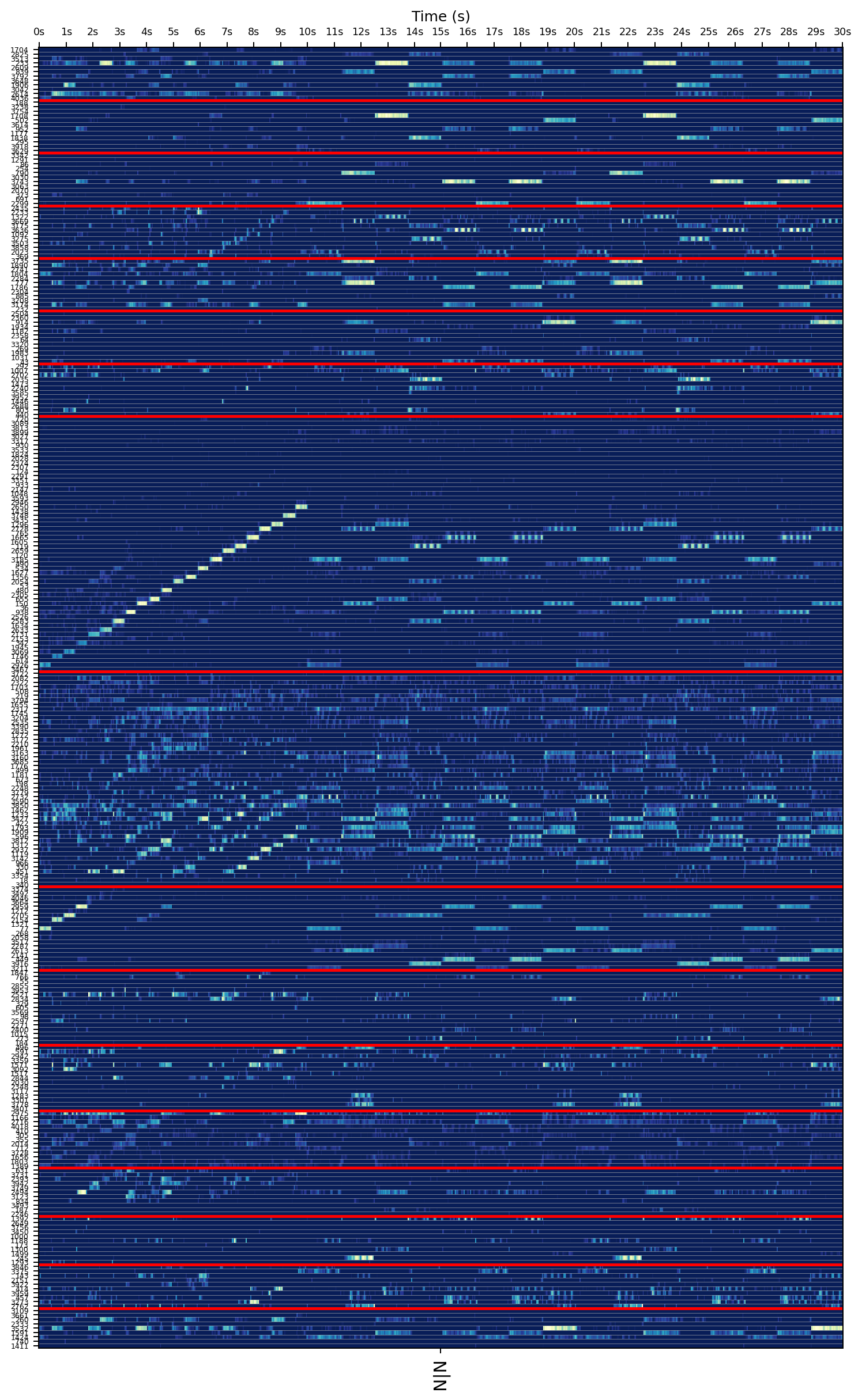}
    \caption{Woodwinds in C minor.}
  \end{subfigure}

\begin{subfigure}[t]{0.23\textwidth}
    \centering
\includegraphics[width=\linewidth]{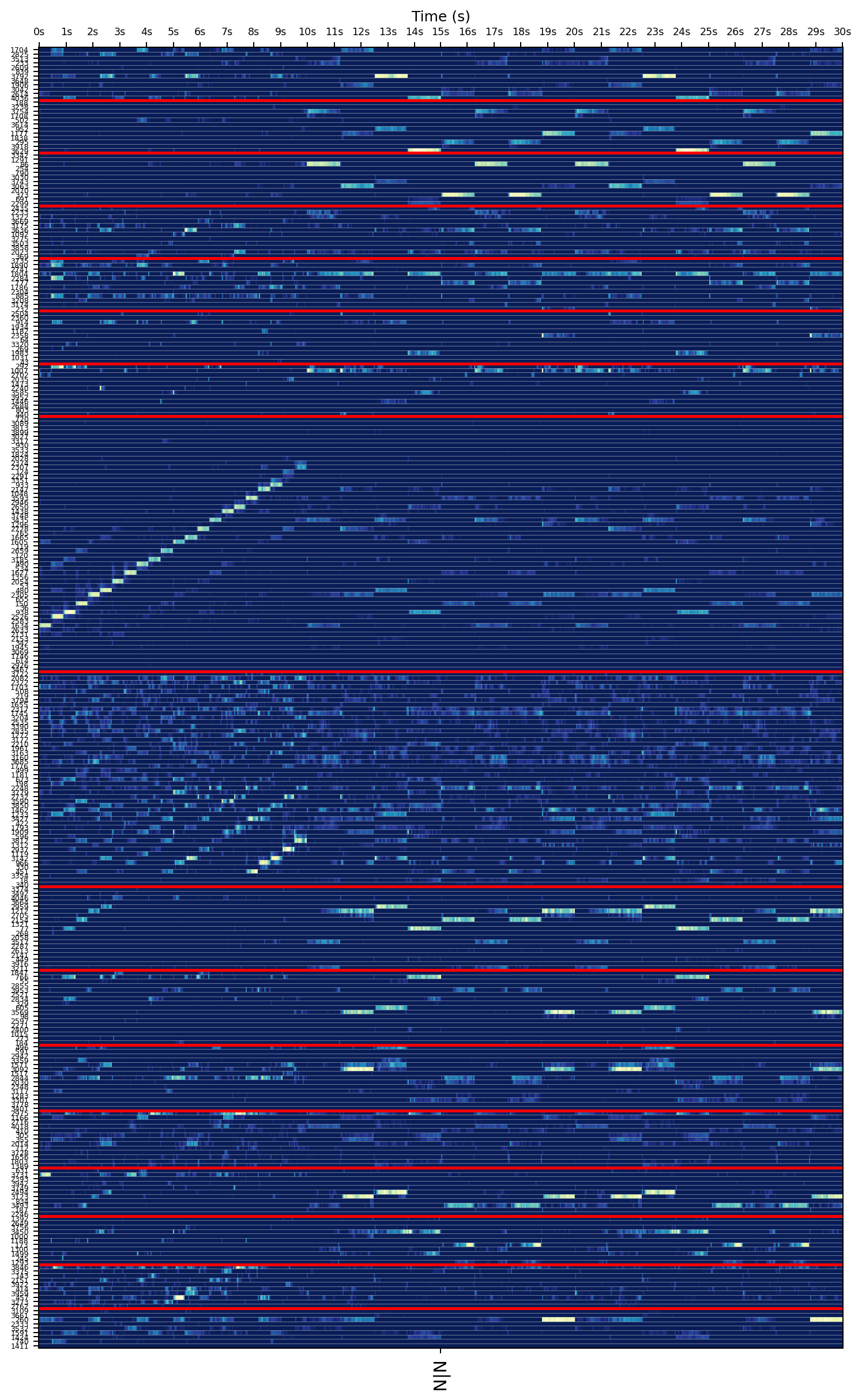}
    \caption{Piano in A minor.}
  \end{subfigure}
\hfill
\begin{subfigure}[t]{0.23\textwidth}
    \centering
\includegraphics[width=\linewidth]{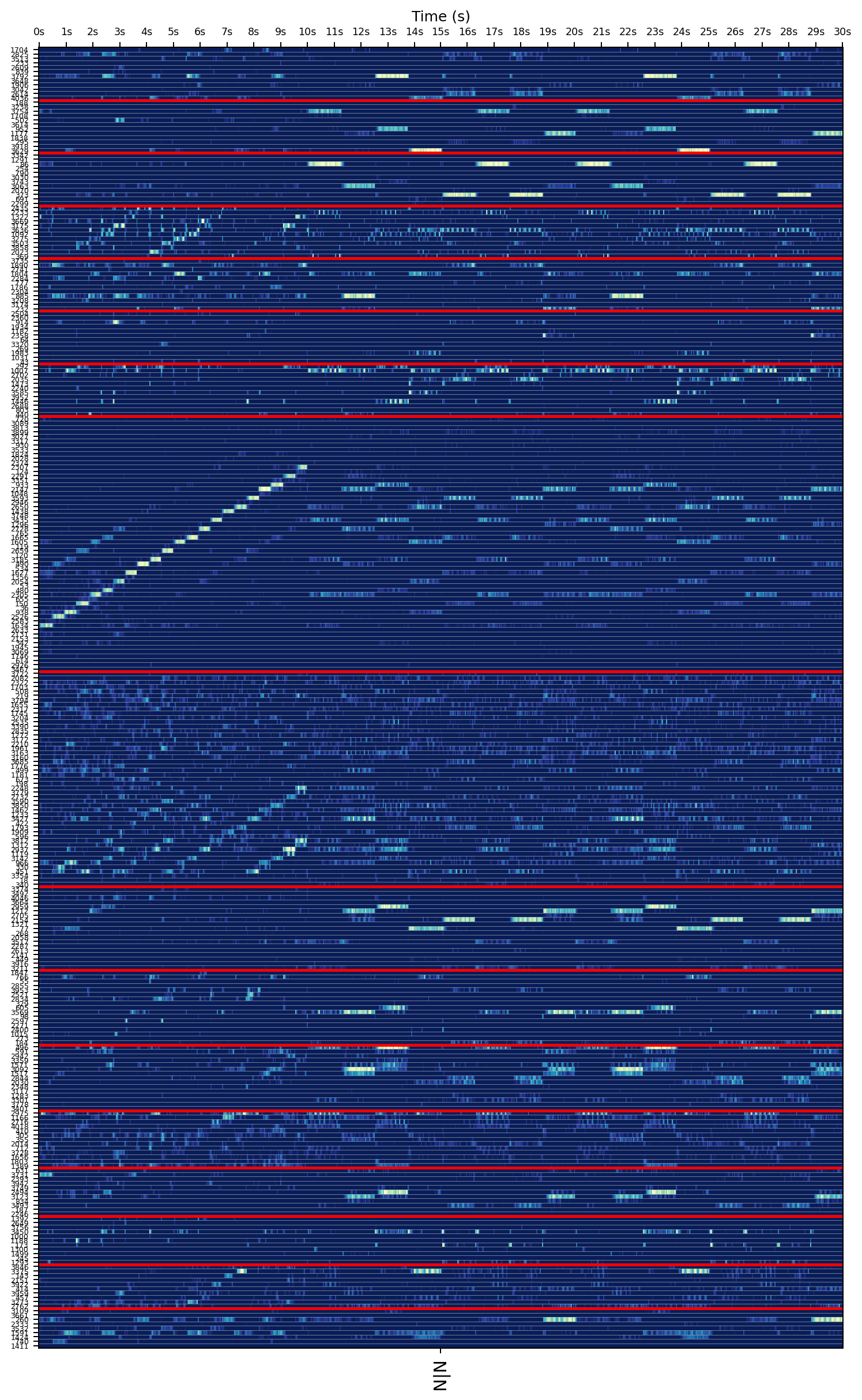}
    \caption{Strings in A minor.}
  \end{subfigure}
\hfill
\begin{subfigure}[t]{0.23\textwidth}
    \centering
\includegraphics[width=\linewidth]{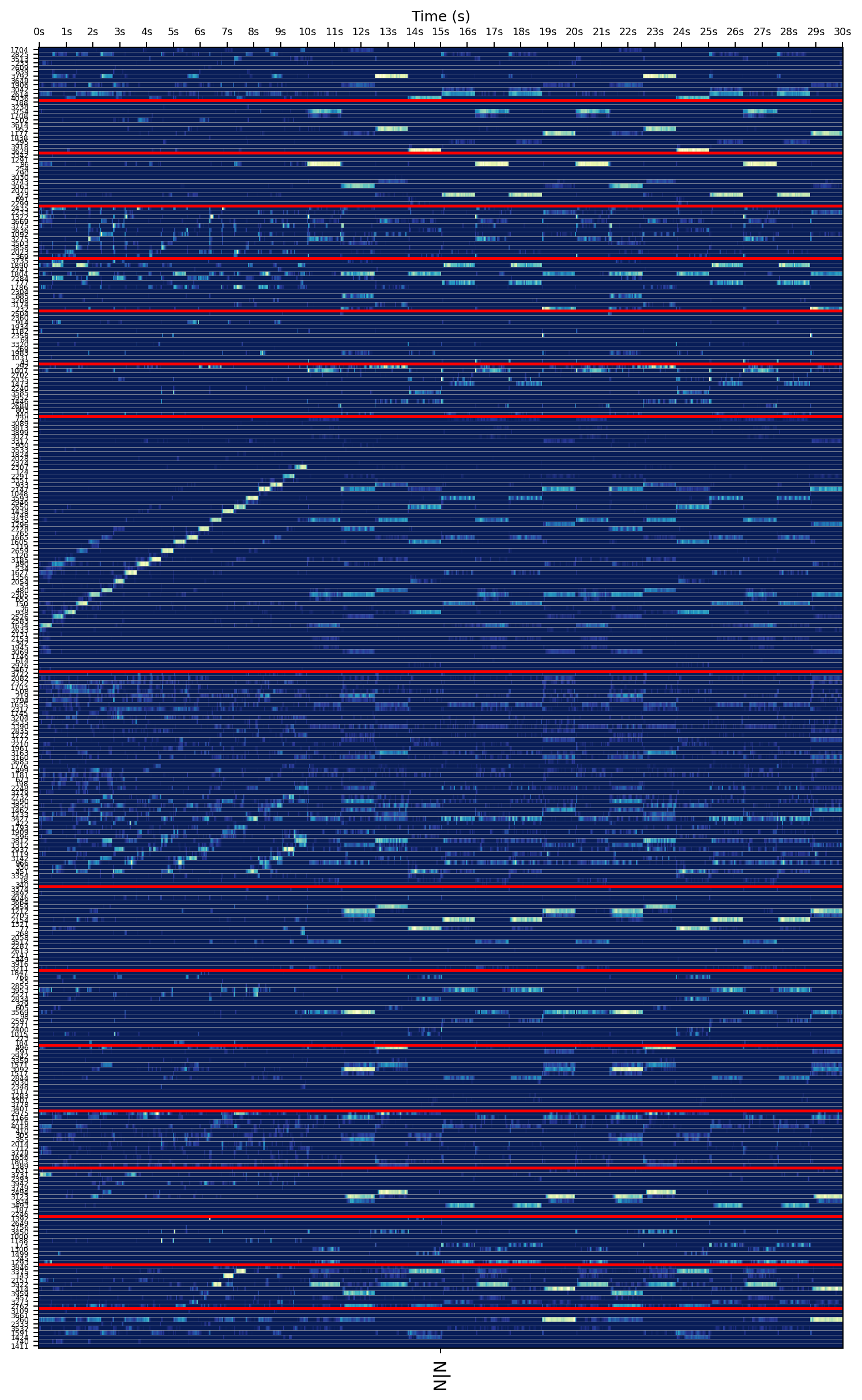}
    \caption{Brass in A minor.}
  \end{subfigure}
\hfill
  \begin{subfigure}[t]{0.23\textwidth}
    \centering
\includegraphics[width=\linewidth]{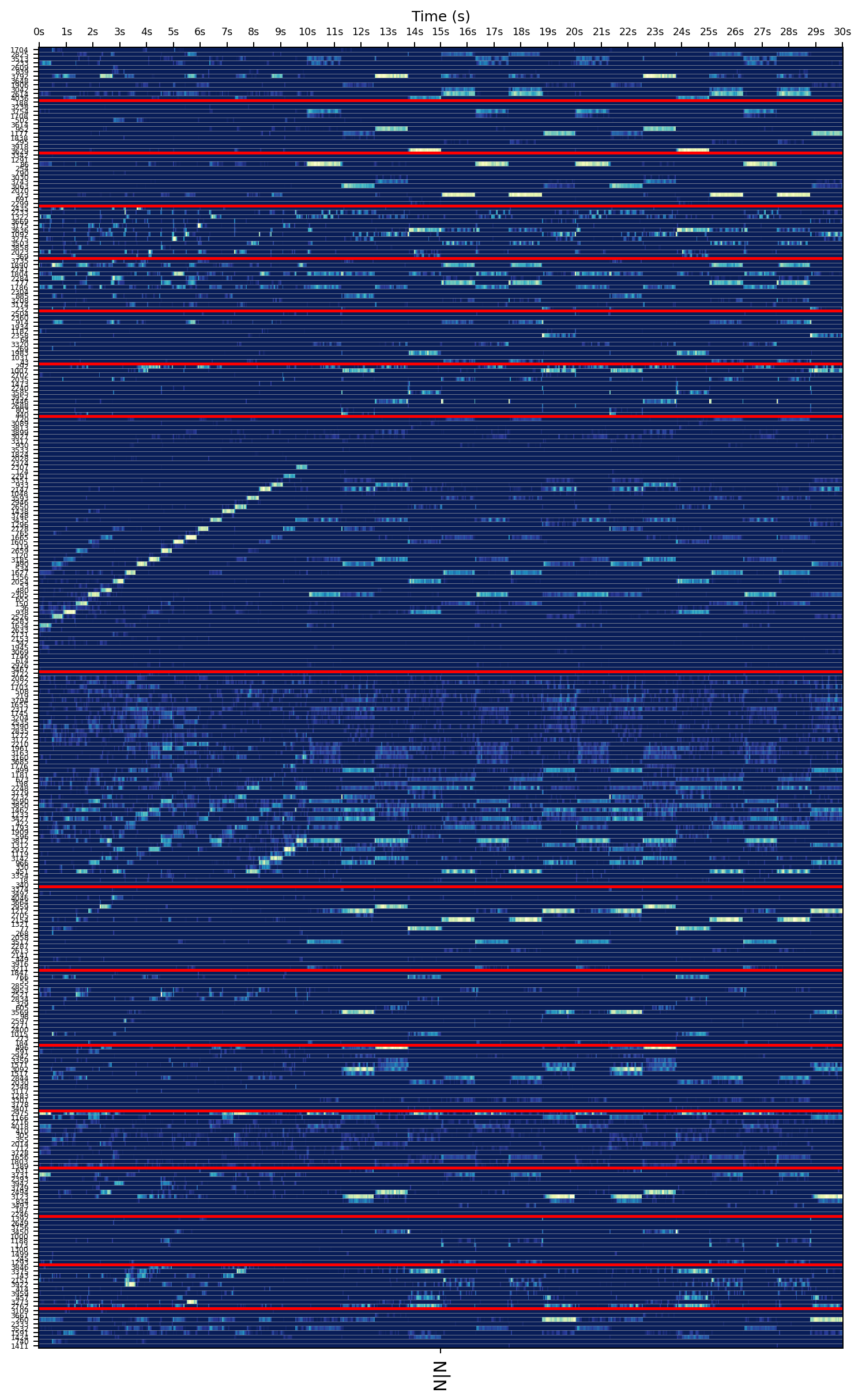}
    \caption{Woodwinds in A minor.}
  \end{subfigure}

\caption{
Frame-level activations of recovered MuQ (2) orbits on controlled synthetic
excerpts in C major and A minor, rendered with piano, strings, brass, and
woodwinds, using $K=32$ and $\tau=0.5$. The comparison isolates the
effects of key and mode across different instrument families.
}
\label{fig:activations_syn}
\end{figure}

\begin{figure}[t]
  \centering
\begin{subfigure}[t]{0.32\textwidth}
    \centering
\includegraphics[width=\linewidth]{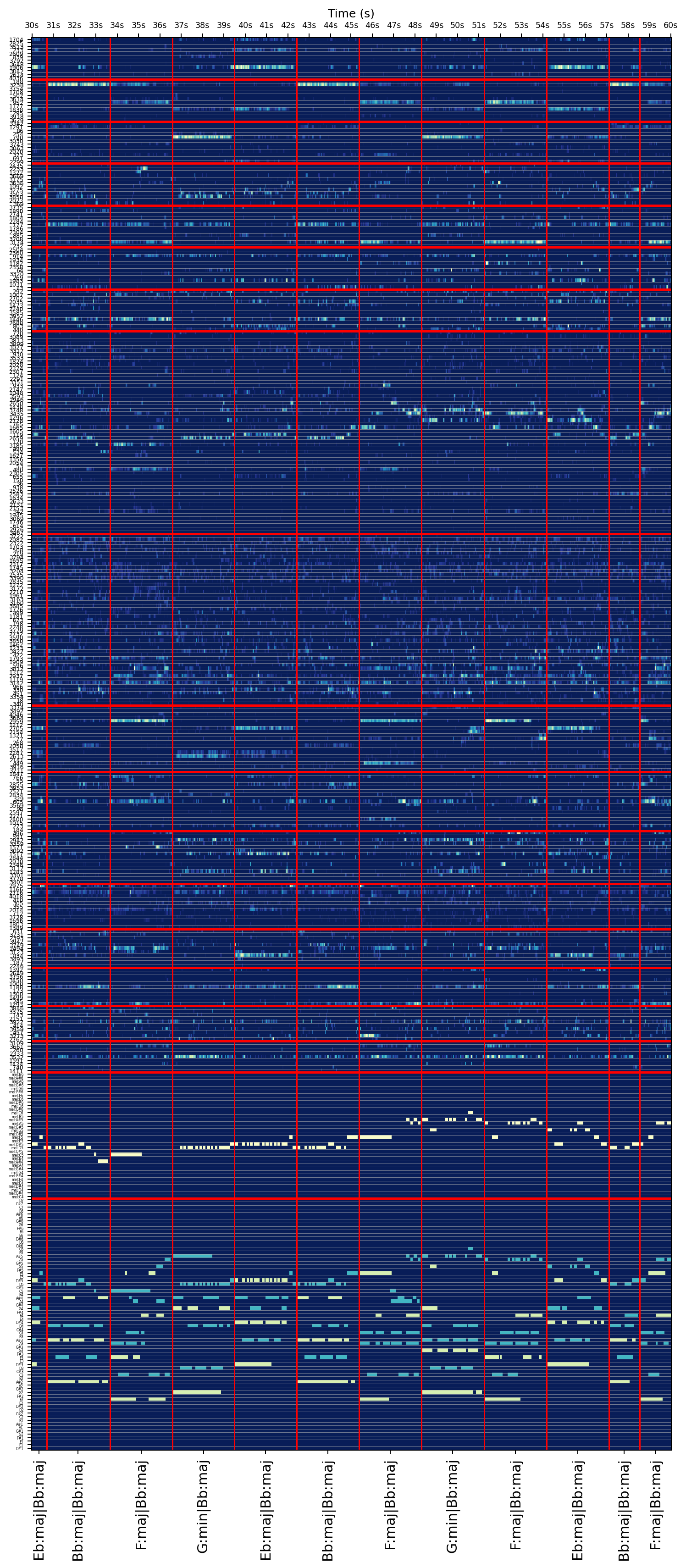}
    \caption{B$\flat$ major.}
  \end{subfigure}
\hfill
\begin{subfigure}[t]{0.32\textwidth}
    \centering
\includegraphics[width=\linewidth]{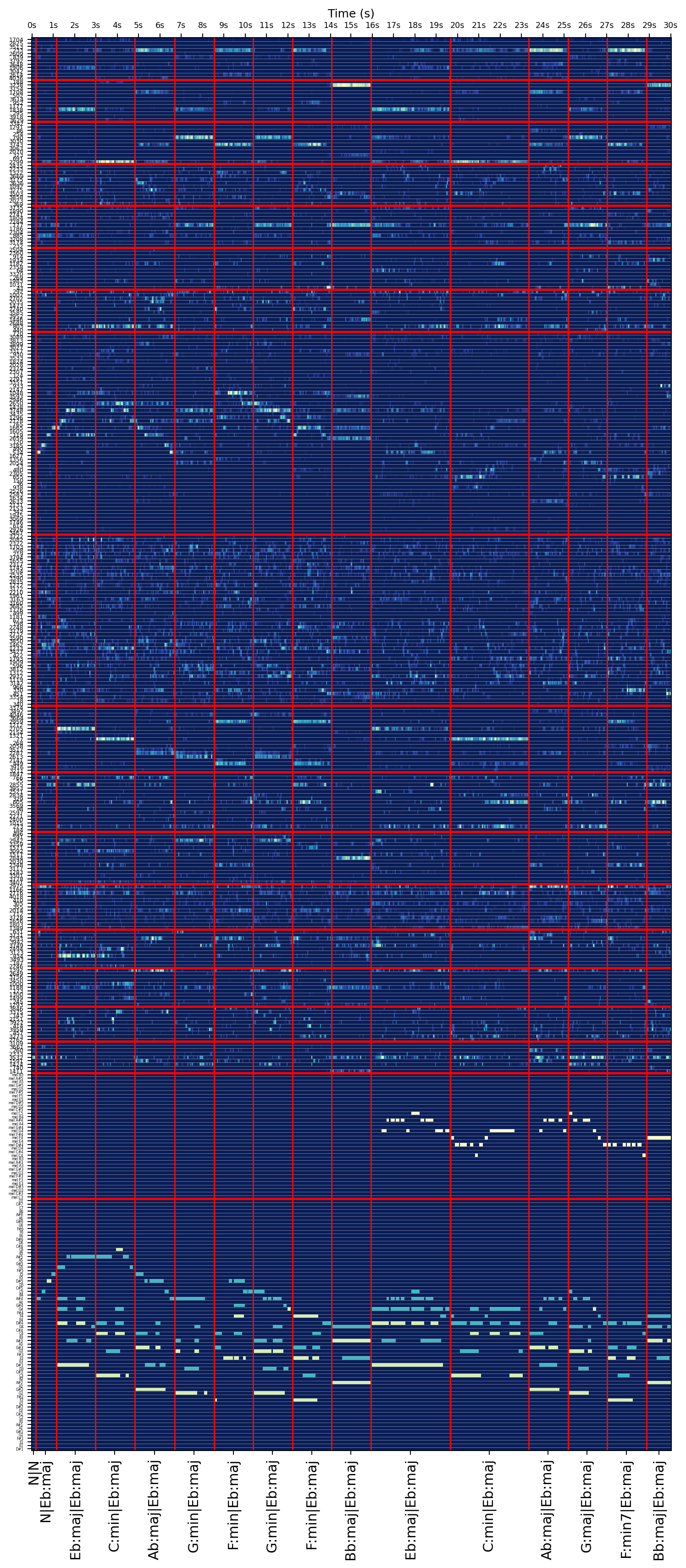}
    \caption{E$\flat$ major.}
  \end{subfigure}
\hfill
\begin{subfigure}[t]{0.32\textwidth}
    \centering
\includegraphics[width=\linewidth]{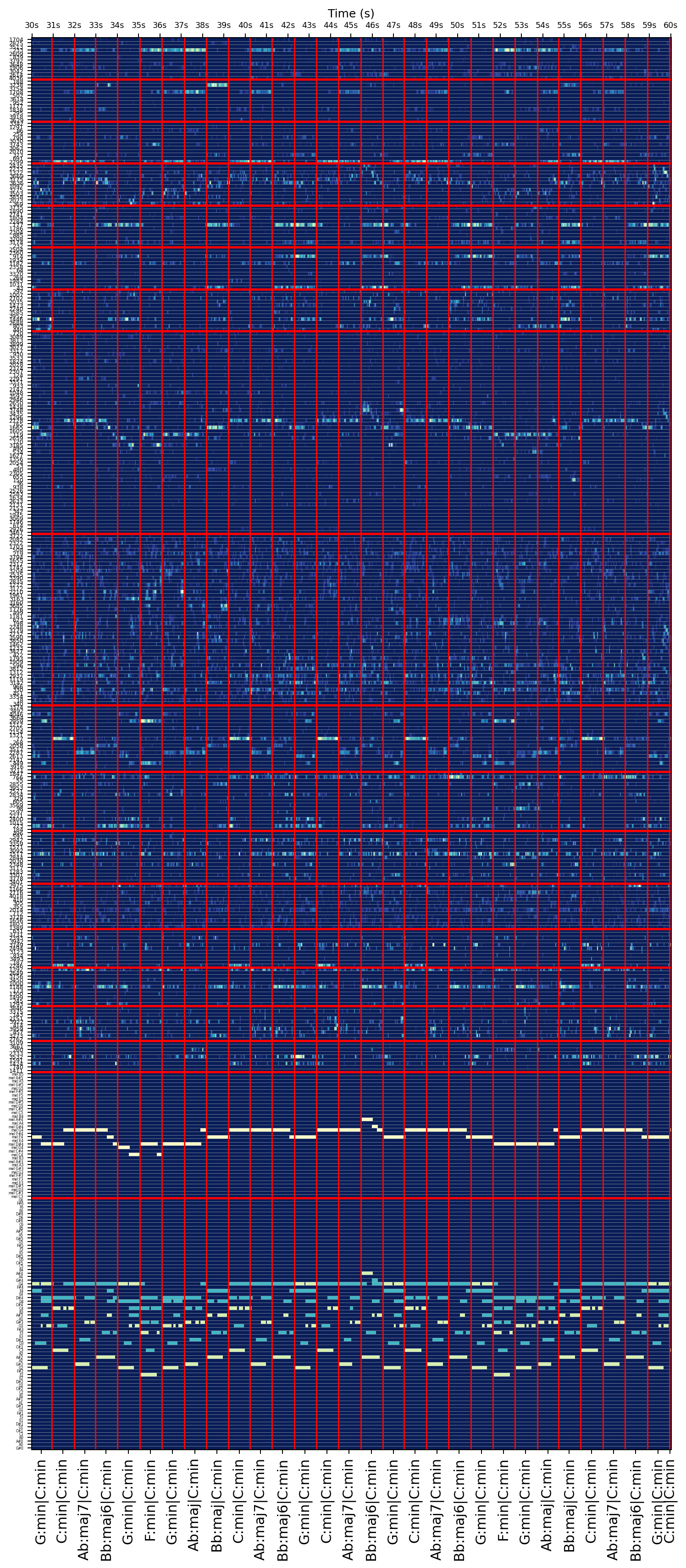}
    \caption{C minor.}
  \end{subfigure}
\caption{
Frame-level activations of all recovered MuQ (2) orbits with more than 10 features
for three POP909 excerpts, using $K=32$ and $\tau=0.5$.
The horizontal axis shows time with chord annotations, while feature rows are
grouped by recovered orbit. All excerpts are real multi-instrument vocal
recordings with complex arrangements rather than synthetic audio.
}
\label{fig:activation_all_orbits}
\end{figure}

\subsection{Orbit Activation Patterns}
\label{app:orbit_activation}

We visualize the frame-level activations of all the recovered orbits shown in Figure~\ref{fig:orbit_muq_2_05} with at least 10 features across
music excerpts. 

\paragraph{Controlled synthetic diagnostic dataset.}
To separate pitch and harmonic effects from instrumentation and arrangement,
we additionally construct a controlled diagnostic dataset. This dataset is
used only for post-hoc analysis of the trained SAE and is not included in SAE
training. It contains 96 audio samples, covering four instrument families,
12 tonic pitch classes, and two modes:
\[
4\ \text{families}
\times 12\ \text{roots}
\times \{\text{major},\text{minor}\}.
\]
The four families are piano, strings, woodwinds, and brass. Piano samples use
an acoustic grand piano, while the ensemble families distribute the musical
parts across multiple instruments: cello, viola, and violins for strings;
bassoon, clarinet, oboe, and flute for woodwinds; and trombone, French horn,
and trumpets for brass.

Each sample lasts exactly 30 seconds. The first 10 seconds contain an ascending
major or natural-minor scale spanning three octave regions. The 22 scale notes
are evenly distributed over this interval. For the ensemble families, the low,
middle, and high regions are assigned to different instruments, whereas all
three regions are performed by the piano in the piano condition.

The remaining 20 seconds contain two repetitions of an eight-chord Canon
progression. Major samples use
$I$--$V$--$vi$--$iii$--$IV$--$I$--$IV$--$V$, while minor samples use
$i$--$v$--$VI$--$III$--$iv$--$i$--$iv$--$V$. The final dominant in the
minor progression is major and therefore introduces the raised seventh scale
degree. Each chord lasts 1.25 seconds and is rendered as a four-note voicing.
For strings, woodwinds, and brass, the four chord voices are distributed
across different instruments within the family. The MIDI files are rendered
to audio using FluidSynth and a General MIDI SoundFont.

\paragraph{Activation patterns in synthetic diagnostic dataset.} Figure~\ref{fig:activations_syn} shows the frame-level activations of recovered MuQ (2) orbits on controlled synthetic
excerpts in \textit{C major}, \textit{C minor}, and \textit{A minor} using $K=32$ and $\tau=0.5$. The comparison isolates the
effects of key and mode across different instrument families. The top three orbits in each subfigure are subdominant, major chord, and minor chord orbits, respectively. Each of these three orbits starts from root \textit{C} to \textit{B}.

Across the ascending and descending scale sections, similar ordered activation
patterns recur over different instrument families. During
the final chord progression, some orbit activations become more sustained and
reflect the simultaneous harmonic content of each block chord. The recurrence
of related patterns indicates that
the recovered ordering is not specific to a single instrumental timbre.
Nevertheless, activation strength and selectivity vary across families,
showing that these features retain some sensitivity to timbre and orchestral
voicing. Interestingly, it is quite easy for a human to distinguish a key from a measure-scale ascending or descending melody, but we can not find evidence that the model does so. Consistent with what we discuss throughout the paper, the activations of subdominant rings reveal that the model does detect the key where there is a submominant chord.

\paragraph{Activation patterns in real recordings.}
In the main text, we presented four selected orbits from three audio excerpts. Here, we expand the visualization to include all recovered
orbits containing more than 10 features under the same MuQ (2) model with
$K=32$ and $\tau=0.5$, as shown in
Figure~\ref{fig:activation_all_orbits}. The three excerpts are taken from
POP909 and are real multi-instrument vocal recordings with complex
arrangements. They therefore contain singing voices, overlapping instruments,
non-chord tones, and production effects, rather than the simplified acoustic
conditions of synthetic examples.

In the \textit{B$\flat$ major} and \textit{E$\flat$ major} excerpts, several orbit blocks exhibit
localized activation changes aligned with chord boundaries. Within these
blocks, different ordered features become prominent as the chord root changes,
supporting the pitch-class interpretation of the recovered structures. A
longer orbit also follows the activity of the piano-roll reference more
directly, suggesting that some recovered structures encode local melody content
rather than chord or key identity alone.

The \textit{C minor} excerpt provides a less ideal example. Its activation patterns are
broader and overlap more strongly across neighboring chord segments, and some
expected orbit--chord correspondences are weak or incomplete. This example
illustrates that the recovered semantics are not perfectly disentangled at
every frame. Dense instrumentation, vocal melody, non-chord tones, and the
ambiguity between minor keys and their relative majors can all affect which
features dominate locally. We retain this example to show both the
interpretability and the limitations of the recovered orbits in realistic
audio. Additional activation examples across musical genres are discussed in
Section~\ref{app:genre_robustness}.

Together, the POP909 and synthetic examples provide complementary evidence.
The POP909 excerpts demonstrate that orbit semantics remain visible in
realistic vocal music with complex arrangements, while the synthetic dataset
isolates their responses to scale degree, mode, register, and instrument
family.

\section{Ablation Studies}
\label{app:ablations}

We study the effects of orbit alignment, foundation-model layer, and SAE
sparsity.





\begin{figure}[t]
  \centering

  \begin{subfigure}[t]{0.49\textwidth}
    \centering
\includegraphics[width=\linewidth]{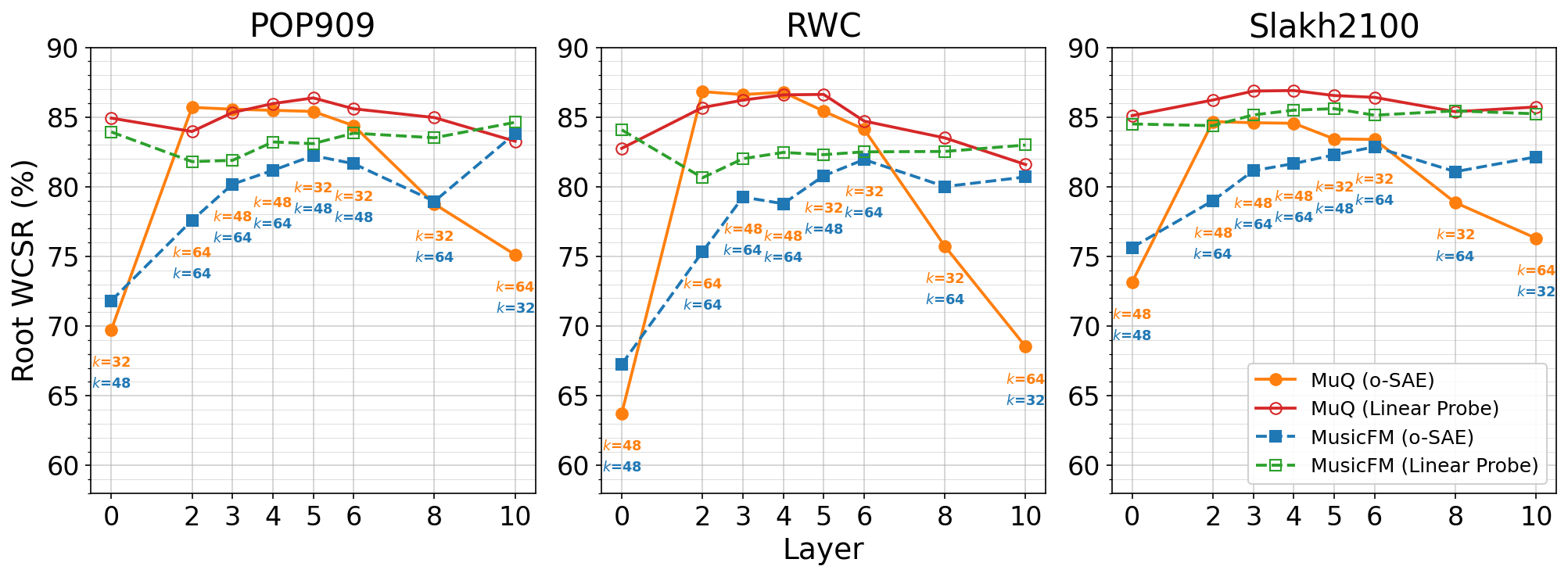}
    \caption{Root WCSR under \textit{Oracle Bd}.}
  \end{subfigure}
  \hfill
  \begin{subfigure}[t]{0.49\textwidth}
    \centering
\includegraphics[width=\linewidth]{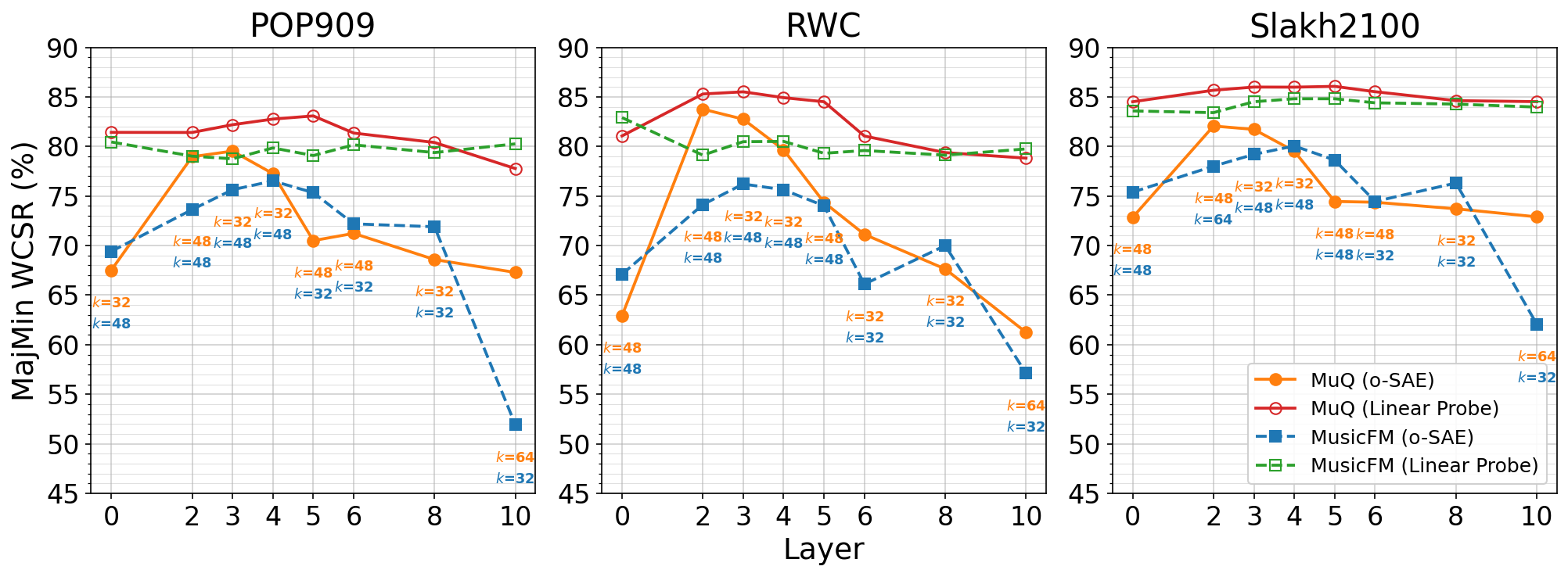}
\caption{Majmin WCSR under \textit{Oracle Bd}.}
  \end{subfigure}
 \begin{subfigure}[t]{0.49\textwidth}
    \centering
\includegraphics[width=\linewidth]{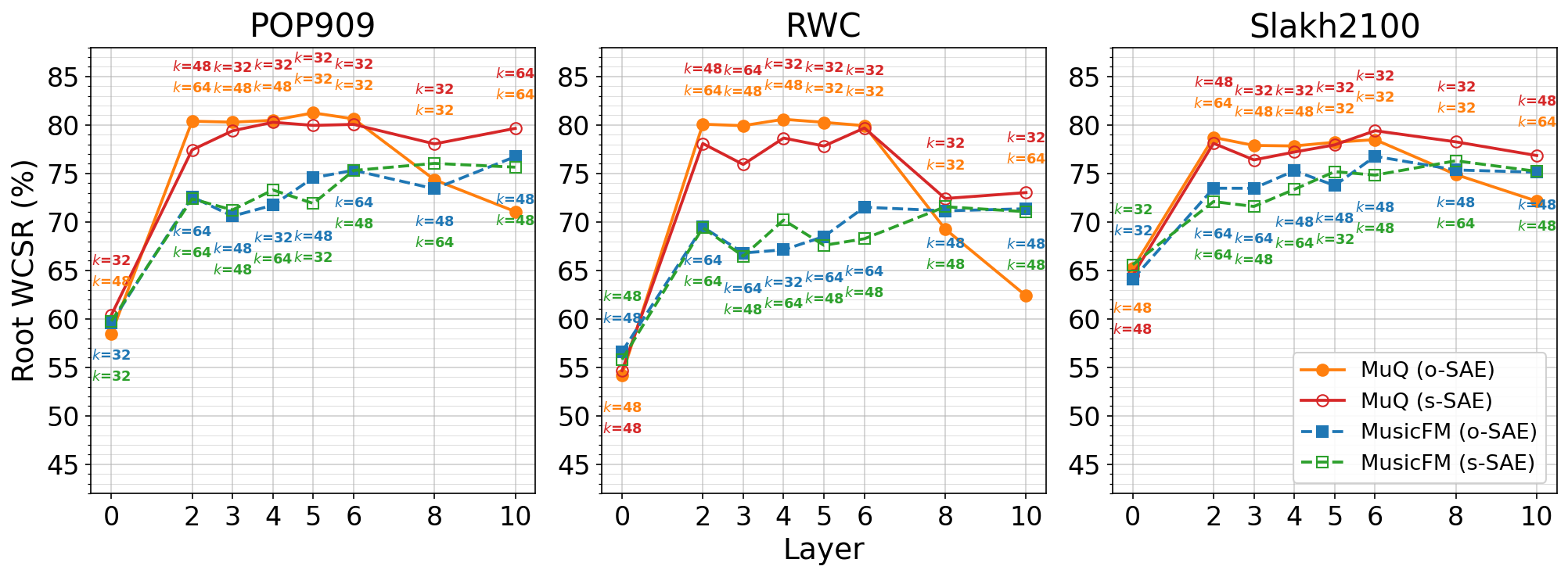}
    \caption{Root WCSR under \textit{Est. Bd}.}
  \end{subfigure}
  \hfill
  \begin{subfigure}[t]{0.49\textwidth}
    \centering
\includegraphics[width=\linewidth]{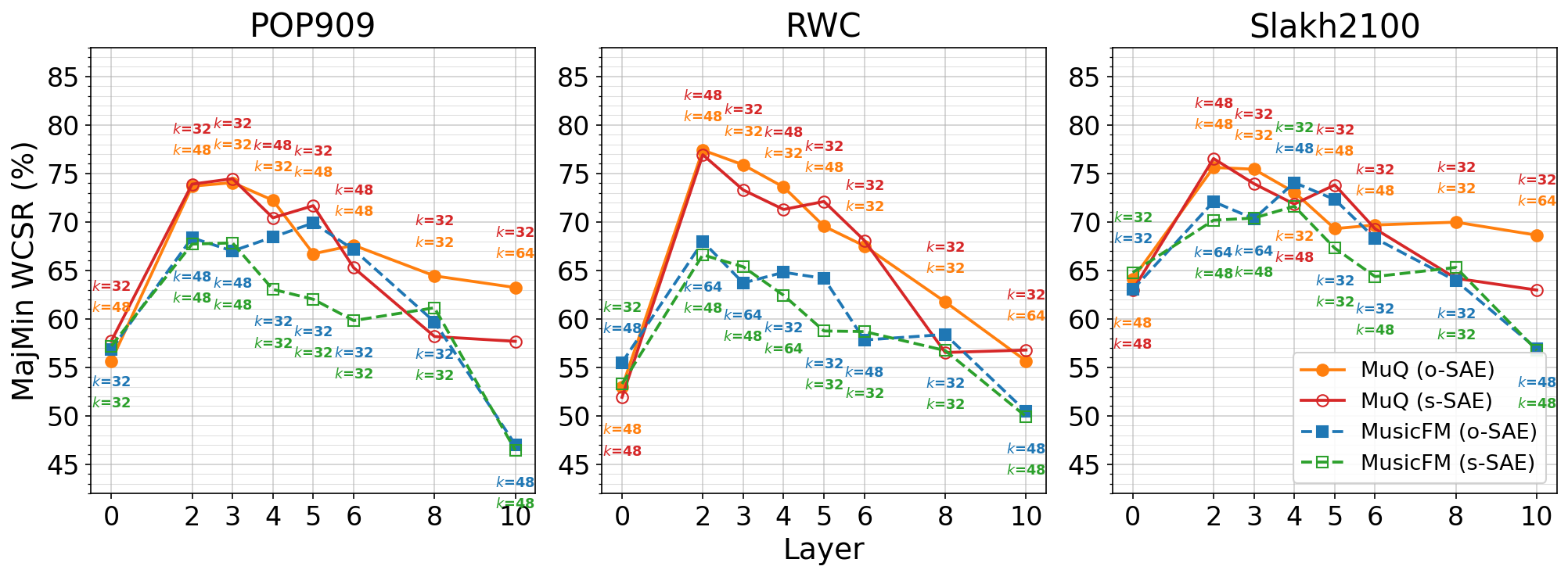}
\caption{Majmin WCSR under \textit{Est. Bd}.}
  \end{subfigure}
  \caption{Chord recognition results.} \label{fig:chord_recognition}
\end{figure}
\begin{figure}[t]
  \centering
  \begin{subfigure}[t]{0.49\textwidth}
    \centering
\includegraphics[width=\linewidth]{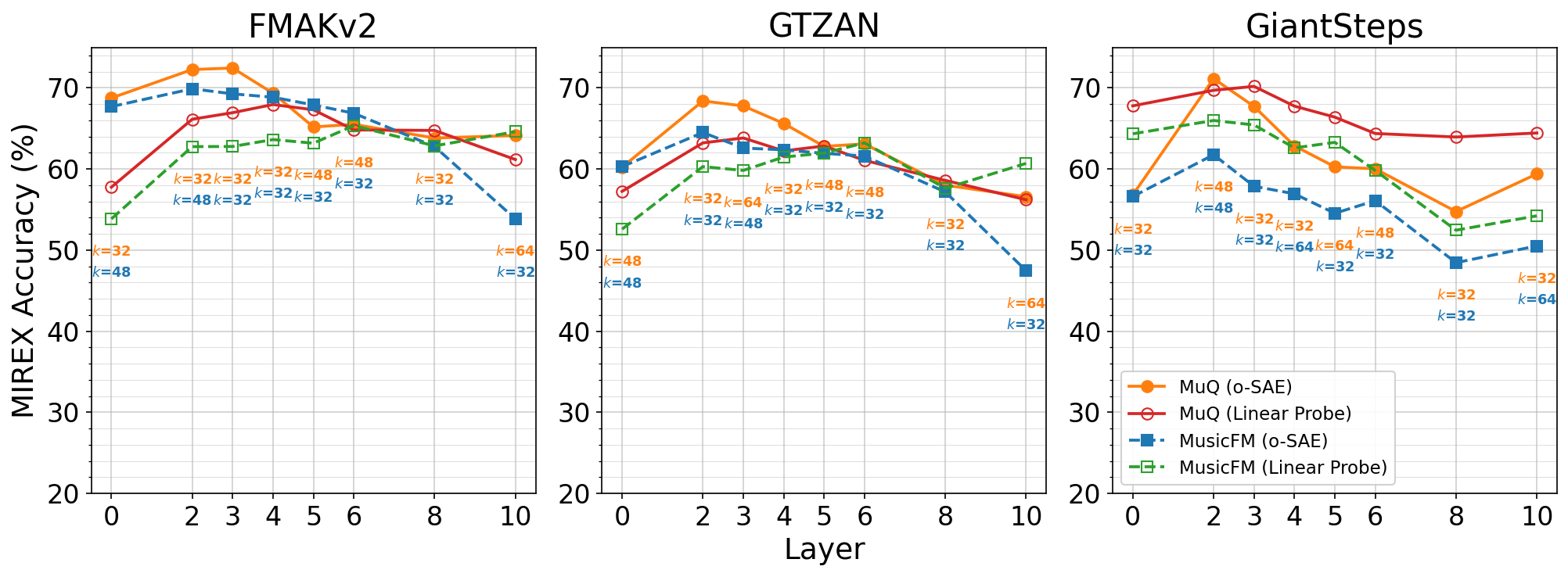}
    \caption{MIREX accuracy.}
  
  \end{subfigure}
  \hfill
  \begin{subfigure}[t]{0.49\textwidth}
    \centering \includegraphics[width=\linewidth]{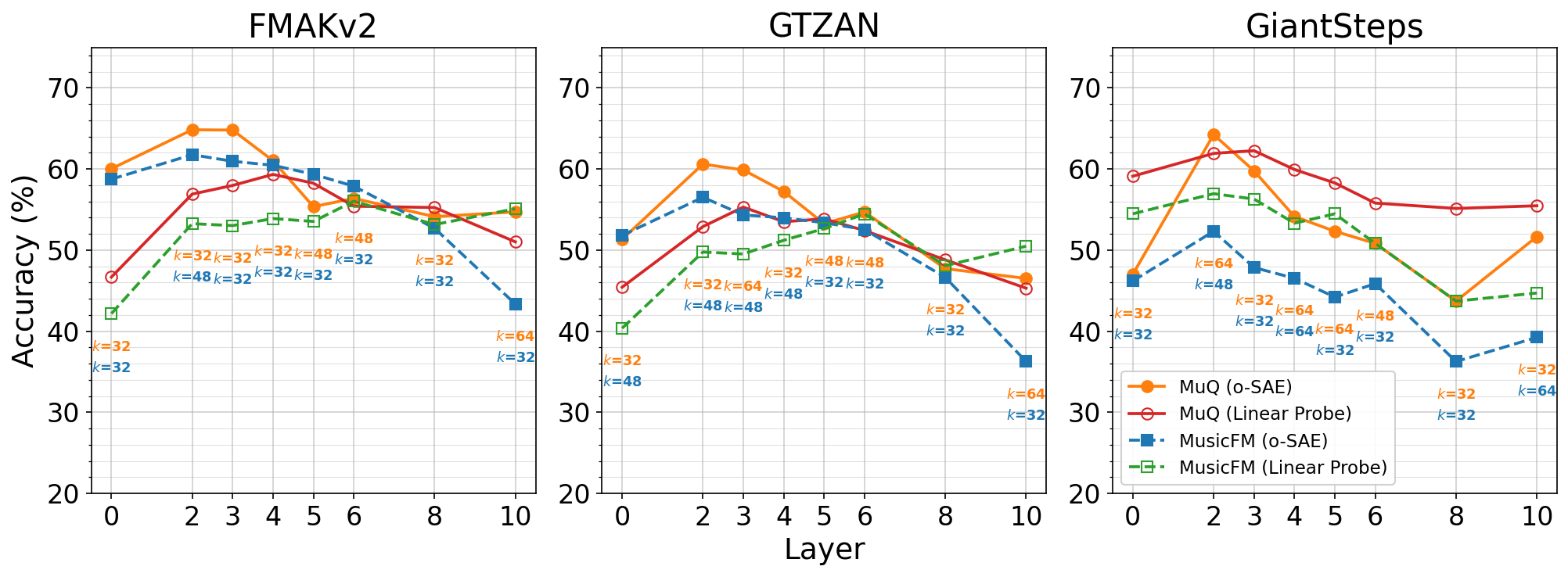}  \caption{Strict key accuracy.}
  \end{subfigure}
  \caption{Layer-wise key detection results derived from major- and
    minor-chord features.} \label{fig:chord-key}

\end{figure}
\begin{figure}[t]
  \centering

  \begin{subfigure}[t]{0.49\textwidth}
    \centering
    \includegraphics[width=\linewidth]{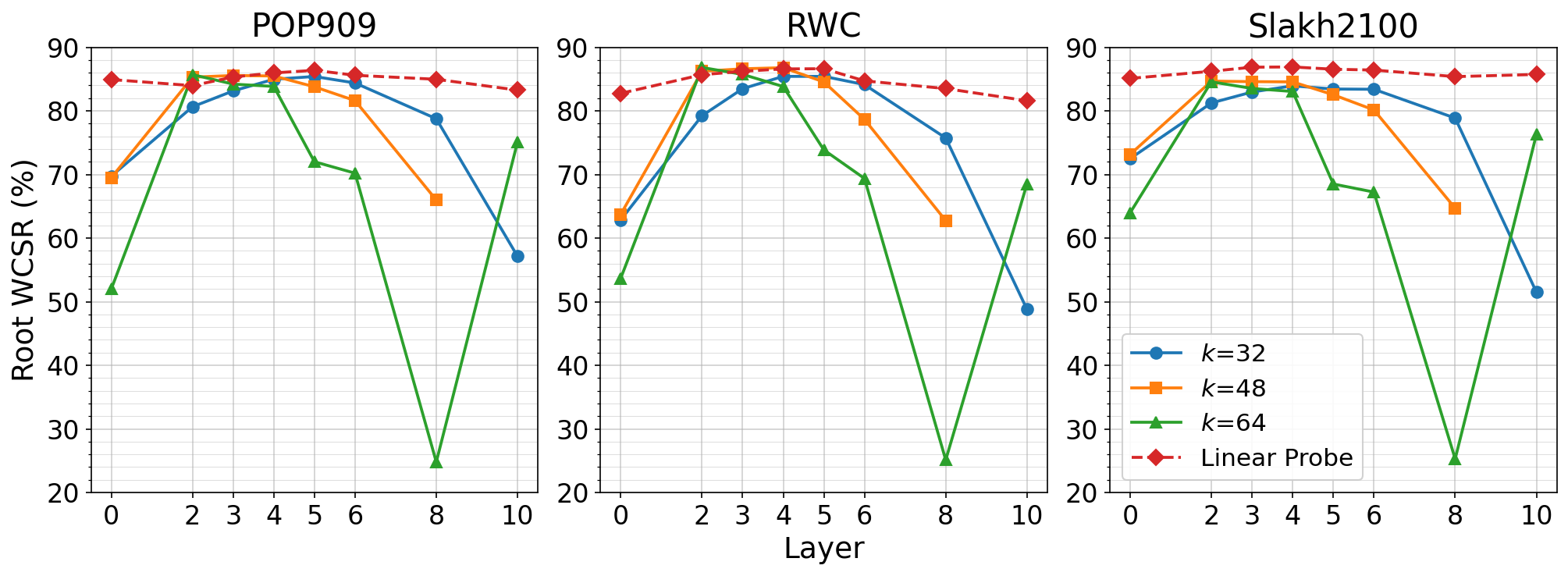}
    \caption{Root WCSR of MuQ o-SAE.}
  \end{subfigure}
  \hfill
  \begin{subfigure}[t]{0.49\textwidth}
    \centering
    \includegraphics[width=\linewidth]{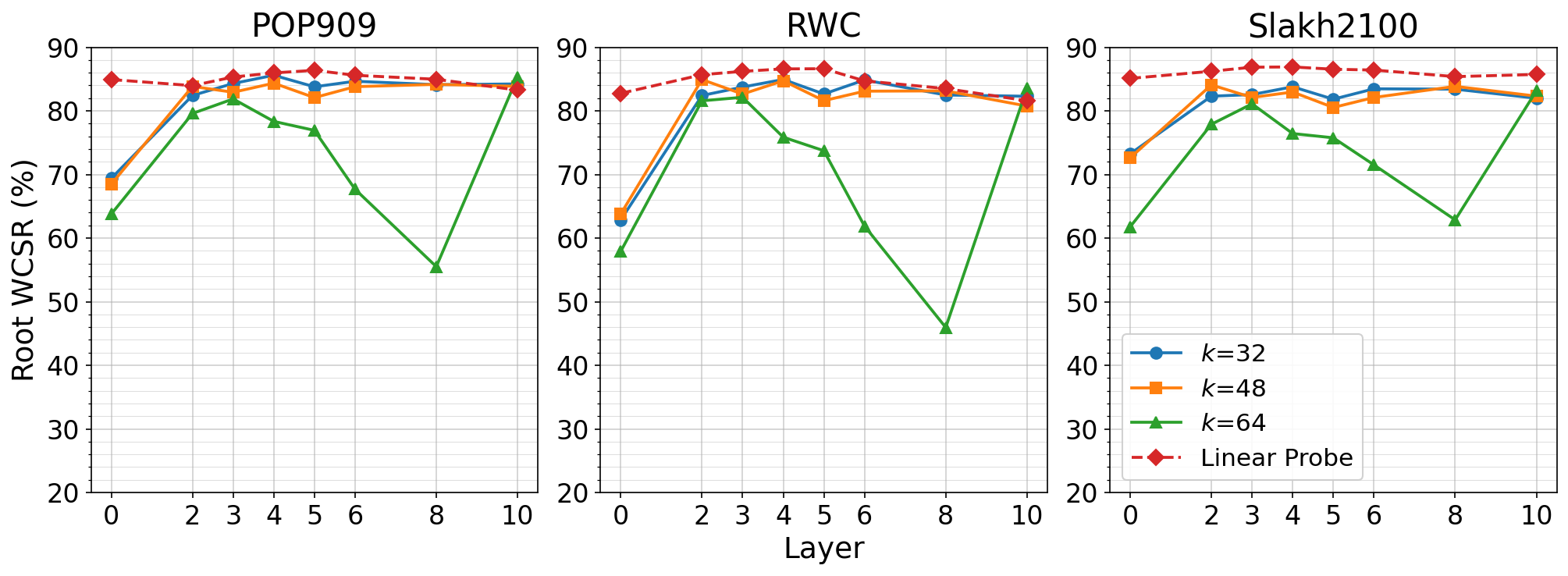}
    \caption{Root WCSR of MuQ s-SAE.}
  \end{subfigure}

  \begin{subfigure}[t]{0.49\textwidth}
    \centering
    \includegraphics[width=\linewidth]{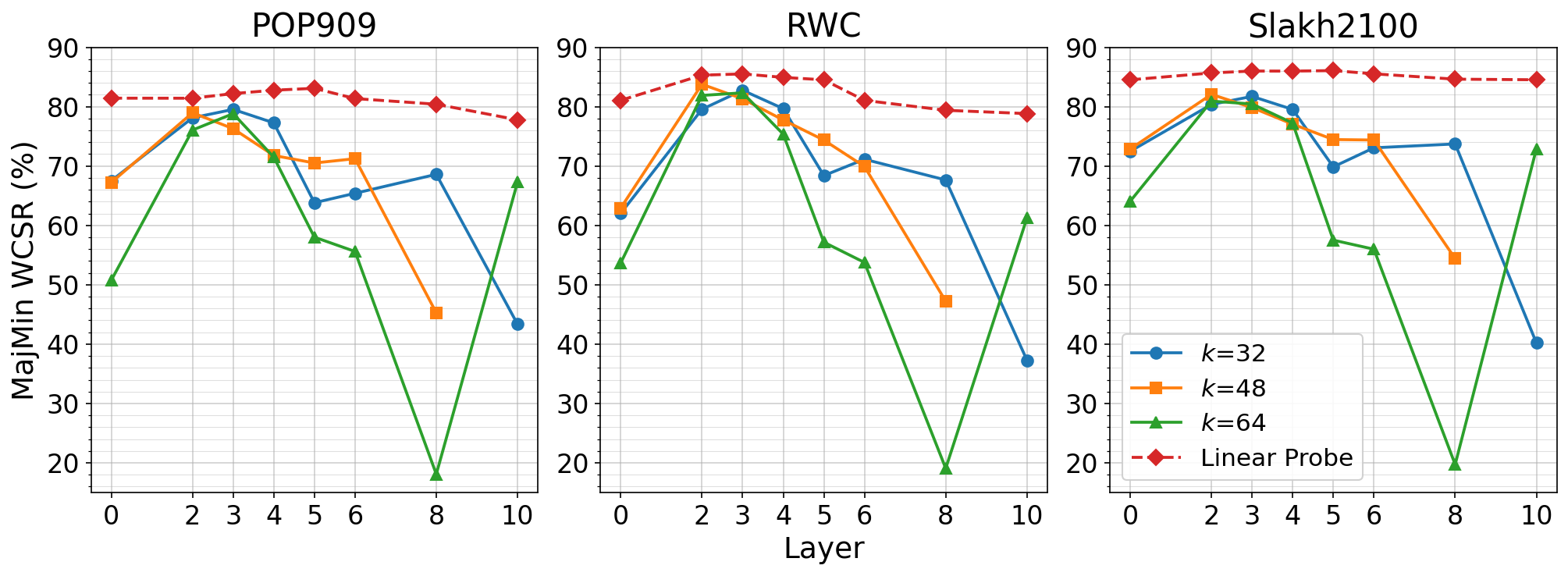}
    \caption{MajMin WCSR of MuQ o-SAE.}
  \end{subfigure}
  \hfill
  \begin{subfigure}[t]{0.49\textwidth}
    \centering
    \includegraphics[width=\linewidth]{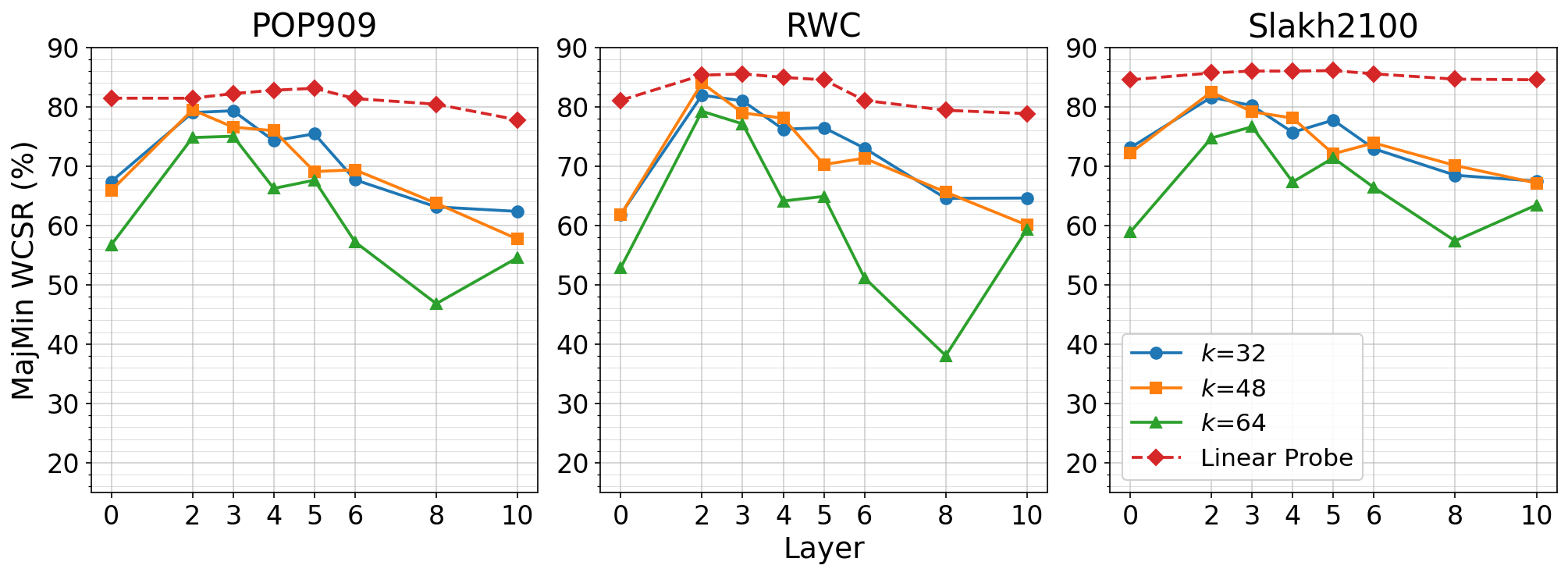}
    \caption{MajMin WCSR of MuQ s-SAE.}
  \end{subfigure}

  \begin{subfigure}[t]{0.49\textwidth}
    \centering
    \includegraphics[width=\linewidth]{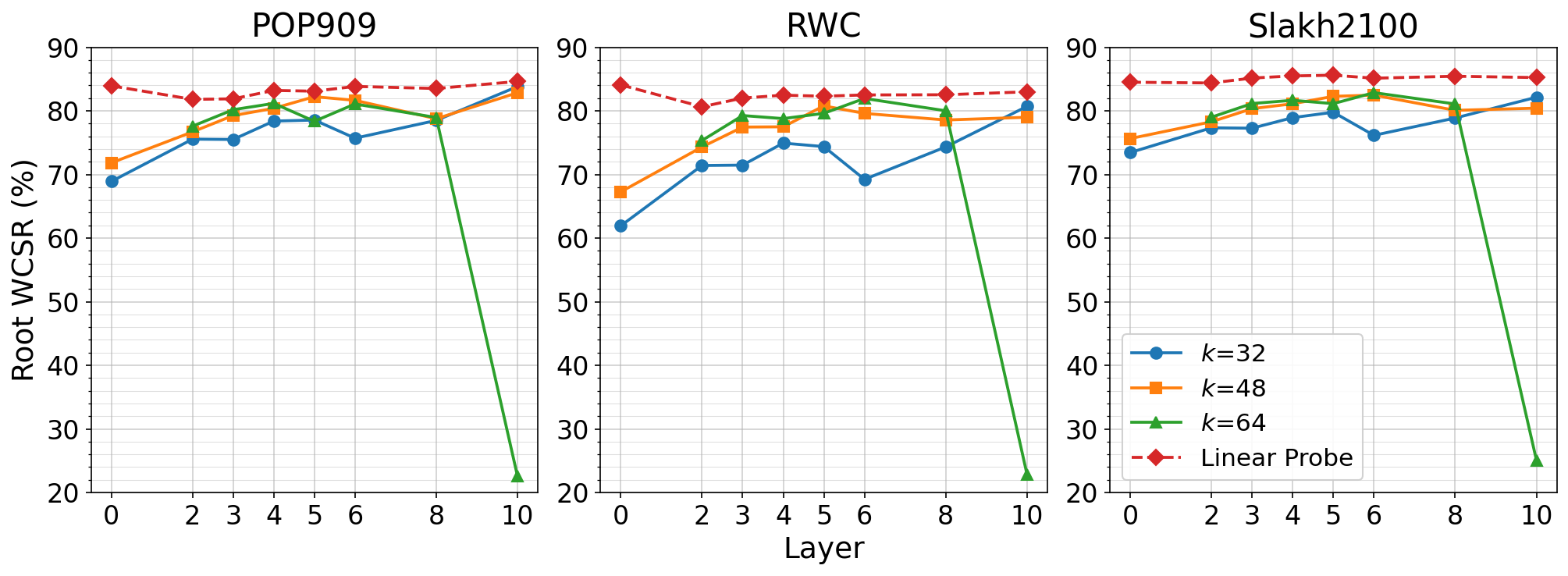}
    \caption{Root WCSR of MusicFM o-SAE.}
  \end{subfigure}
  \hfill
  \begin{subfigure}[t]{0.49\textwidth}
    \centering
    \includegraphics[width=\linewidth]{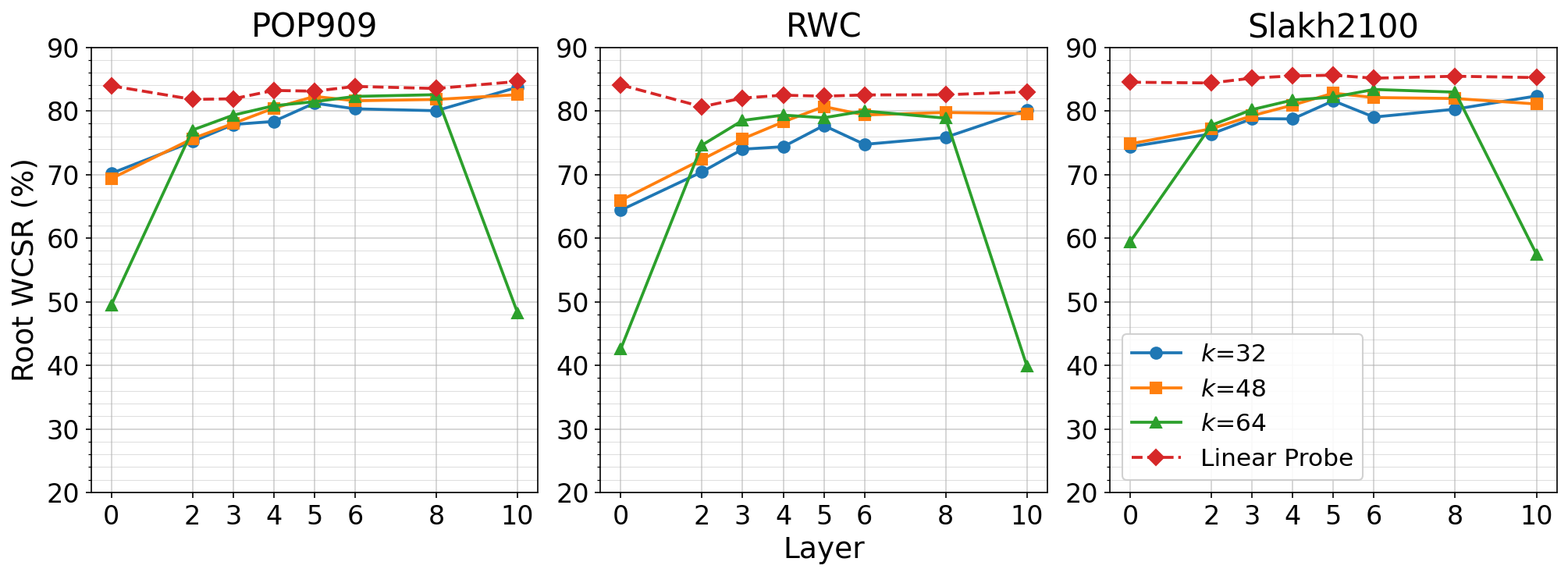}
    \caption{Root WCSR of MusicFM s-SAE.}
  \end{subfigure}

  \begin{subfigure}[t]{0.49\textwidth}
    \centering
    \includegraphics[width=\linewidth]{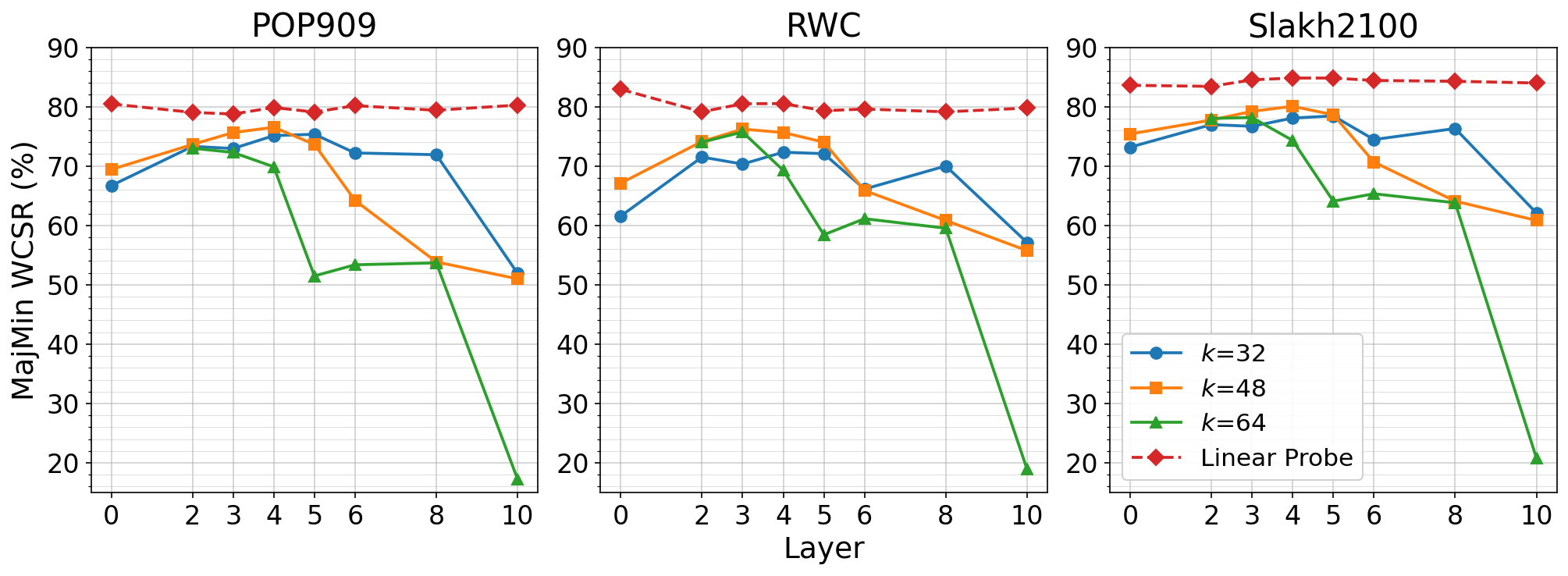}
    \caption{MajMin WCSR of MusicFM o-SAE.}
  \end{subfigure}
  \hfill
  \begin{subfigure}[t]{0.49\textwidth}
    \centering
    \includegraphics[width=\linewidth]{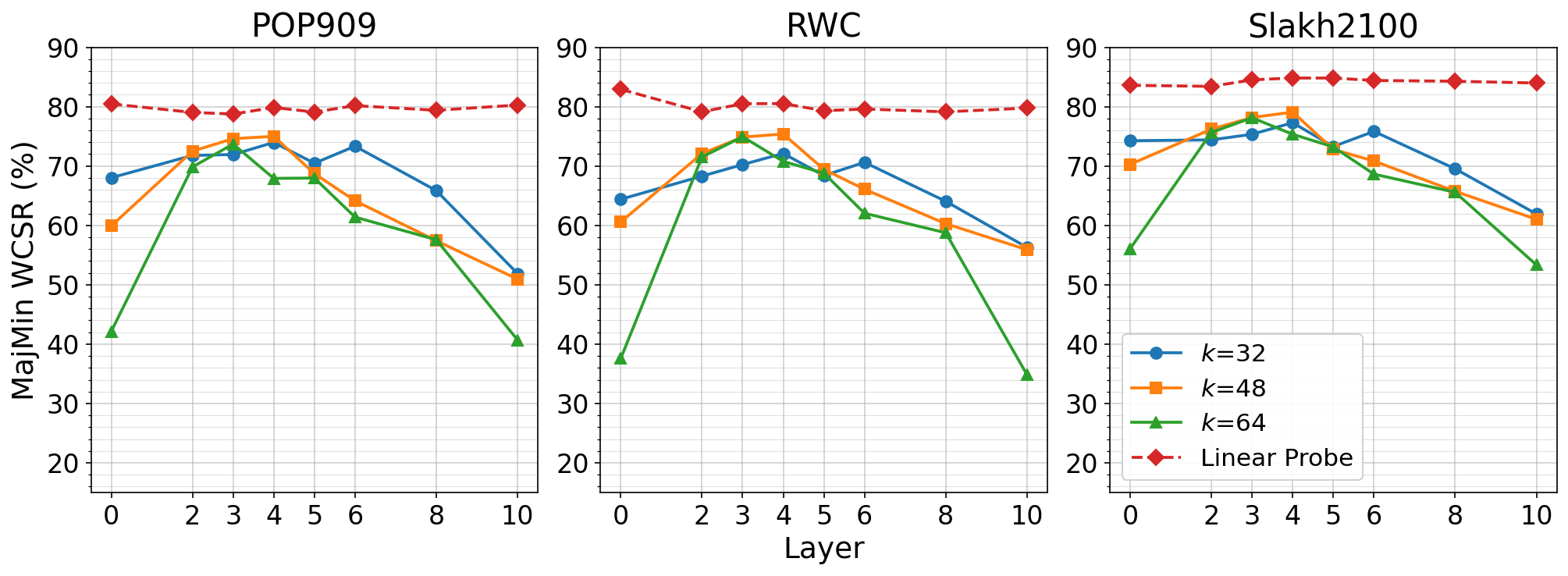}
    \caption{MajMin WCSR of MusicFM s-SAE.}
  \end{subfigure}

  \caption{
  Effect of Top-$K$ sparsity on layer-wise chord-recognition performance
  with \textit{Oracle Bd}. Results are shown for
  $K\in\{32,48,64\}$, with the linear probe included as a reference.
  }
  \label{fig:chord_topk}
\end{figure}

\subsection{s-SAE and Layer-Wise Comparison}
\label{app:ablation_ssae}

As an ablation of the proposed orbit-based SAE (o-SAE), we train a standard
single Top-$K$ SAE, denoted as s-SAE, using the same representations and
training data. Unlike o-SAE, s-SAE does not use pitch-shifted views or
shared-support alignment and therefore does not directly produce
transposition-induced orbits.

For downstream inference, we retrieve semantic features from s-SAE using
frozen linear probes. Let $w_c$ denote the weight vector of probe output $c$
and $e_i$ the encoder vector of SAE feature $i$. We select
\begin{equation}
i^\star(c)
=
\arg\max_i \operatorname{cos}(w_c,e_i).
\label{eq:probe_feature_retrieval}
\end{equation}
The retrieved features are ordered according to the probe outputs and processed
using the same inference procedure as the corresponding o-SAE orbits. This
matching is performed only after training and does not update the SAE
parameters.

Because different representations may perform best at different sparsity
levels, each layer-wise figure reports the best observed result across
$K\in\{32,48,64\}$ for each representation, SAE type, layer, and metric. This
provides a direct comparison across layers; results for all individual
sparsity levels are reported in Section~\ref{app:ablation_topk}.

\paragraph{Chord Recognition.}
Figure~\ref{fig:chord_recognition} compares the linear probe, s-SAE, and
o-SAE using Root and MajMin WCSR on POP909, RWC-POP-100, and the Slakh2100
test split. For s-SAE, the 12 major- and 12 minor-chord probe outputs retrieve
two pitch-class-ordered sets of SAE features, which are used in place of the
recovered o-SAE chord rings.

\paragraph{Key Detection via the Major- and Minor-Chord Rings.}
Figure~\ref{fig:chord-key} compares key detection from chord-feature
activations. For o-SAE, predictions are derived from the recovered major- and
minor-chord rings; for s-SAE, they are derived from the two feature sets
retrieved using the chord probe. We report MIREX and strict key accuracy on
FMAKv2, GTZAN-Key, and GiantSteps.

\begin{figure}[t]
  \centering

  \begin{subfigure}[t]{0.49\textwidth}
    \centering
    \includegraphics[width=\linewidth]{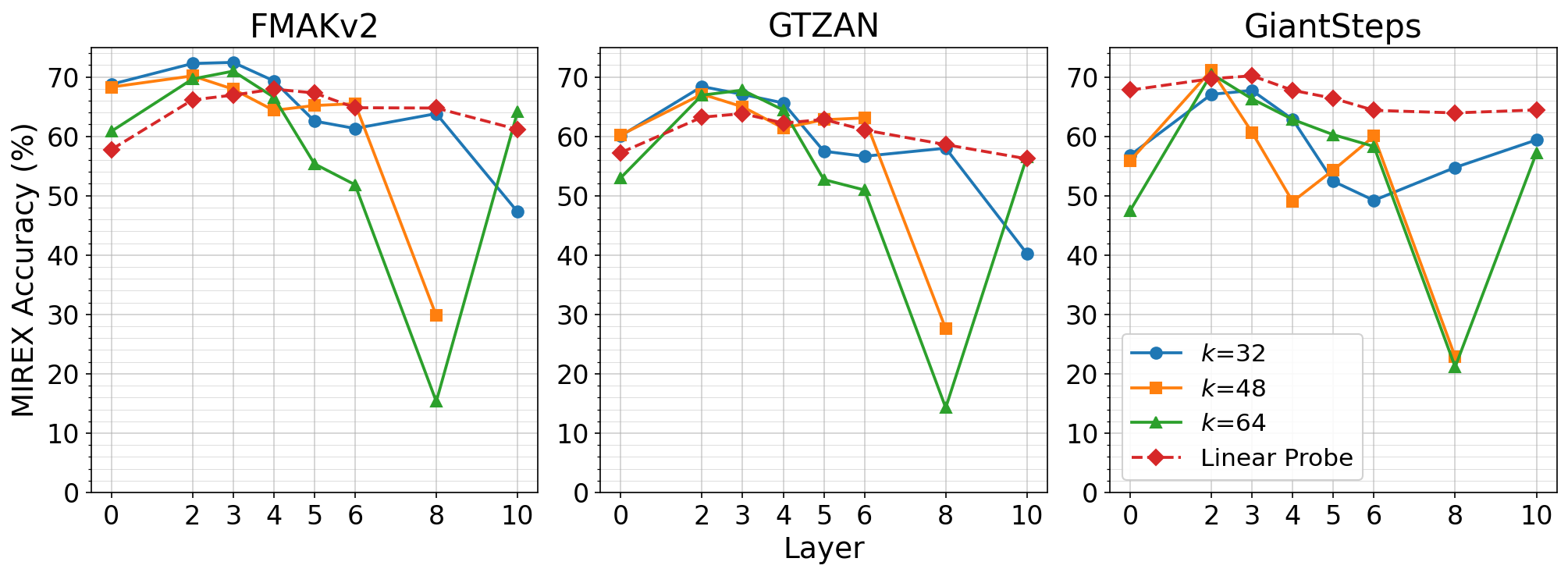}
    \caption{MIREX accuracy of MuQ o-SAE.}
  \end{subfigure}
  \hfill
  \begin{subfigure}[t]{0.49\textwidth}
    \centering
    \includegraphics[width=\linewidth]{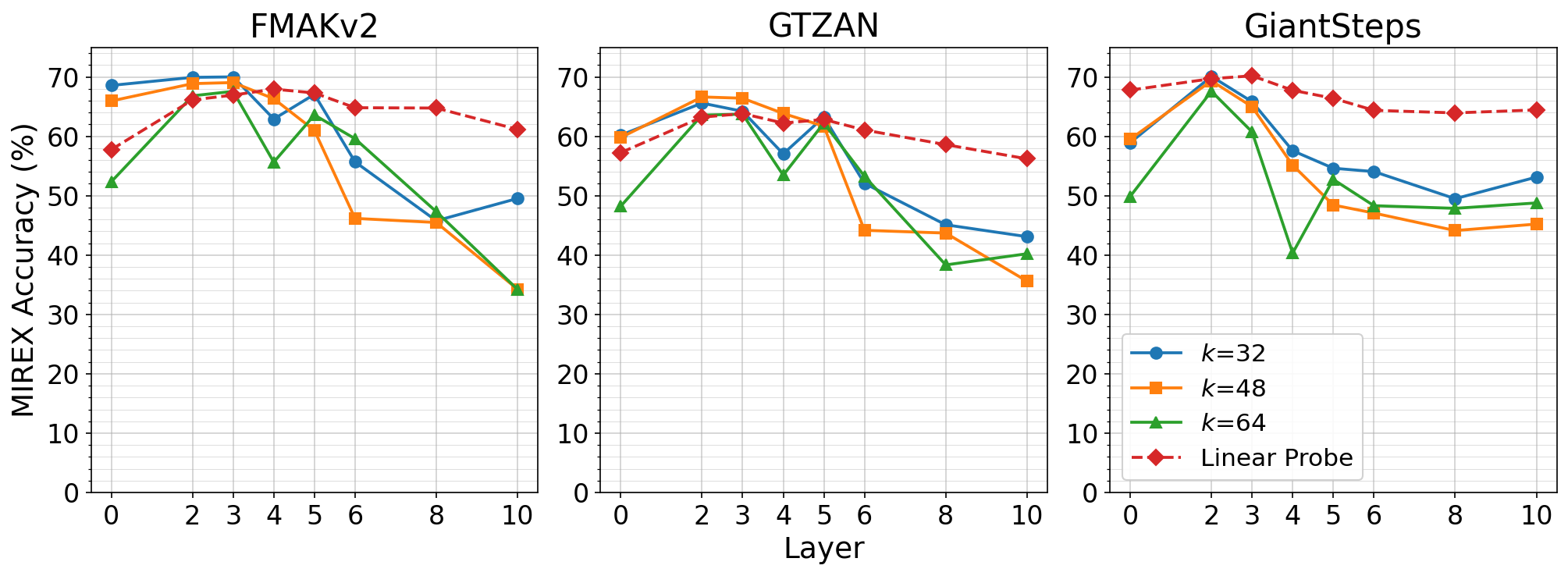}
    \caption{MIREX accuracy of MuQ s-SAE.}
  \end{subfigure}

  \begin{subfigure}[t]{0.49\textwidth}
    \centering
    \includegraphics[width=\linewidth]{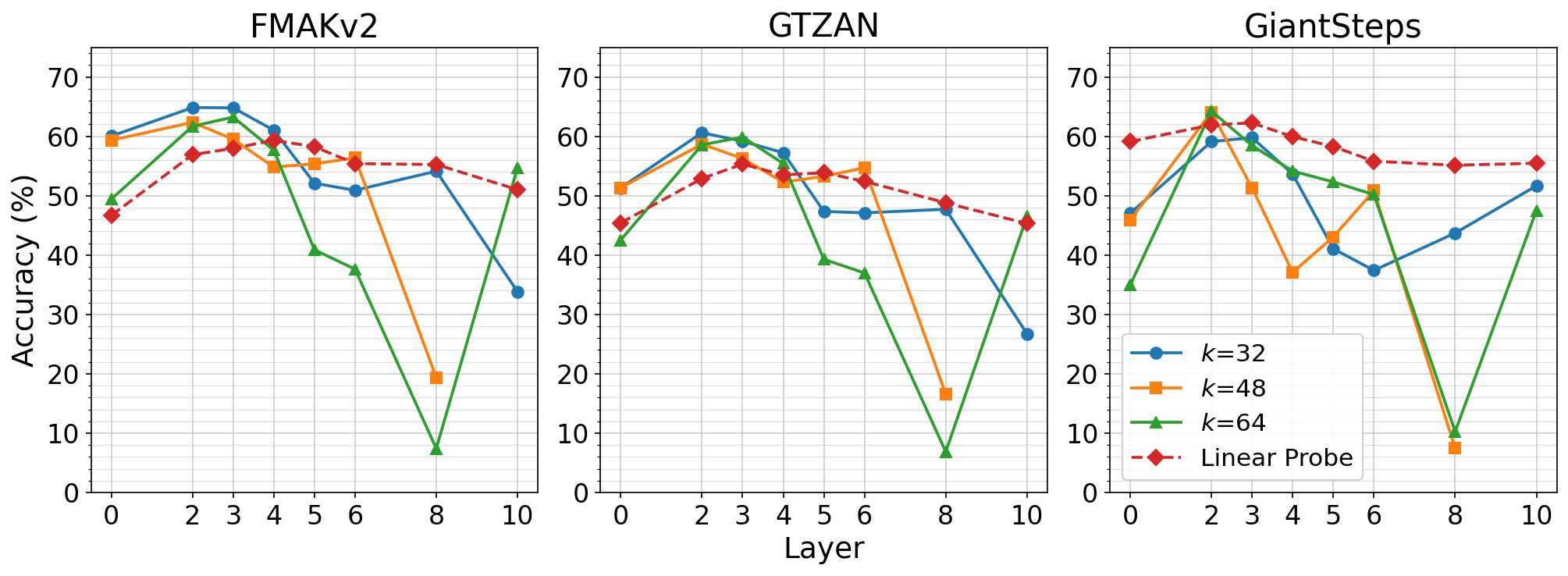}
    \caption{Strict key accuracy of MuQ o-SAE.}
  \end{subfigure}
  \hfill
  \begin{subfigure}[t]{0.49\textwidth}
    \centering
    \includegraphics[width=\linewidth]{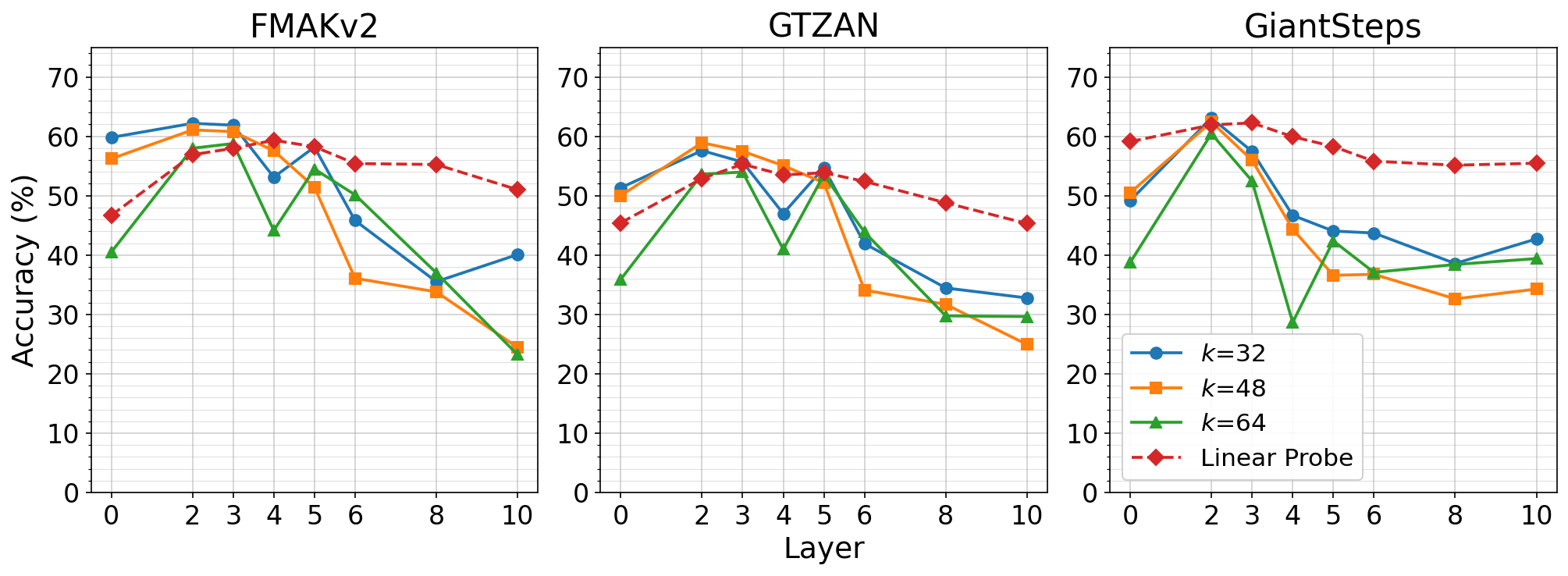}
    \caption{Strict key accuracy of MuQ s-SAE.}
  \end{subfigure}

  \begin{subfigure}[t]{0.49\textwidth}
    \centering
    \includegraphics[width=\linewidth]{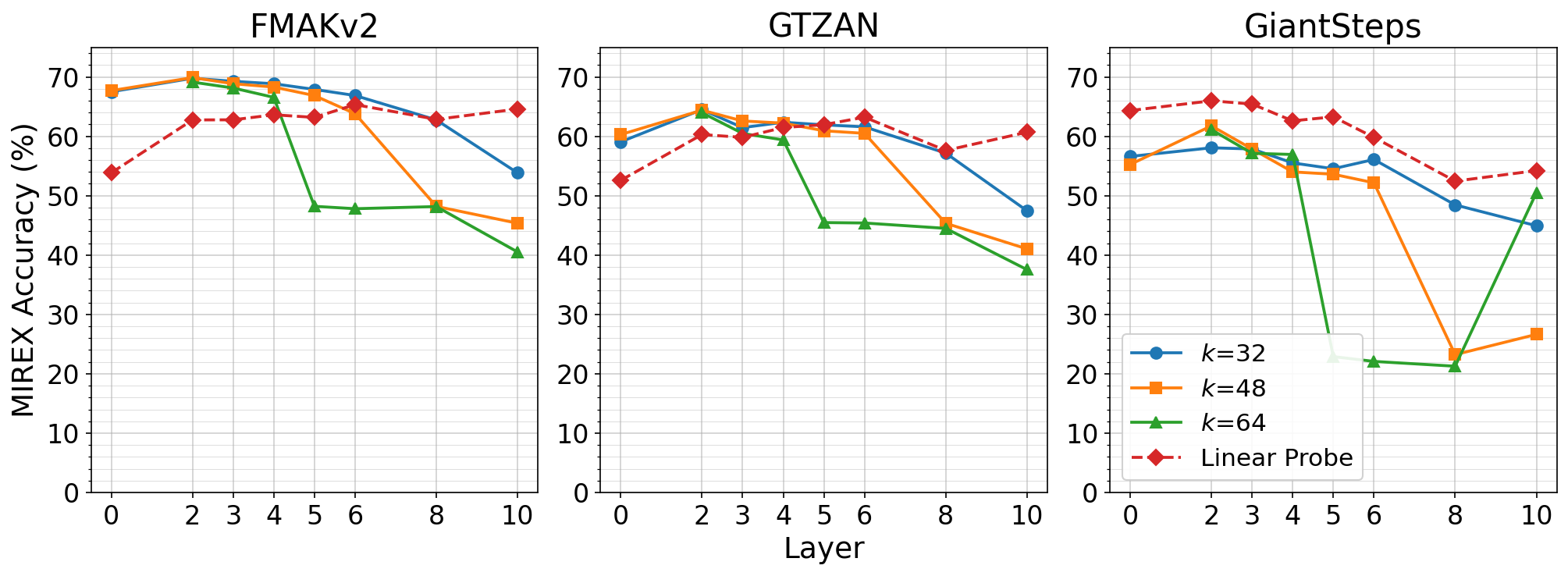}
    \caption{MIREX accuracy of MusicFM o-SAE.}
  \end{subfigure}
  \hfill
  \begin{subfigure}[t]{0.49\textwidth}
    \centering
    \includegraphics[width=\linewidth]{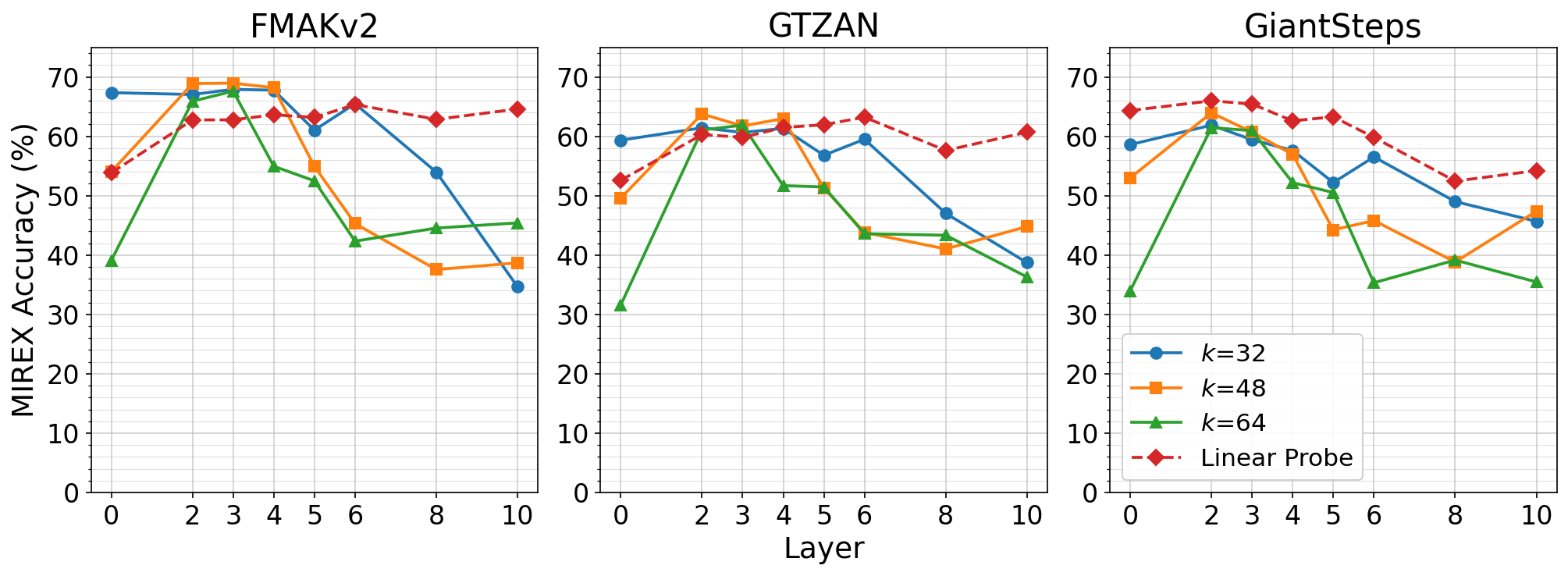}
    \caption{MIREX accuracy of MusicFM s-SAE.}
  \end{subfigure}

  \begin{subfigure}[t]{0.49\textwidth}
    \centering
    \includegraphics[width=\linewidth]{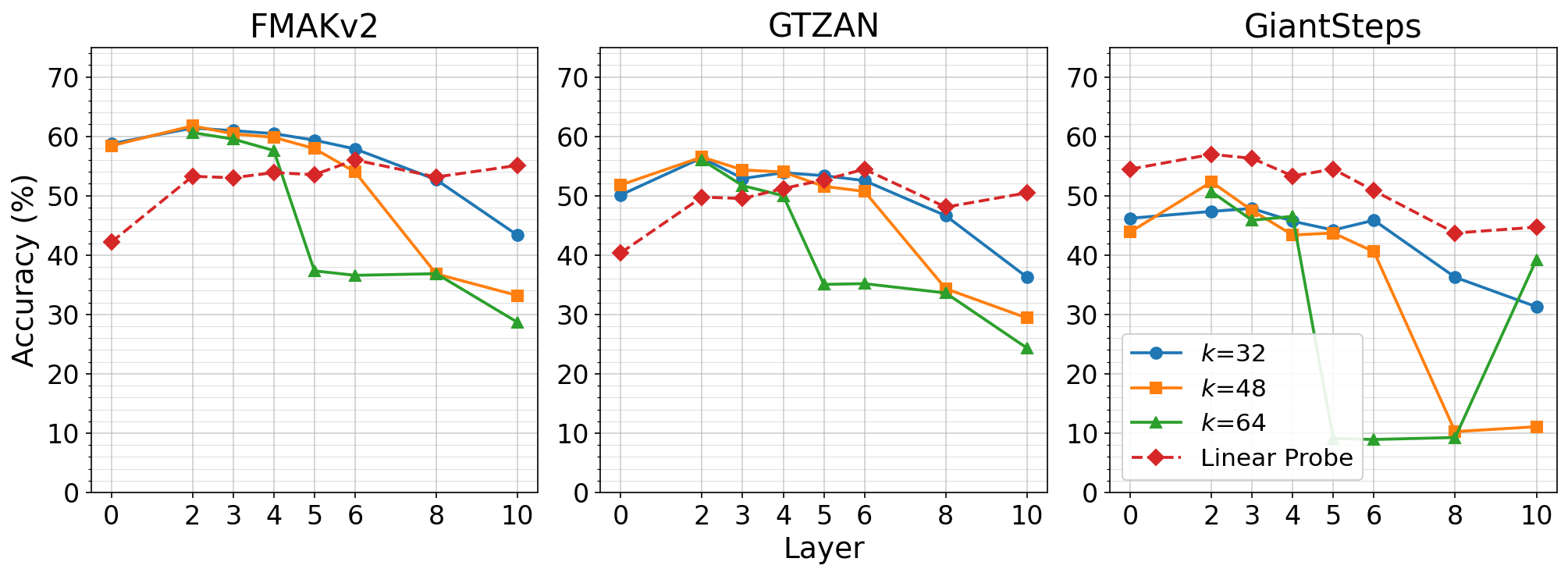}
    \caption{Strict key accuracy of MusicFM o-SAE.}
  \end{subfigure}
  \hfill
  \begin{subfigure}[t]{0.49\textwidth}
    \centering
    \includegraphics[width=\linewidth]{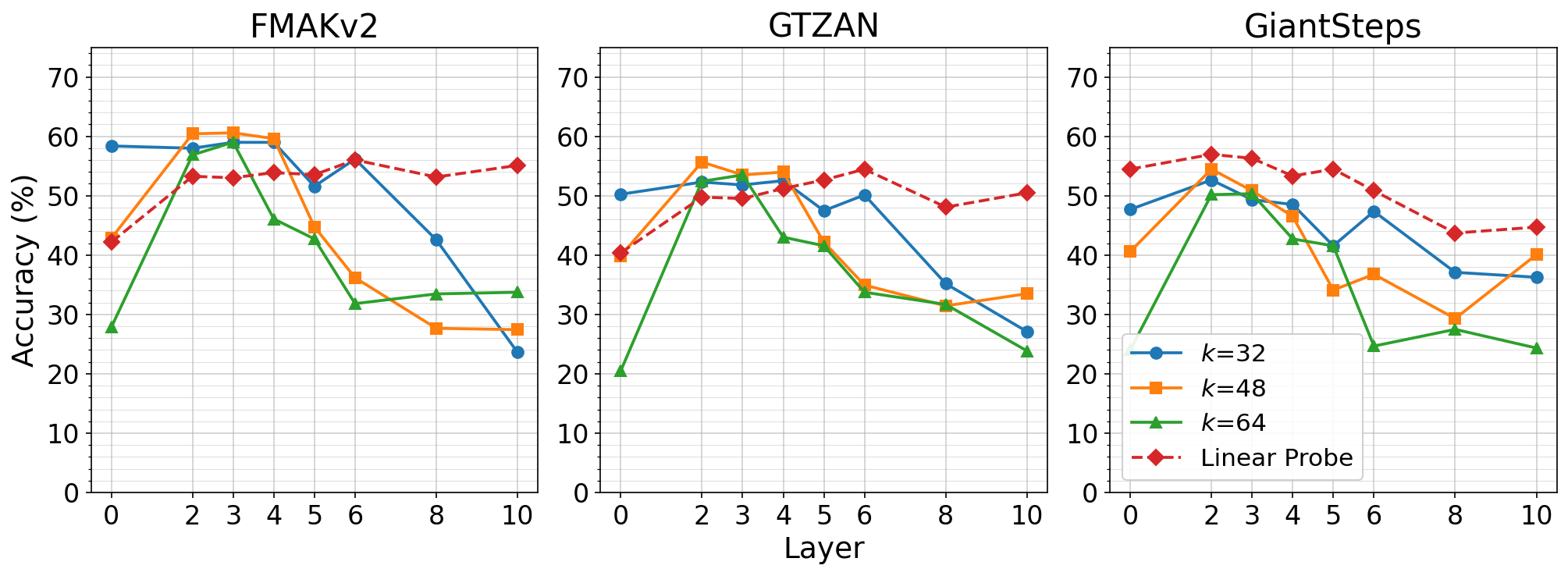}
    \caption{Strict key accuracy of MusicFM s-SAE.}
  \end{subfigure}

  \caption{
  Effect of Top-$K$ sparsity on layer-wise key-detection
  performance derived from chord features. Results are shown for
  $K\in\{32,48,64\}$, with the linear probe included as a reference.
  }
  \label{fig:chord_key_topk}
\end{figure}

\subsection{Effect of Top-$K$ Sparsity}
\label{app:ablation_topk}

We study the effect of SAE sparsity by varying
$K\in\{32,48,64\}$ while keeping the remaining training and inference
settings fixed. Unlike the layer-wise comparison above, which reports the
best observed result across $K$, the following figures show every sparsity
setting separately. The corresponding linear probes are included as
references and are independent of $K$.

\paragraph{Chord Recognition.}
Figure~\ref{fig:chord_topk} reports Root and MajMin WCSR with oracle chord
boundaries for o-SAE and s-SAE across MuQ and MusicFM layers. Each curve
corresponds to one value of $K$, allowing us to compare how chord-feature
quality changes with SAE sparsity across representations, layers, and
datasets.

\paragraph{Key Detection via the Major- and Minor-Chord Rings.}
Figure~\ref{fig:chord_key_topk} reports major--minor key detection derived
from chord-feature activations. Results are shown separately for MuQ and
MusicFM, o-SAE and s-SAE, and all three values of $K$. We report both strict
key accuracy and MIREX accuracy.

\begin{figure}[t]
  \centering

  \begin{subfigure}[t]{0.49\textwidth}
    \centering
    \includegraphics[width=\linewidth]{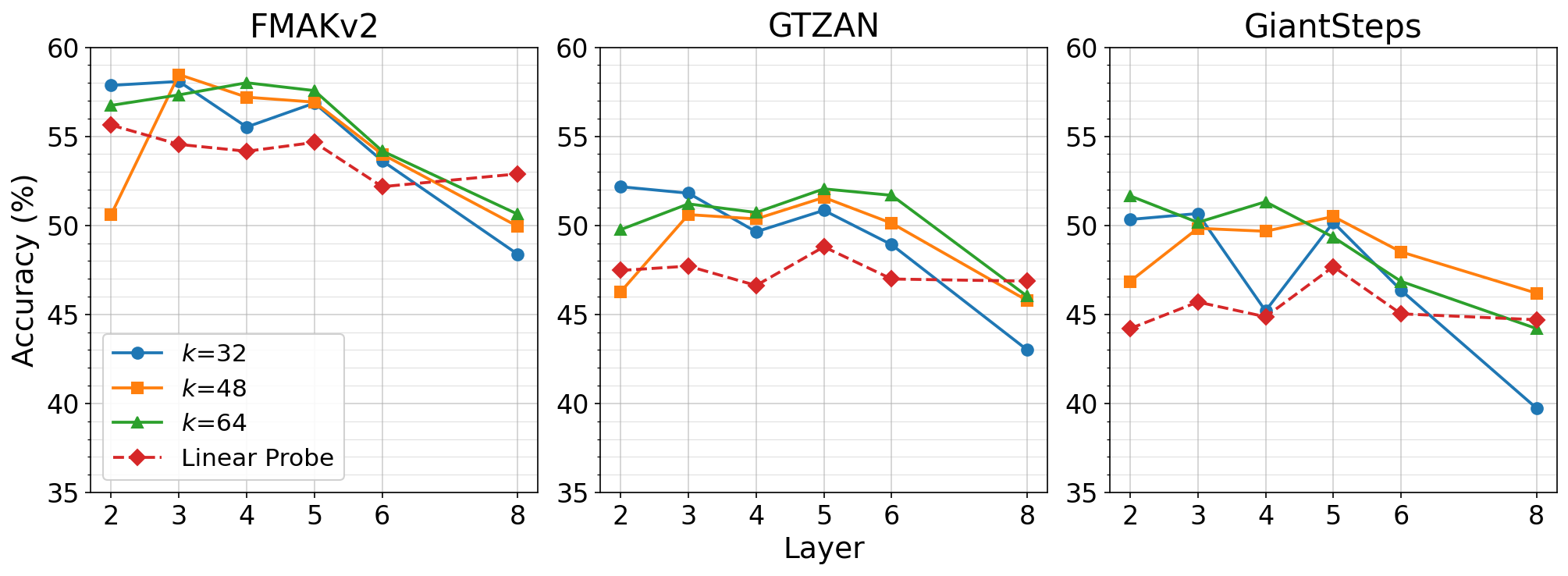}
    \caption{Key-signature accuracy of MuQ o-SAE.}
  \end{subfigure}
  \hfill
  \begin{subfigure}[t]{0.49\textwidth}
    \centering
    \includegraphics[width=\linewidth]{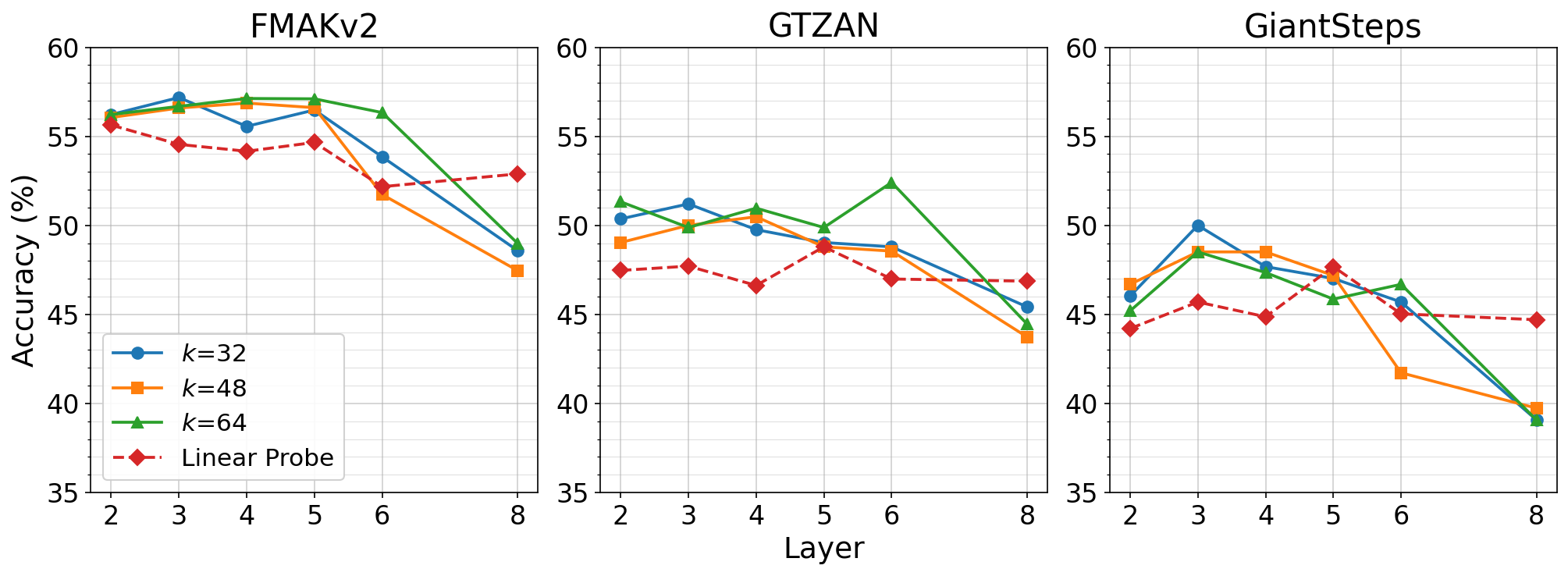}
    \caption{Key-signature accuracy of MuQ s-SAE.}
  \end{subfigure}

  \caption{
  Effect of Top-$K$ sparsity on layer-wise key-signature detection from
  subdominant features in MuQ. Results are shown for
  $K\in\{32,48,64\}$, with the linear probe included as a reference.
  }
  \label{fig:subdominant_key_topk}
\end{figure}

\paragraph{Key Detection via the Subdominant Ring.}
Figure~\ref{fig:subdominant_key_topk} shows key-signature detection from
subdominant features for MuQ o-SAE and s-SAE. We report the three values of
$K$ separately across model layers, together with the corresponding
subdominant probe as a reference.

 

\begin{table}[t]
  \centering
  \small
\renewcommand{\arraystretch}{0.88}
  \setlength{\tabcolsep}{4.5pt}
  \caption{Performance of the model trained on POP909}\label{tab:pop909_performance}
  
  \begin{subtable}[t]{0.43\textwidth}
    \centering
    \caption{Subdominant orbit: song-level key accuracy}
\label{tab:subdominant_orbit}
    \begin{tabular}{lc}
      \toprule
      \textbf{Dataset} & \textbf{Accuracy (\%)} \\
      \midrule
      GTZAN & 54.81 \\
      GiantSteps & 49.83 \\
      FMAKv2 & 59.26 \\
      \bottomrule
    \end{tabular}
  \end{subtable}
  \begin{subtable}[t]{0.56\textwidth}
    \centering
    \caption{Major- and minor-chord orbits: WCSR}
\label{tab:chord_orbits}
    \begin{tabular}{lcccc}
      \toprule
      & \multicolumn{2}{c}{\textbf{Oracle Bd}} & \multicolumn{2}{c}{\textbf{Est. Bd}} \\
      \cmidrule(lr){2-3} \cmidrule(lr){4-5}
      \textbf{Dataset} & \textbf{Root} & \textbf{Majmin} & \textbf{Root} & \textbf{Majmin} \\
      \midrule
      RWC & 86.58 & 75.60 & 79.40 & 69.37 \\
      Slakh2100 & 84.31 & 75.91 & 78.95 & 71.17 \\
      \bottomrule
    \end{tabular}
  \end{subtable}
\end{table}

\begin{figure}[t]
  \centering

    \includegraphics[width=0.7\linewidth]{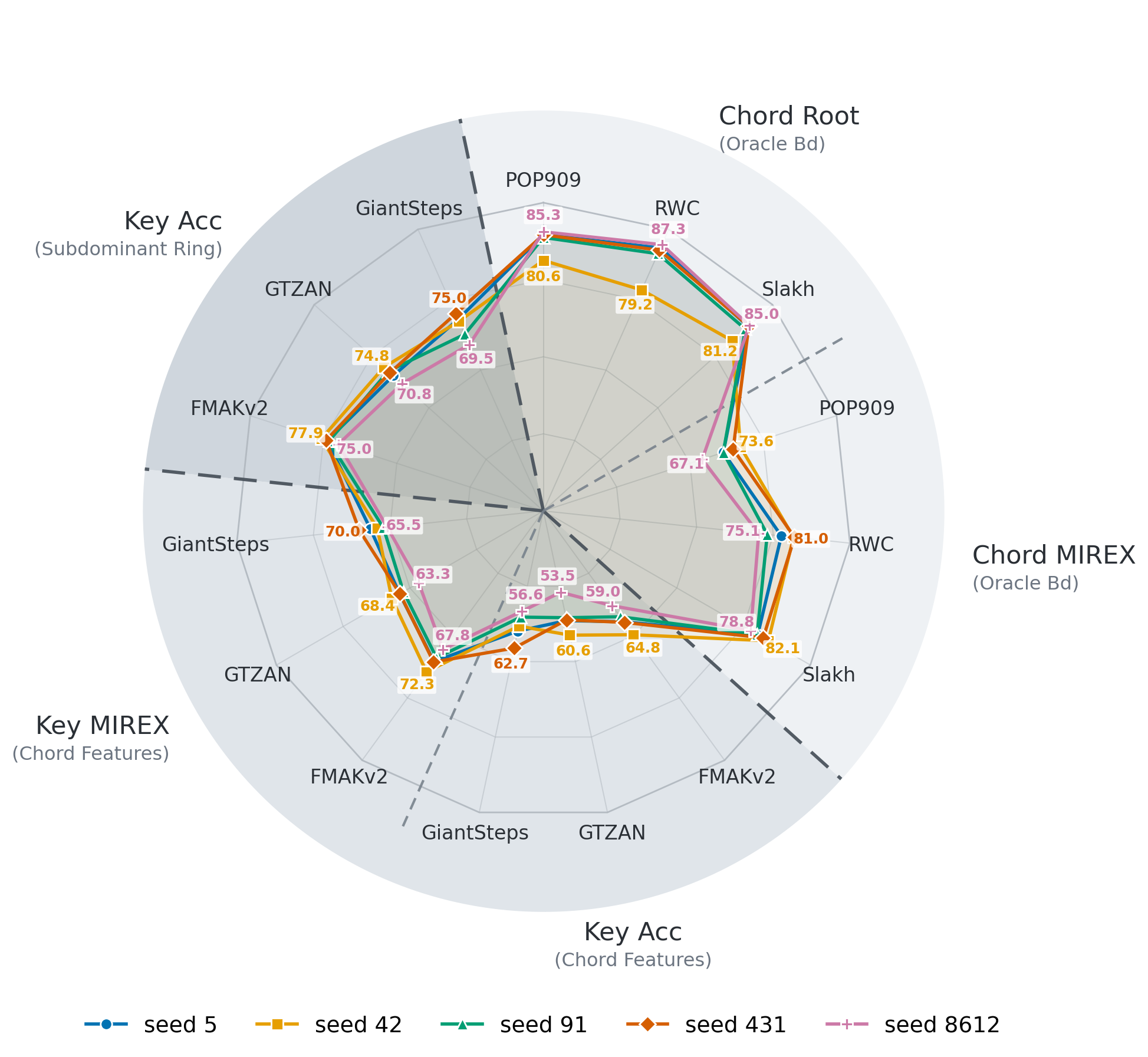}

  \caption{
  Downstream performance across random seeds for MuQ (2) o-SAE at $K=48$.
  }
  \label{fig:seed_robustness}
\end{figure}

\begin{figure}[t]
  \centering

  \begin{subfigure}[t]{0.49\textwidth}
    \centering
    \includegraphics[width=0.9\linewidth]{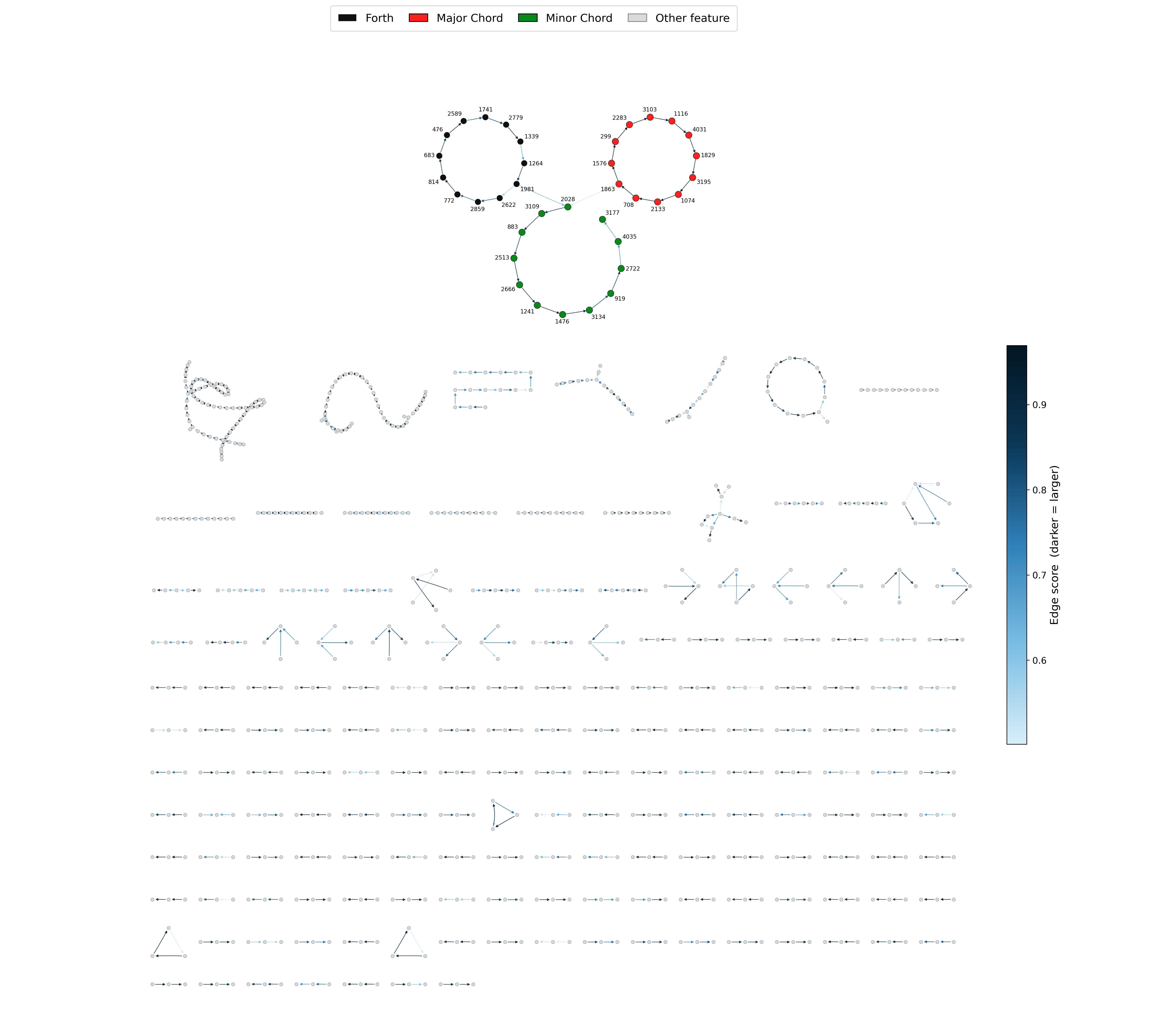}
    \caption{Random Seed 5.}

  \end{subfigure}
\hfill
  \begin{subfigure}[t]{0.49\textwidth}
    \centering
    \includegraphics[width=0.9\linewidth]{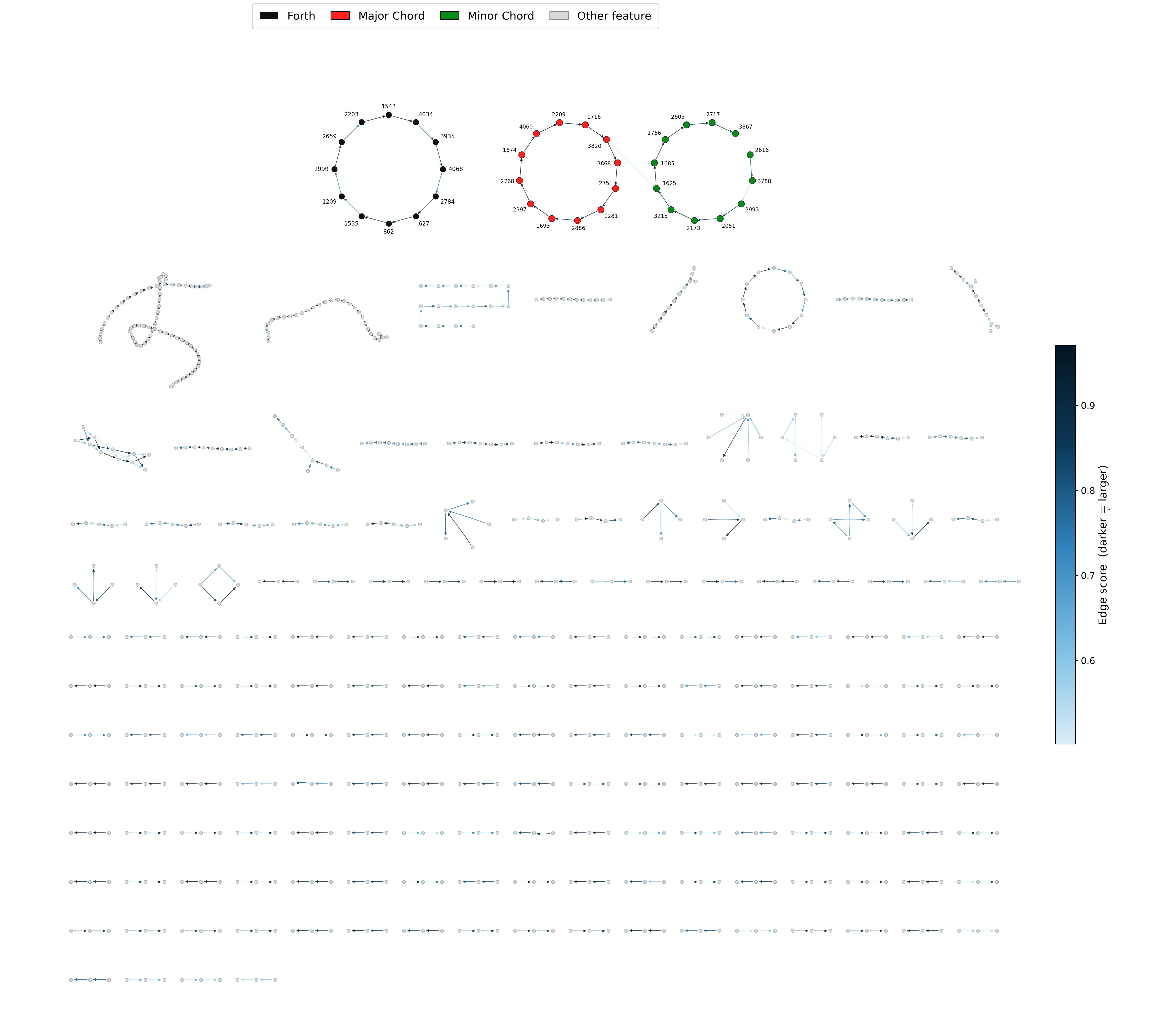}
    \caption{Random Seed 91.}
  \end{subfigure}

  \begin{subfigure}[t]{0.49\textwidth}
    \centering
    \includegraphics[width=0.9\linewidth]{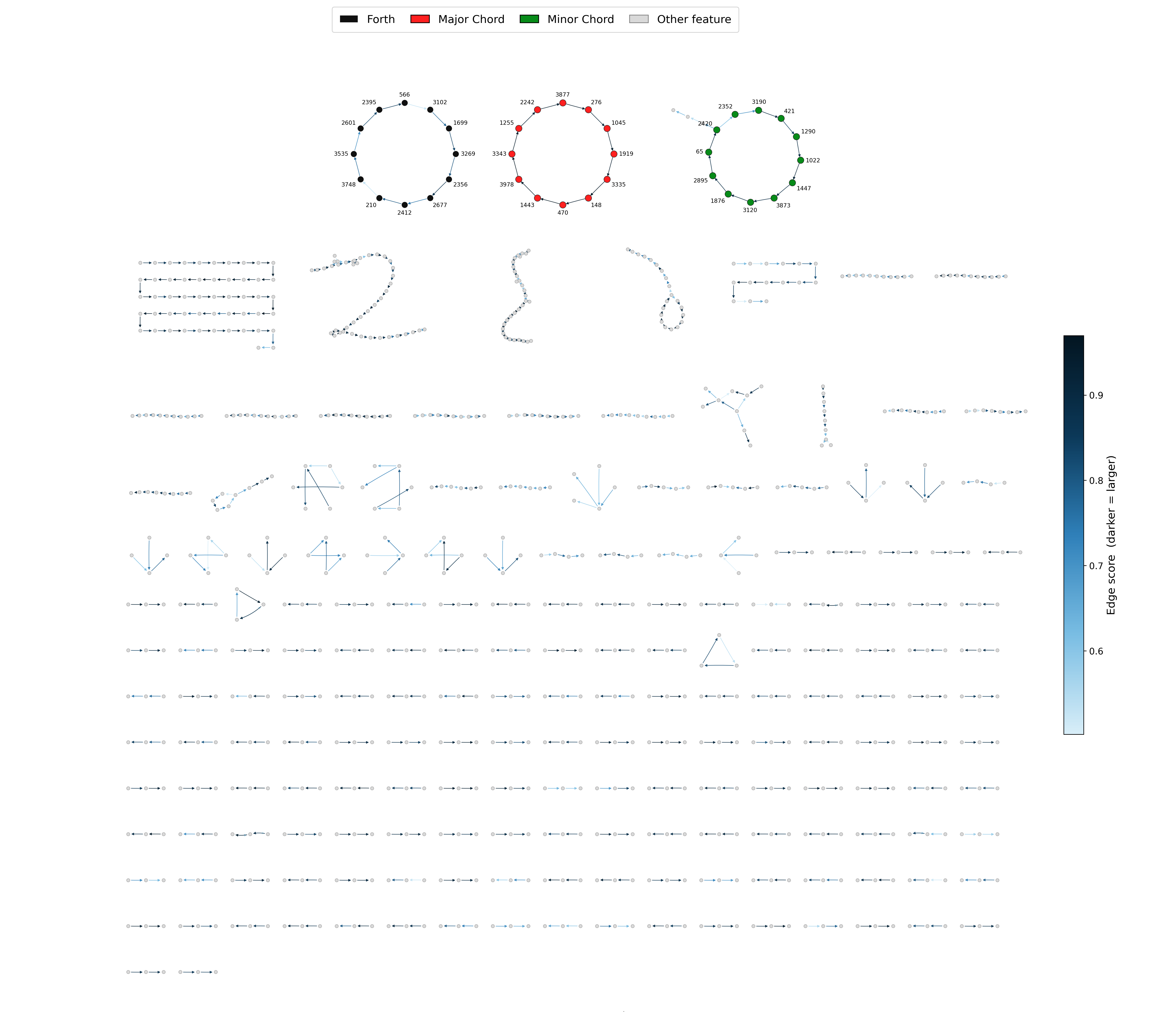}
    \caption{Random Seed 431.}
  \end{subfigure}
\hfill
  \begin{subfigure}[t]{0.49\textwidth}
    \centering
    \includegraphics[width=0.9\linewidth]{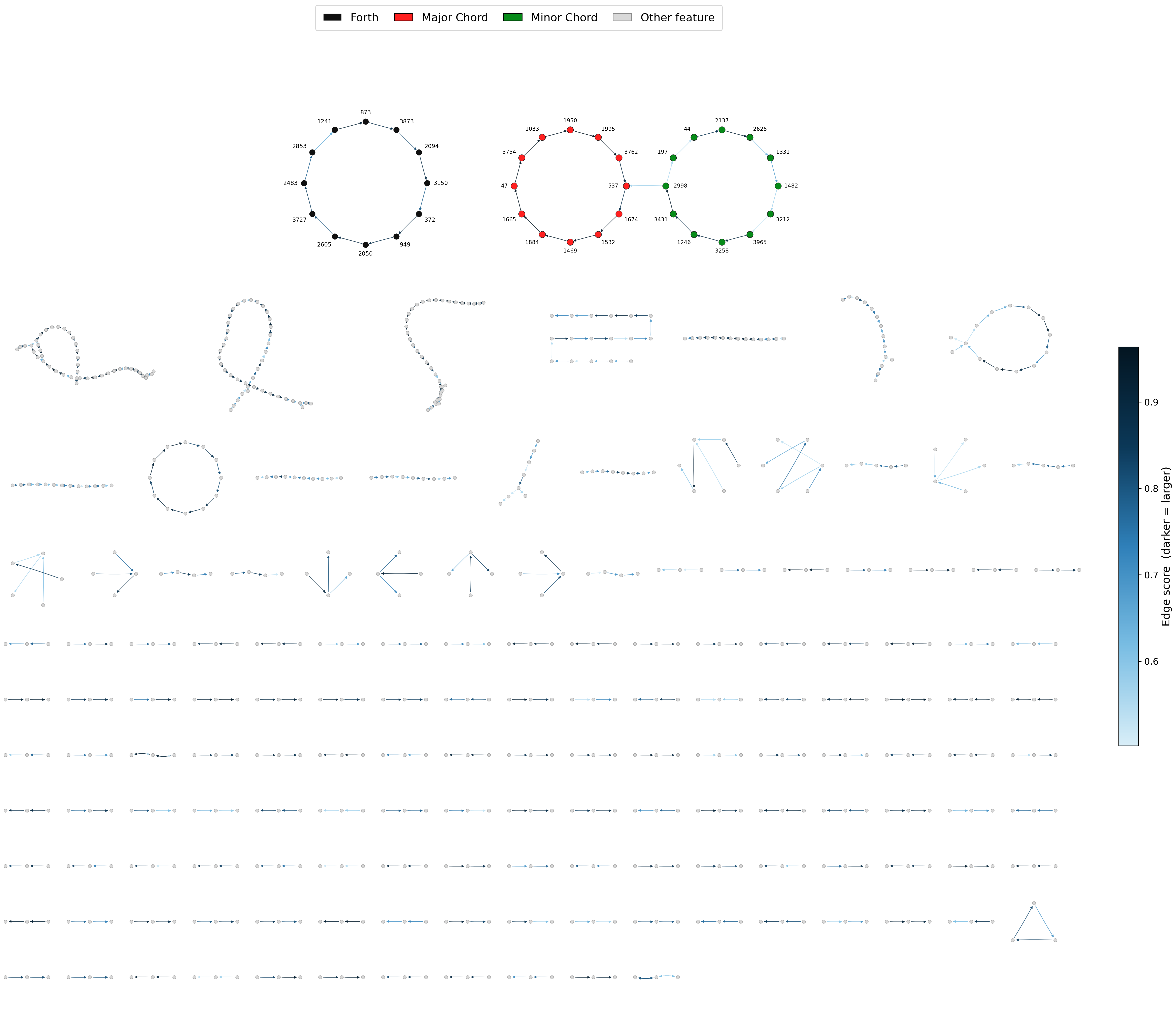}
    \caption{Random Seed 8612.}
  \end{subfigure}

\caption{
Directed orbit graphs recovered from MuQ across random seeds at $K=48$,
$n_e=10$, and $\tau=0.5$. Fixed points and components with fewer than three
nodes are omitted. All seeds are sampled randomly one-shot.
}
\label{fig:orbit_graphs_robust}
 
\end{figure}
  
\subsection{Training Dataset Ablation}
We additionally trained our aligned SAEs on a subset of real-world POP909 recordings, which contain vocals and complex production acoustics.We again recover the three core orbits: major chord, minor chord, and subdominant, under a setting of MuQ layer 2 and $k=32$. The corresponding downstream results are shown in Table~\ref{tab:pop909_performance}. All three core orbits still emerge and achieve comparable downstream performance, supporting the robustness of our discovery procedure across SAE-training distributions.

\section{Robustness Analysis}
\label{app:robustness}

We further examine whether the discovered structures are robust to random
initialization and variations in musical style.

\subsection{Orbit-Aware Initialization}
\label{app:orbit_initialization}

Each SAE decoder is initialized with Kaiming-uniform weights and normalized
column-wise, while the corresponding encoder is initialized as the transpose
of the decoder. All sub-SAEs share a common pre-encoding bias, which is
initialized to the geometric median of sampled training representations.

For the multi-SAE model, the first sub-SAE serves as a randomly initialized
anchor. We apply the anchor SAE to aligned frames from the original and
pitch-shifted views. Suppose that anchor feature $i$ has the largest activation
for the original view and anchor feature $j$ has the largest activation for a
shifted view. The decoder vector of feature $j$ is then accumulated to
initialize feature $i$ in the corresponding shifted sub-SAE, weighted by the
geometric mean of their activation magnitudes. Features with insufficient
observations fall back to the corresponding anchor decoder vector with a small
random perturbation. Finally, all decoder vectors are normalized, and the
encoder weights are again initialized as their transposes.

This initialization does not assign fixed semantic identities to individual
features, since the anchor dictionary remains randomly initialized. Instead,
it establishes an initial correspondence among the sub-SAEs from paired
pitch-shifted observations. Note that this initialization is not necessary for orbit emergence; rather,
it improves the initial correspondence among sub-SAEs and reduces sensitivity
to random initialization. Consequently, feature IDs may differ across runs,
while the relative transposition structure is less sensitive to independent
random initialization.

\begin{figure}[t]
  \centering

  \begin{subfigure}[t]{\textwidth}
    \centering
    \includegraphics[width=\linewidth]{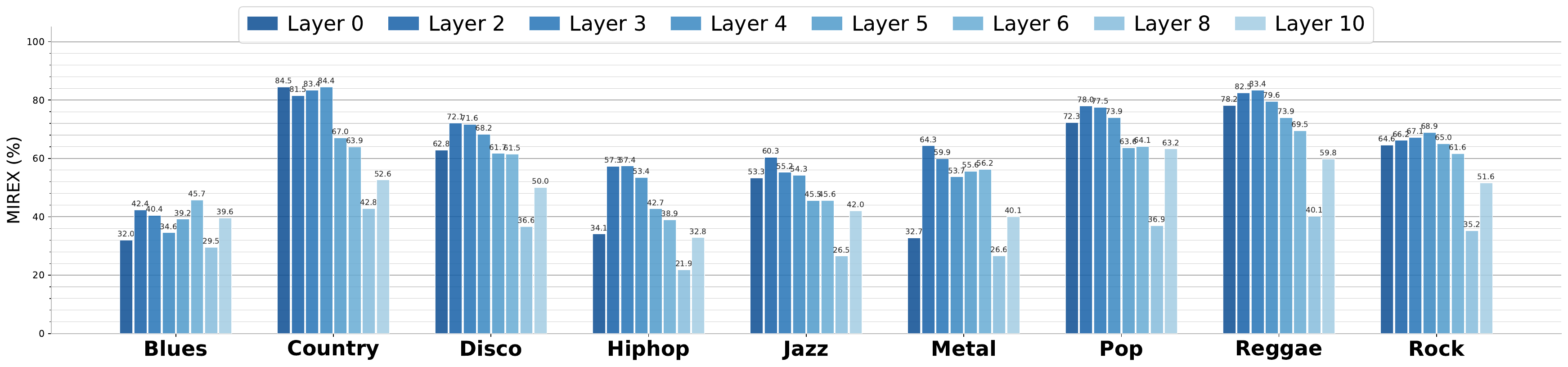}
    \caption{Chord-derived MIREX accuracy with MuQ.}
  \end{subfigure}

  \begin{subfigure}[t]{\textwidth}
    \centering
    \includegraphics[width=\linewidth]{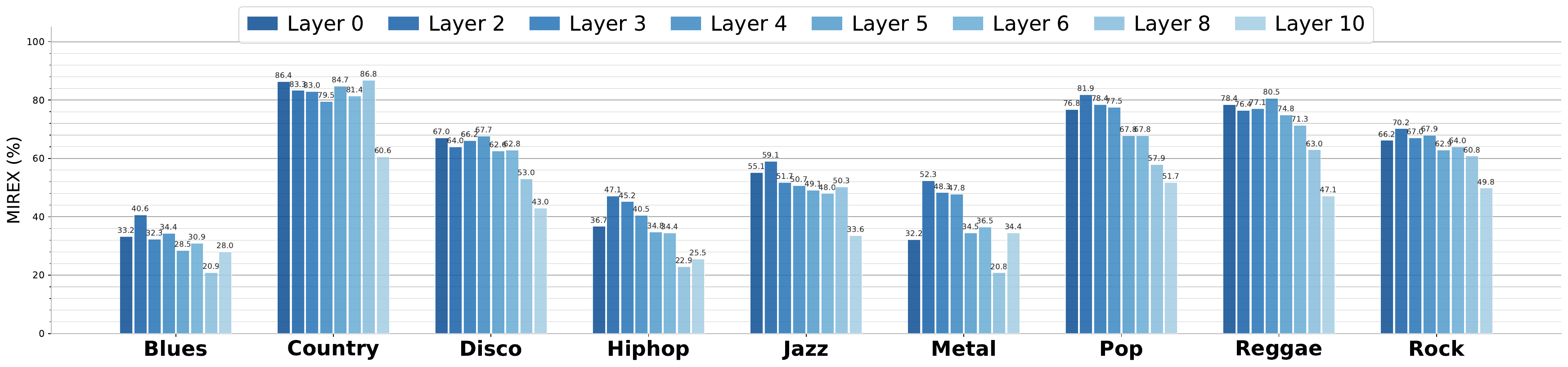}
    \caption{Chord-derived MIREX accuracy with MusicFM.}
  \end{subfigure}

  \begin{subfigure}[t]{\textwidth}
    \centering
    \includegraphics[width=\linewidth]{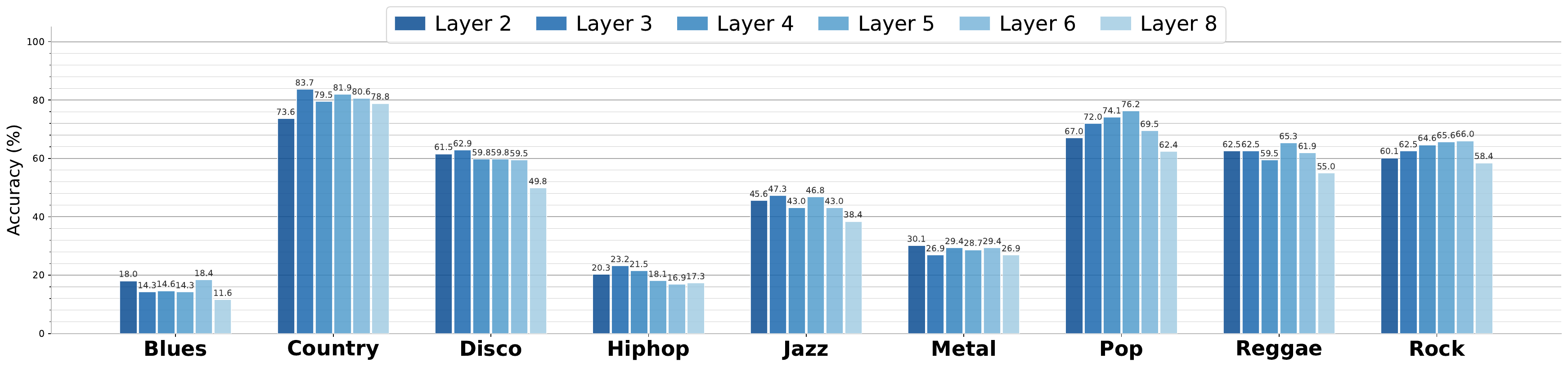}
    \caption{Key-signature accuracy from the subdominant ring with MuQ.}
  \end{subfigure}

  \caption{Genre-wise key-detection results on GTZAN-Key.}
  \label{fig:genre_robustness}
\end{figure}

\subsection{Robustness across Random Seeds}
\label{app:seed_robustness}

We train the same o-SAE configuration using multiple random seeds and compare
its downstream performance on chord recognition, chord-derived key detection,
and key-signature detection from the subdominant ring.
Figure~\ref{fig:seed_robustness} summarizes the results for MuQ layers 2 at $K=48$. Each polygon represents one random seed, and each axis corresponds
to one metric--dataset pair. The
similar performance profiles across seeds indicate that the downstream
utility of the discovered features is not tied to a particular
initialization.

We additionally examine structural robustness by visualizing the directed
orbit graphs recovered from four random seeds. As shown in
Figure~\ref{fig:orbit_graphs_robust}, the exact feature IDs and local graph
connections vary across runs, as expected under random initialization.
Nevertheless, the major-chord, minor-chord, and subdominant structures remain
identifiable in the graphs recovered with $n_e=10$ and $\tau=0.5$. This
suggests that the core orbit semantics are reproducible up to feature
permutation, even when the strict one-edge recovery procedure does not yield
an identical graph for every seed.

\begin{figure}[p]
  \centering

  \begin{subfigure}[t]{0.32\textwidth}
    \centering
\includegraphics[width=0.8\linewidth]{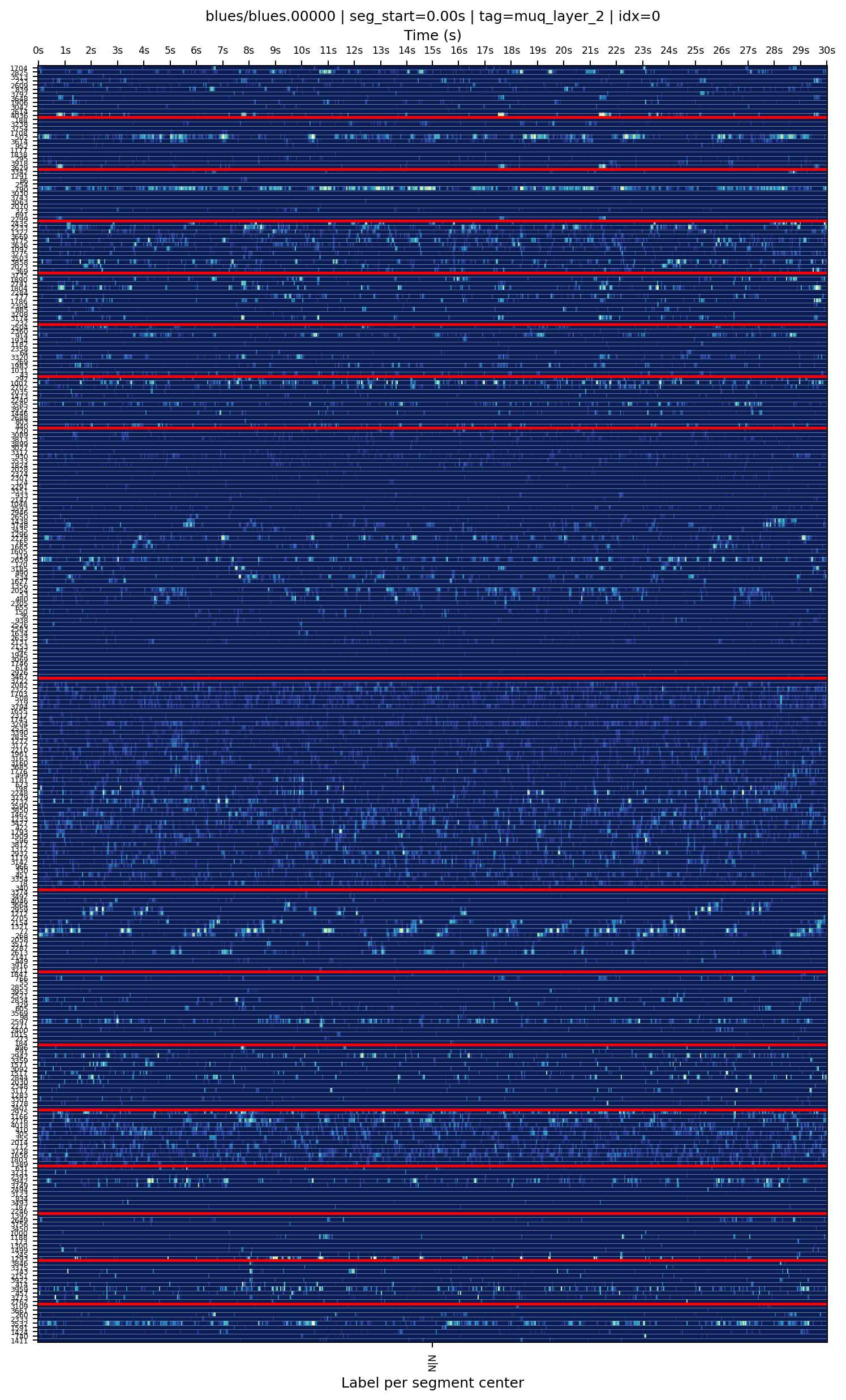}
    \caption{Blues.}
  \end{subfigure}
\hfill
  \begin{subfigure}[t]{0.32\textwidth}
    \centering
\includegraphics[width=0.8\linewidth]{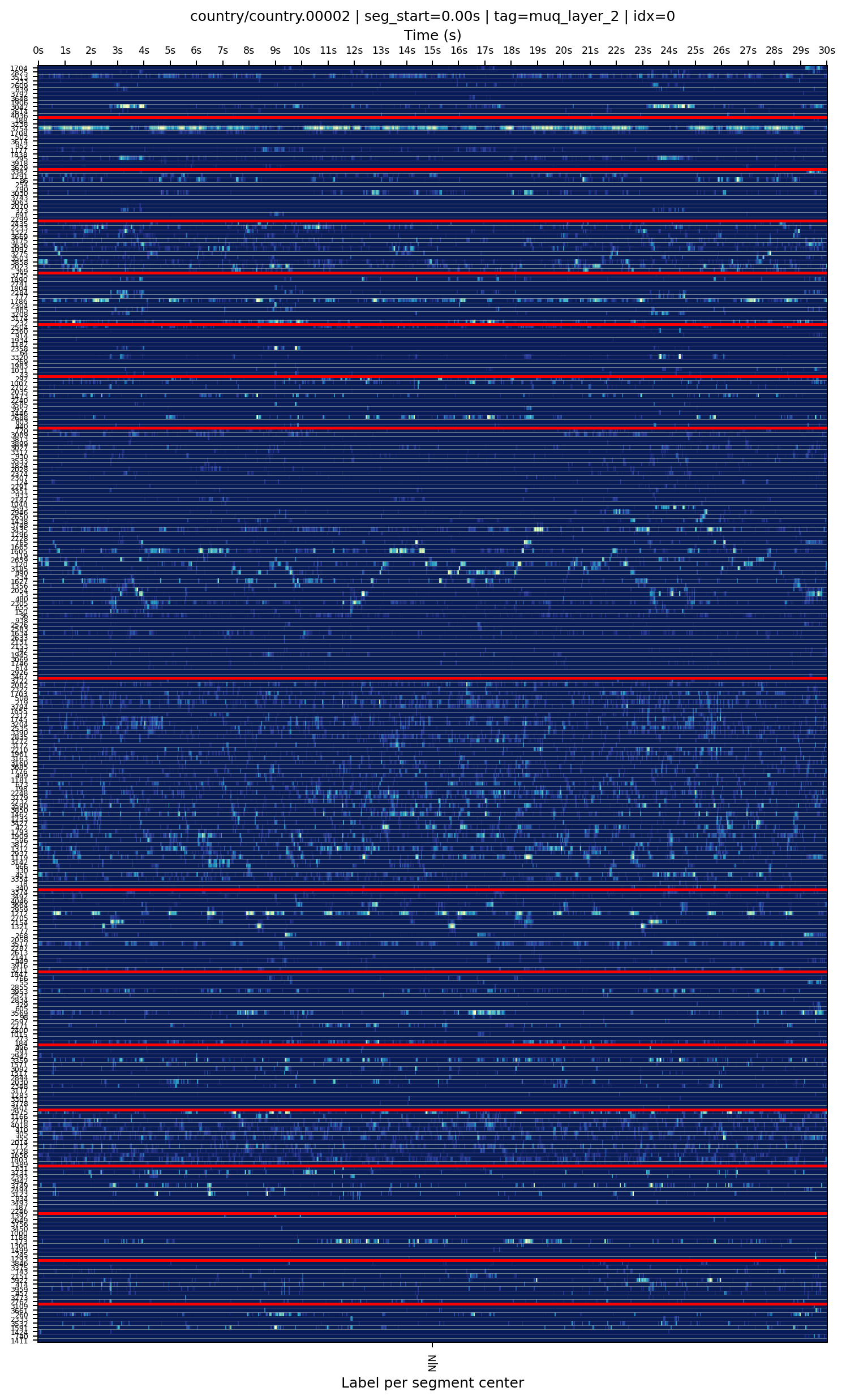}
    \caption{Country.}
  \end{subfigure}
\hfill
 \begin{subfigure}[t]{0.32\textwidth}
    \centering
\includegraphics[width=0.8\linewidth]{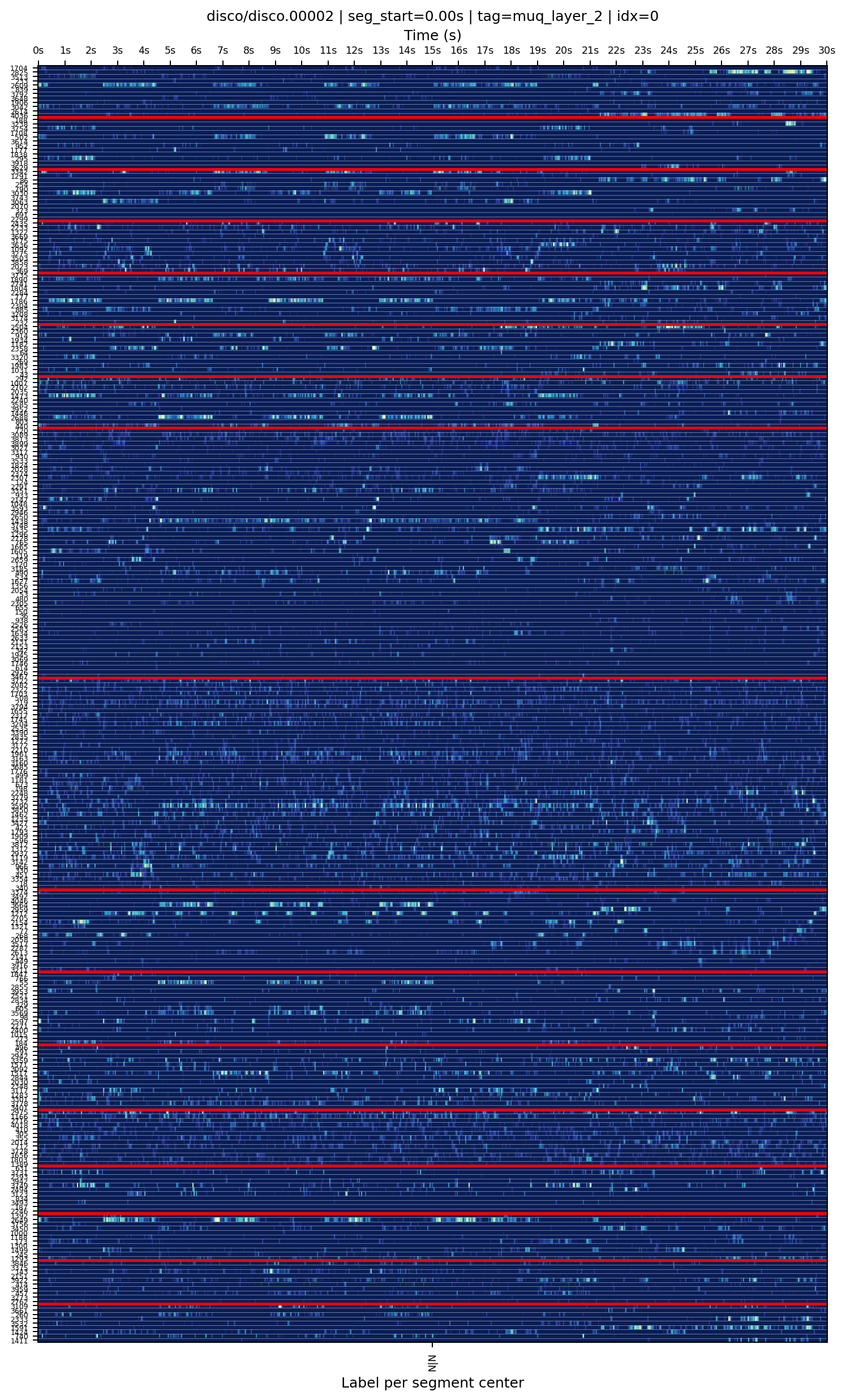}
    \caption{Disco.}
  \end{subfigure}
\begin{subfigure}[t]{0.32\textwidth}
    \centering
\includegraphics[width=0.8\linewidth]{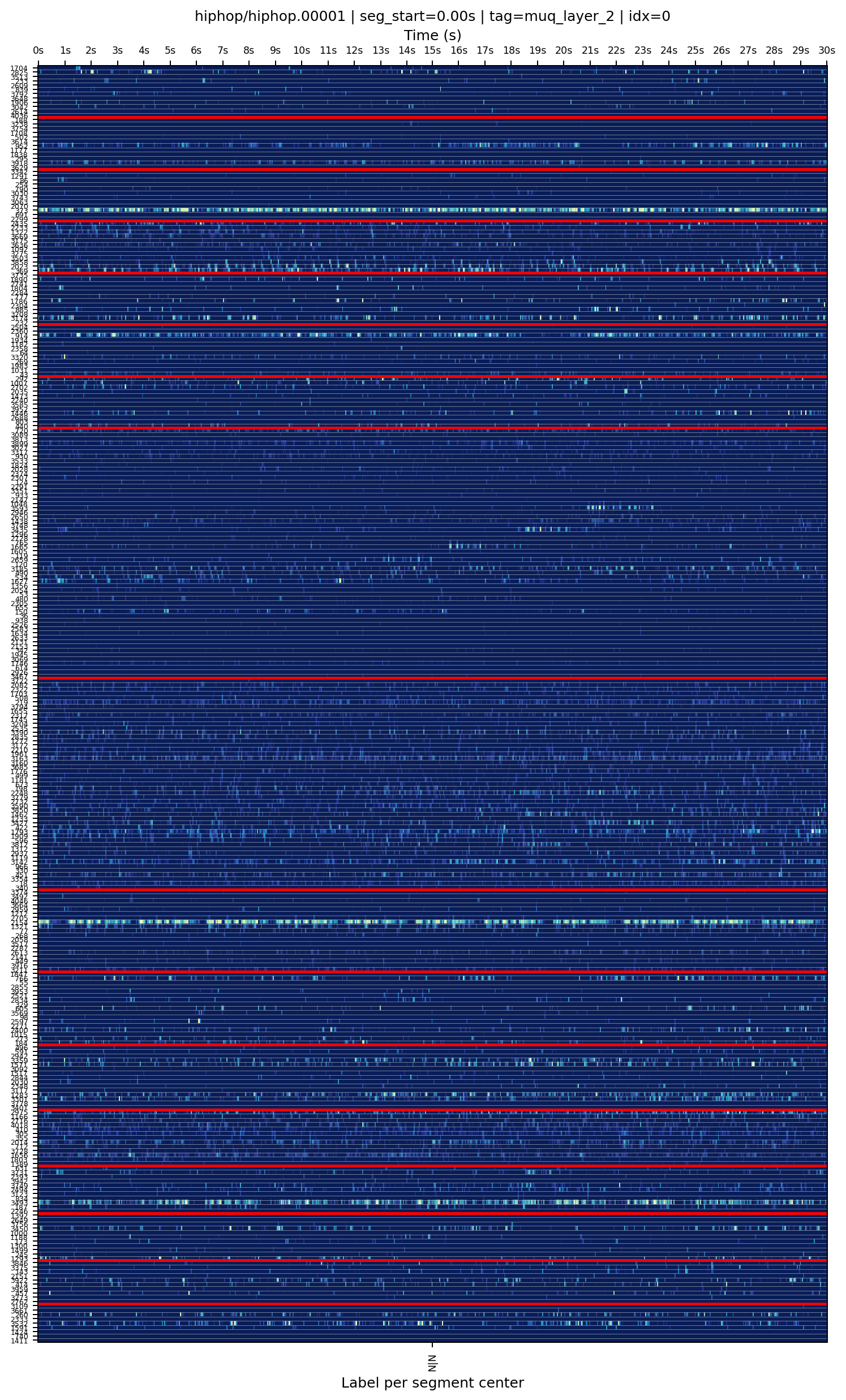}
    \caption{Hiphop.}
  \end{subfigure}
\hfill
  \begin{subfigure}[t]{0.32\textwidth}
    \centering
\includegraphics[width=0.8\linewidth]{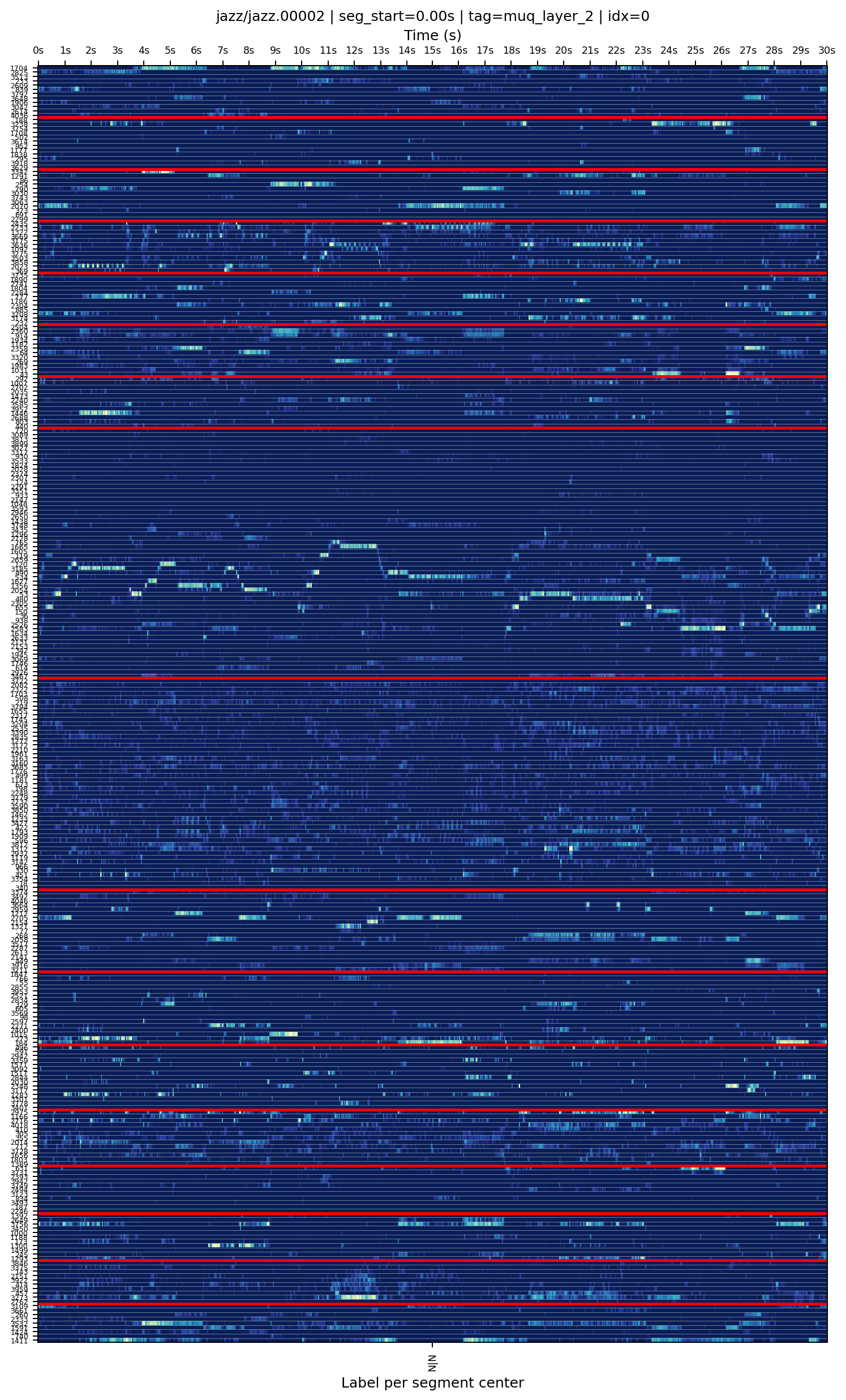}
    \caption{Jazz.}
  \end{subfigure}
\hfill
 \begin{subfigure}[t]{0.32\textwidth}
    \centering
\includegraphics[width=0.8\linewidth]{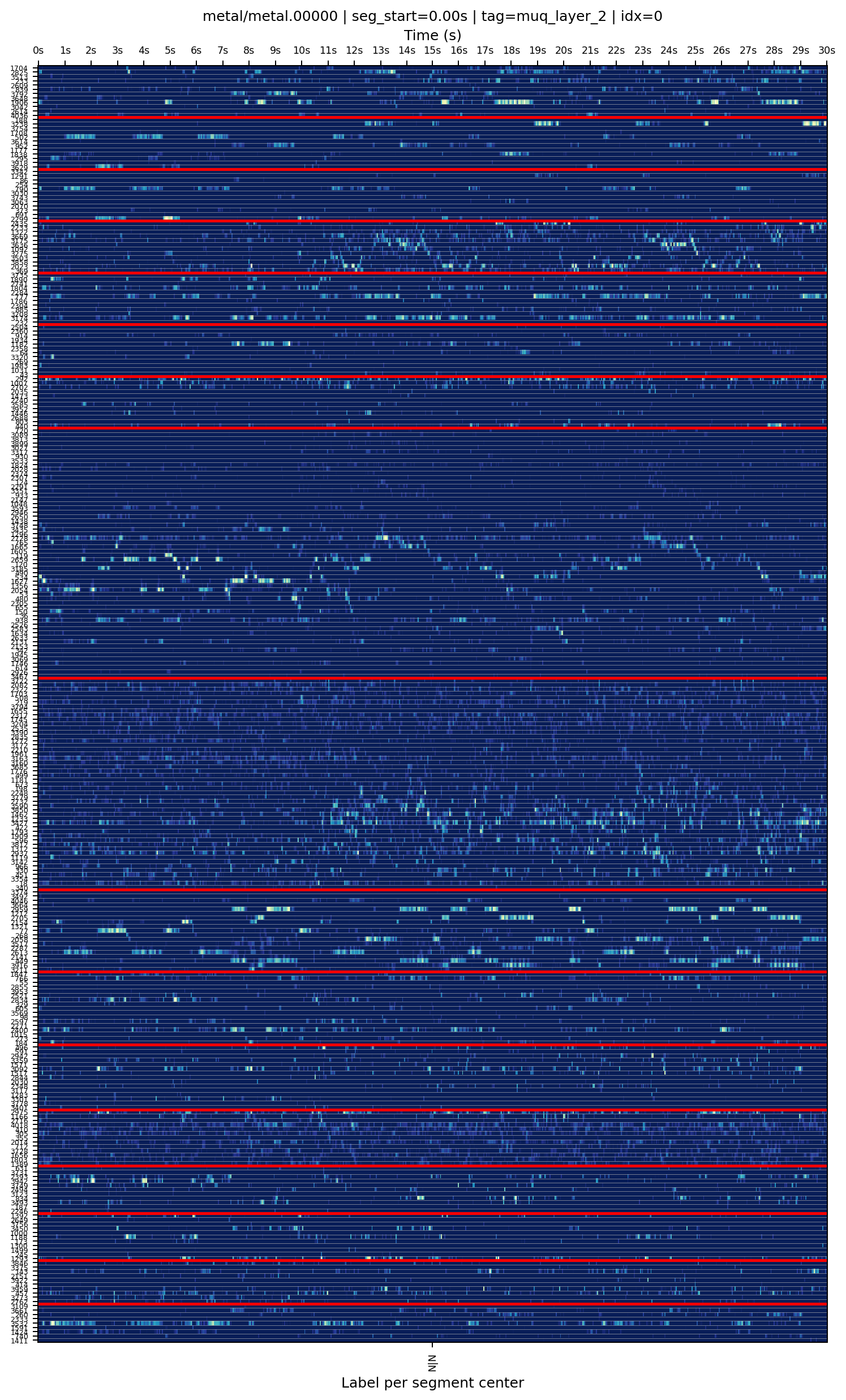}
    \caption{Metal.}
  \end{subfigure}

  \begin{subfigure}[t]{0.32\textwidth}
    \centering
\includegraphics[width=0.8\linewidth]{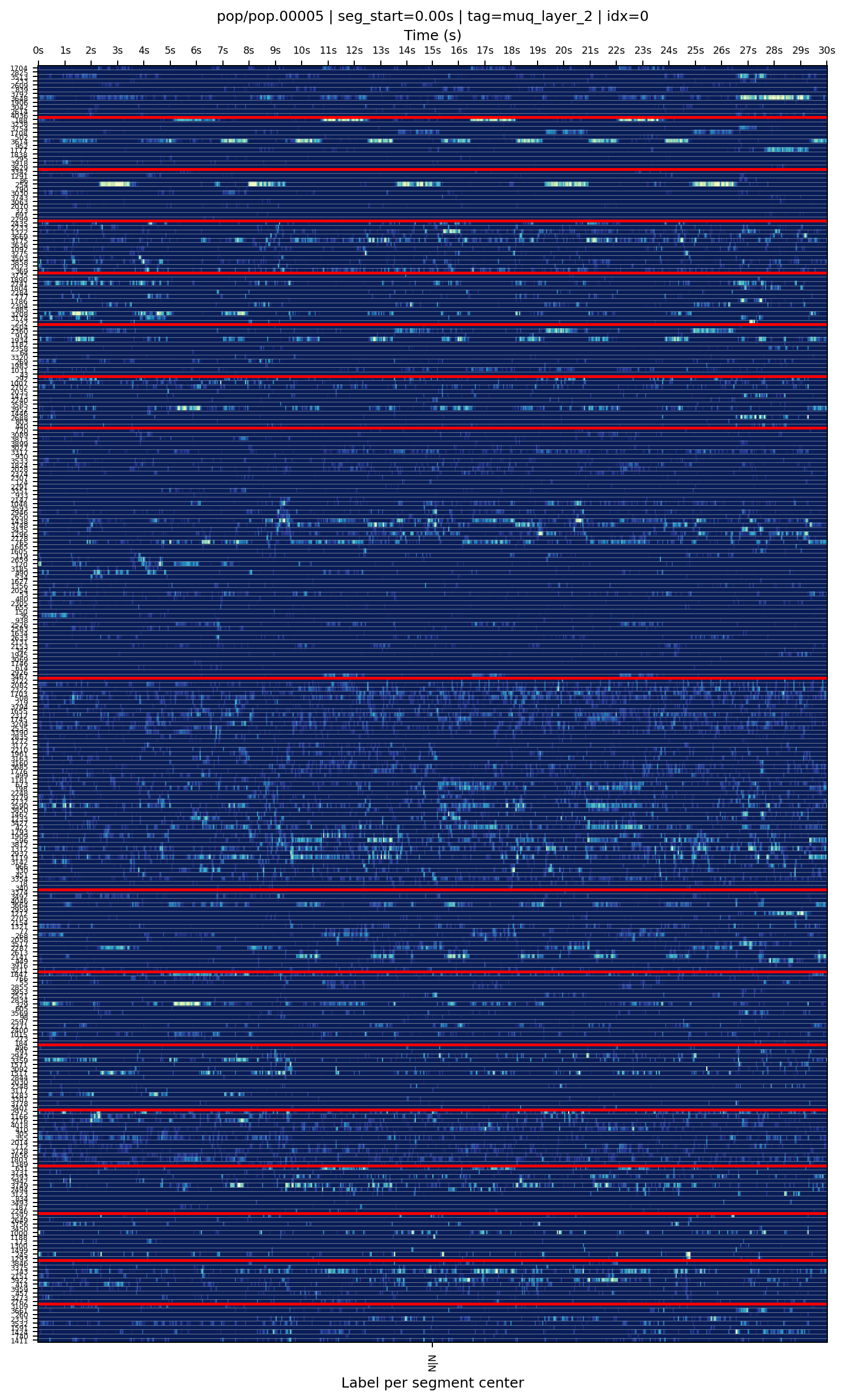}
    \caption{Pop.}
  \end{subfigure}
\hfill
  \begin{subfigure}[t]{0.32\textwidth}
    \centering
\includegraphics[width=0.8\linewidth]{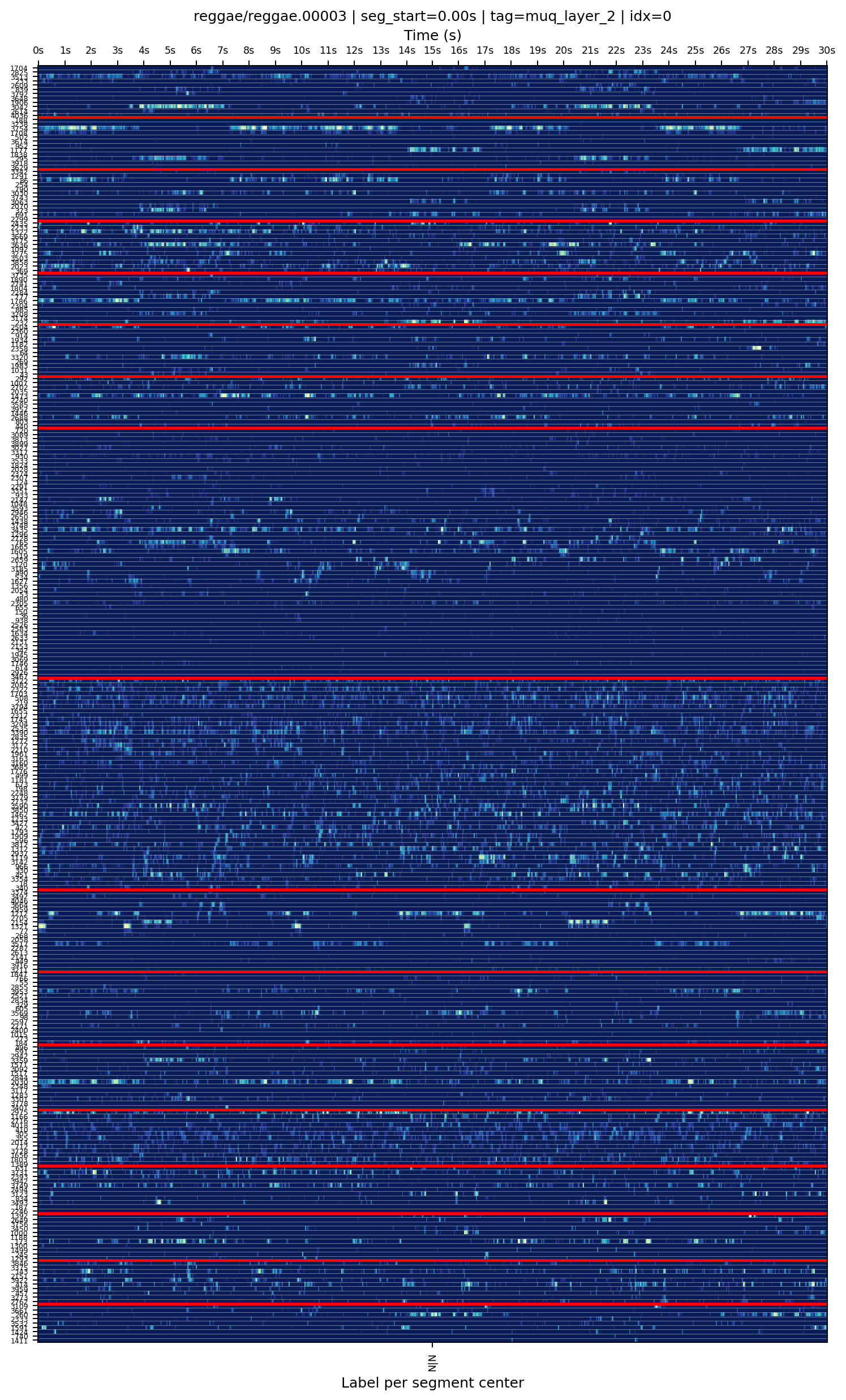}
    \caption{Reggae.}
  \end{subfigure}
\hfill
 \begin{subfigure}[t]{0.32\textwidth}
    \centering
\includegraphics[width=0.8\linewidth]{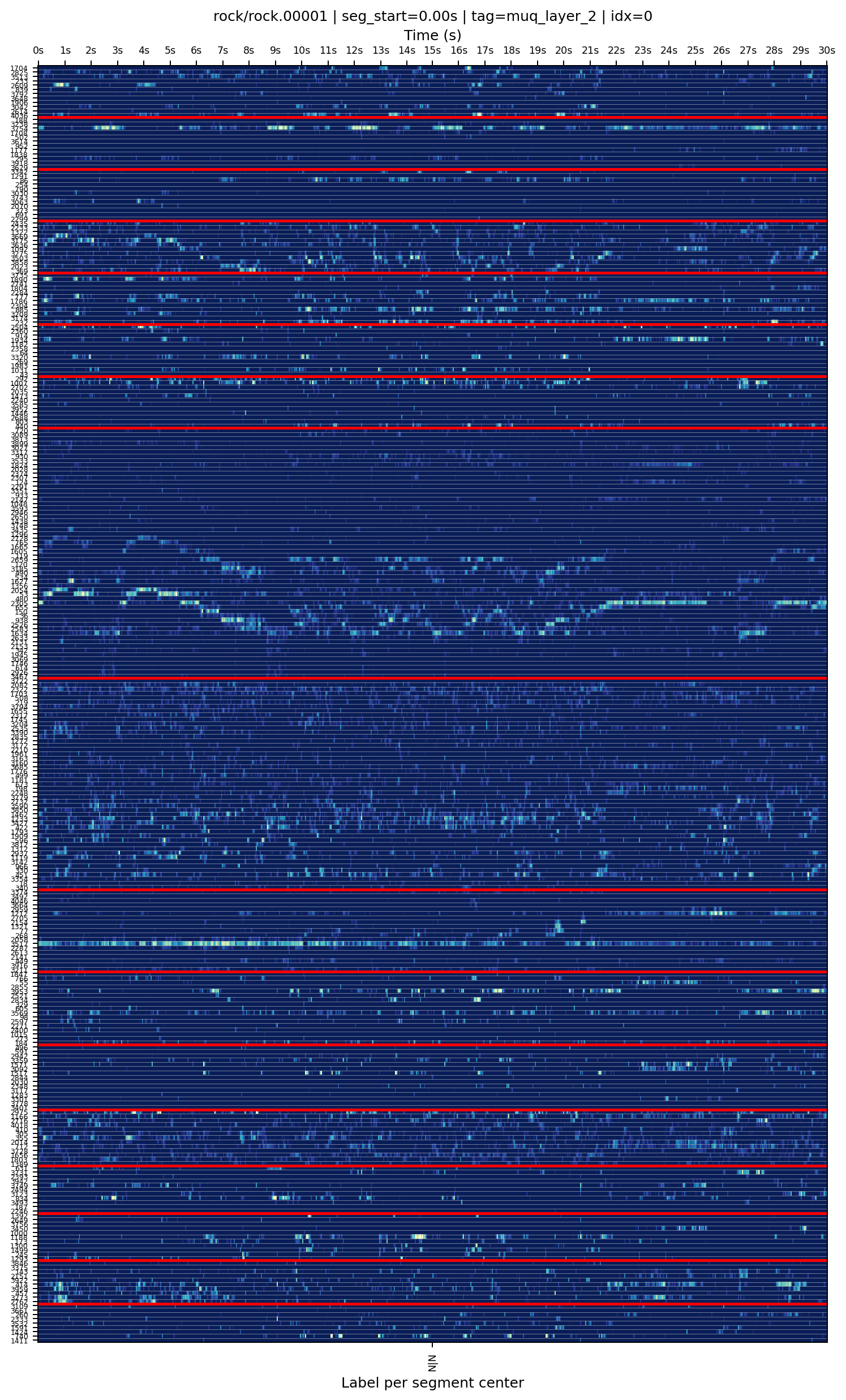}
    \caption{Rock.}
  \end{subfigure}

\caption{
Activations of songs across different genres.
}
\label{fig:activation_genres}
 
\end{figure}

\subsection{Genre Sensitivity}
\label{app:genre_robustness}

We examine whether the discovered structures remain useful across different
musical styles. We group the GTZAN-Key predictions by genre and compute the
evaluation metrics independently for each group. We keep the same trained models,
recovered orbits, and inference settings for all genres, without
genre-specific feature selection or parameter tuning.

Figure~\ref{fig:genre_robustness} reports chord-derived MIREX accuracy for MuQ
and MusicFM, together with key-signature accuracy obtained from the MuQ
subdominant ring. The recovered structures remain informative across all 9
GTZAN genres, although performance varies. Blues shows the
lowest overall performance, whereas pop, country, and reggae generally achieve
strong results. 

To complement these aggregate results, Figure~\ref{fig:activation_genres}
visualizes frame-level activations for excerpts from Gtzan Dataset. All examples use the same MuQ (2) o-SAE with $K=32$ and the same
orbits recovered at $\tau=0.5$. We found that the activations of blues and hip-hop are harder to interpret than other genres, which is consistent with the key detection performance. Together, the quantitative and
qualitative results suggest that the recovered orbit semantics generalize
across musical styles, while remaining sensitive to genre-dependent acoustic
and harmonic complexity.


\end{document}